\documentclass[aps,floats]{revtex4}
\usepackage{amsmath,amssymb}
\usepackage{graphicx,epsfig}
\usepackage[greek,english]{babel}
\usepackage{bbold}

\begin{document}
\bibliographystyle {plain}

\pdfoutput=1
\def\oppropto{\mathop{\propto}} 
\def\opsimeq{\mathop{\simeq}}
\def\opoverderline{\mathop{\overline}}
\def\operarrow{\mathop{\longrightarrow}}
\def\opsim{\mathop{\sim}}

\def\fig#1#2{\includegraphics[height=#1]{#2}}
\def\figx#1#2{\includegraphics[width=#1]{#2}}


\title{ Large deviations of the Lyapunov exponent in 2D matrix Langevin dynamics  \\
  with applications to one-dimensional Anderson Localization models } 


\author{ C\'ecile Monthus }
 \affiliation{Institut de Physique Th\'{e}orique, 
Universit\'e Paris Saclay, CNRS, CEA,
91191 Gif-sur-Yvette, France}

\begin{abstract}
For the 2D matrix Langevin dynamics that corresponds to the continuous-time limit of the product of some $2 \times 2$ random matrices, the finite-time Lyapunov exponent can be written as an additive functional of the associated Riccati process submitted to some Langevin dynamics on the infinite periodic ring. Its large deviations properties can be thus analyzed from two points of view that are equivalent in the end by consistency but give different perspectives. In the first approach, one starts from the large deviations at level 2.5 for the joint probability of the empirical density and of the empirical current of the Riccati process and one performs the appropriate Euler-Lagrange optimization in order to compute the cumulant generating function of the Lyapunov exponent. In the second approach, this cumulant generating function is obtained from the spectral analysis of the appropriate tilted Fokker-Planck operator. The associated conditioned process obtained via the generalization of Doob's h-transform allows to clarify the equivalence with the first approach. Finally, applications to one-dimensional Anderson Localization models are described in order to obtain explicitly the first cumulants of the finite-size Lyapunov exponent.

\end{abstract}

\maketitle


\section{Introduction   }

Products of random matrices and their continuous counterparts play a major role 
in Probability Theory and in Statistical Physics, 
with important applications in the field of random dynamical systems and 
in the field of disordered systems, either classical or quantum
(see \cite{bougerol,lifbook,luckbook,crisantibook,tourigny_scatter,tourigny_Lyapunov,tourigny_continuous,tourigny_houches,comets,c_largedevdisorder} and references therein).
The Lyapunov exponent that measures their exponential growth is an essential observable.
Since its typical value appears with probability one in the thermodynamical limit, 
the main goal has been usually to compute this typical value
in various models via the Dyson-Schmidt invariant measure method.
However, more recent works have analyzed in great detail 
the finite-size fluctuations of the Lyapunov exponent via its first cumulants
or via its full large deviations properties,
either for dynamical models with noise  \cite{titov,zillmer,pumir},
 for Anderson Localization models \cite{titovLloyd,ramola,fyodorov},
 and for products of random matrices \cite{vanneste},
 in particular for random matrices of the group $SL(2,R)$ \cite{texier_jstat,texier_epl,texier_comtet}
 that are related to various Localization models.

In the present paper, we consider the 2D matrix Langevin dynamics that corresponds to the 
continuous-time limit of the product of some $2 \times 2$ random matrices.
Since the Lyapunov exponent can be rewritten as an additive functional 
of the associated Riccati process,
it is interesting to revisit the problem of its large deviations 
 in relation with the recent progresses made in the field of 
large deviations for non-equilibrium stochastic processes
(see the reviews with different scopes \cite{derrida-lecture,harris_Schu,searles,harris,mft,sollich_review,lazarescu_companion,lazarescu_generic,jack_review}, the PhD Theses \cite{fortelle_thesis,vivien_thesis,chetrite_thesis,wynants_thesis} 
 and the HDR Thesis \cite{chetrite_HDR}).
 The theory of large deviations (see the reviews \cite{oono,ellis,review_touchette} and references therein)
 is usually based on the classification into three levels, namely Level 1 for empirical observables, Level 2 for the empirical measure, and Level 3 for the empirical process. 
 More recently for non-equilibrium Markovian processes, the intermediate 'Level 2.5' 
 concerning the joint distribution of the empirical measure and of the empirical flows
 has emerged as the simplest level
 where one can write explicit expressions for the large deviations.
This large deviation analysis at Level 2.5 has been applied to various settings,
in particular Markov Chains (discrete-space and discrete-time) \cite{fortelle_thesis,fortelle_chain,review_touchette,c_largedevdisorder,c_reset},
Markov Jump processes (discrete-space and continuous-time)
\cite{fortelle_thesis,fortelle_jump,maes_canonical,maes_onandbeyond,wynants_thesis,chetrite_formal,BFG1,BFG2,chetrite_HDR,c_ring,c_interactions,c_open,barato_periodic,chetrite_periodic,c_reset}
and Diffusion processes (continuous-space and continuous-time) 
\cite{wynants_thesis,maes_diffusion,chetrite_formal,engel,chetrite_HDR,c_reset},
while the recent generalization to the Lindblad dynamics 
\cite{previousquantum2.5doob,quantum2.5doob} offers new possibilities for open quantum systems.
Since any time-additive observable of the dynamical trajectory
 can be reconstructed via its decomposition in terms of the empirical density and of the empirical flows,
 its large deviation properties can be obtained via the appropriate contraction of the large deviations at level 2.5 for the empirical density and flows.
Another standard method to analyze time-additive observables of stochastic processes
goes back to the famous Feynman-Kac formula \cite{feynman,kac,c_these,review_maj}
and consists in studying the appropriate 'tilted' dynamical process.
This approach has been used extensively in the field of non-equilibrium processes recently 
 \cite{derrida-lecture,sollich_review,lazarescu_companion,lazarescu_generic,jack_review,vivien_thesis,lecomte_chaotic,lecomte_thermo,lecomte_formalism,lecomte_glass,kristina1,kristina2,jack_ensemble,simon1,simon2,simon3,Gunter1,Gunter2,Gunter3,Gunter4,chetrite_canonical,chetrite_conditioned,chetrite_optimal,chetrite_HDR,touchette_circle,touchette_langevin,touchette_occ,touchette_occupation,derrida-conditioned,derrida-ring,bertin-conditioned,touchette-reflected,previousquantum2.5doob,quantum2.5doob},
 with the formulation of the corresponding 'conditioned' process via the generalization of Doob's h-transform.
Even if these two approaches to analyze additive functionals 
have to be equivalent in the end by consistency,
 they provide different perspectives, and it will be thus interesting to consider
  both methods to characterize the large deviations properties of the finite-time Lyapunov exponent
  in 2D matrix Langevin dynamics.

The paper is organized as follows.
In section \ref{sec_matrixLangevin}, we introduce the 2D matrix Langevin dynamics
and its finite-time Lyapunov exponent that can be rewritten as an additive functional of the Riccati process.
In section \ref{sec_steady}, we describe the steady state of the Riccati process
that allows to compute the typical Lyapunov exponent.
We then turn to the study of large deviations properties.
In section \ref{sec_optimization}, we describe the analysis by contraction
from the large deviations at level 2.5 of the Riccati process.
In section \ref{sec_tilted}, the link with the alternative analysis in terms of the tilted dynamics
and of the conditioned dynamics
is discussed in detail in order to obtain another interesting perspective.
Finally, we apply this large deviation analysis to the Lyapunov exponent
in one-dimensional Anderson localization models with random scalar potential
(section \ref{sec_lochalperin}) and with random supersymmetric potential (section \ref{sec_locsusy}).
Our conclusions are summarized in section \ref{sec_conclusion}.
In the appendices, the Euler-Lagrange optimization procedure described in section \ref{sec_optimization}
is solved perturbatively to obtain explicitly the first cumulants, 
both when the Riccati steady state is 
a non-equilibrium state with current (Appendix \ref{sec_pernoneq})
and when the steady state is an equilibrium state without current  (Appendix \ref{sec_pereq}).


\section{Lyapunov exponent for two-dimensional matrix Langevin dynamics   }

\label{sec_matrixLangevin}

In this section, we introduce the notions and the notations that will be useful in the whole paper.
The two-dimensional matrix Langevin dynamics is defined in cartesian coordinates in subsection \ref{sub_carte},
translated into polar coordinates in subsection \ref{sub_polar}, while the Riccati variable is introduced in subsection
\ref{sub_ric}. Then we discuss the finite-time Lyapunov exponent $\lambda_T$ in subsection \ref{sub_lyap}, 
the other Lyapunov exponent in subsection \ref{sub_other}, and the density of zeroes of the first cartesian component in subsection \ref{sub_dos}. Finally, the notations for the
large deviation analysis of the finite-time Lyapunov exponent $\lambda_T$ are introduced in subsection \ref{sub_large},
with the rate function $I(\lambda)$ and the scaled cumulant generating function $\mu(k)$, while their possible Gallavotti-Cohen
symmetry is discussed in subsection \ref{sub_GCsym}.

\subsection{ Cartesian coordinates $(y_1(t),y_2(t))$ }

\label{sub_carte}

We consider the following Langevin dynamics
for the two-dimensional vector of real components $y_1(t)$ and $y_2(t)$
\begin{eqnarray}
\frac{d}{dt}
\begin{pmatrix} 
 y_1(t) 
 \\  y_2(t) 
  \end{pmatrix}
  = \left( M + \eta(t) W  \right)
  \begin{pmatrix} 
 y_1(t) 
 \\  y_2(t) 
  \end{pmatrix}
= \begin{pmatrix} 
M_{11} + \eta(t) W_{11} &  M_{12} + \eta(t) W_{12} \\
M_{21} + \eta(t) W_{21} & M_{22} + \eta(t) W_{22} 
 \end{pmatrix} 
\begin{pmatrix} 
 y_1(t) 
 \\  y_2(t) 
  \end{pmatrix}
\label{langevinmatrix}
\end{eqnarray}
where $M$ and $W$ are two given $2 \times 2$ real matrices, while $\eta(t)$ is a Gaussian white noise
\begin{eqnarray}
<\eta(t)> && =0
\nonumber \\
<\eta(t)\eta(t') > && = \delta(t-t')
 \label{whitenoise}
\end{eqnarray}
Since the noise $\eta(t)$ is multiplicative in the stochastic differential Eq. \ref{langevinmatrix},
we need to specify that we will use the Stratonovich interpretation.

Eq. \ref{langevinmatrix} is interesting on its own as a first-order linear dynamical system perturbed by noise.
It can also appear as a reformulation of second-order differential equations in one dimension (see sections \ref{sec_lochalperin} and \ref{sec_locsusy} on Anderson localization models).
Finally, Eq. \ref{langevinmatrix} corresponds to the continuous-time limit of the product of some $2 \times 2$ random matrices $ \Omega_{p=1,2,..,n}$
\begin{eqnarray}
\begin{pmatrix} 
 y_1(n \Delta t) 
 \\  y_2(n \Delta t) 
  \end{pmatrix}
  = \Omega_n
  \begin{pmatrix} 
 y_1((n-1) \Delta t) 
 \\  y_2((n-1) \Delta t) 
  \end{pmatrix}
  =  \Omega_n  \Omega_{n-1} ... \Omega_2  \Omega_{1}
  \begin{pmatrix} 
 y_1(0) 
 \\  y_2(0) 
  \end{pmatrix}
\label{matrixproduct}
\end{eqnarray}


\subsection{ Polar coordinates $(e^{\xi(t)}, \theta(t))$ }

\label{sub_polar}

Via the change of variables towards polar coordinates with modulus $e^{\xi(t)} = \sqrt{y_1^2(t)+y_2^2(t)}$ and angle $\theta(t)$
\begin{eqnarray}
y_1(t)+i y_2(t) && \equiv e^{ \xi(t) + i \theta(t) } = e^{\xi(t) } ( \cos \theta(t) + i \sin \theta(t) )
\nonumber \\
\dot y_1(t)+i  \dot y_2(t) && = [ \dot  \xi(t) + i \dot  \theta(t) ]  e^{ \xi(t) + i \theta(t) } 
 \label{polarcomplex}
\end{eqnarray}
one obtains that the polar angle $\theta(t)$ evolves according to the following Langevin equation independent of $\xi(t)$
\begin{eqnarray}
\dot \theta(t)
&& = \left[ M_{21} \cos^2 \theta(t) -M_{12} \sin^2 \theta(t) + ( M_{22}-M_{11}) \cos \theta(t) \sin \theta(t) \right]
\nonumber \\
&& + \eta(t) \left[ W_{21} \cos^2 \theta(t) -W_{12} \sin^2 \theta(t) + ( W_{22}-W_{11}) \cos \theta(t) \sin \theta(t) \right]
 \label{langevintheta}
\end{eqnarray}
while the Langevin equation for $\xi(t)$ involves only the angle $\theta(t)$ and the noise $\eta(t)$ on the right hand-side
\begin{eqnarray}
\dot \xi(t)
&& = \left[ M_{11} \cos^2 \theta(t) +M_{22} \sin^2 \theta(t) + ( M_{12}+M_{21}) \cos \theta(t) \sin \theta(t) \right]
\nonumber \\
&& + \eta(t) \left[ W_{11} \cos^2 \theta(t) +W_{22} \sin^2 \theta(t) + ( W_{12}+W_{21}) \cos \theta(t) \sin \theta(t) \right]
 \label{langevinxi}
\end{eqnarray}
The two right-hand sides of Eqs \ref{langevintheta} and \ref{langevinxi} can be written in terms of the double angle $(2 \theta (t))$ only,
and it is thus convenient to use instead the Riccati variable as we now recall.


\subsection{ Langevin dynamics for the Riccati variable $R(t)= \tan \theta(t) = \frac{y_2(t)}{y_1(t)}$ }

\label{sub_ric}

Although the angle $\theta(t)$ has the advantage to be periodic on the finite ring $ [-\frac{\pi}{2},\frac{\pi}{2}]$ without singularities,
one usually prefers to work instead with the Riccati variable 
\begin{eqnarray}
R(t) \equiv \tan \theta(t) = \frac{y_2(t)}{y_1(t)} 
 \label{ricdef}
\end{eqnarray}
that has the disadvantages to live on the periodic infinite ring $]-\infty,+\infty[$ 
and to become infinite $R=\infty$ when the first coordinate vanishes $y_1(t)=0$,
because its Langevin dynamics
\begin{eqnarray}
\dot R(t)  = a[R(t)] + \eta(t) b[R(t)]
 \label{langevinric}
\end{eqnarray}
involves two functions that are simply polynomials of degree 2 in $R$
\begin{eqnarray}
a[R]&& \equiv  M_{21}  + ( M_{22}-M_{11}) R  -M_{12} R^2
\nonumber \\
b[R]&& \equiv W_{21}  + ( W_{22}-W_{11}) R  -W_{12} R^2
 \label{ABric}
\end{eqnarray}
It is thus technically simpler to work with the Riccati variable $R$,
but one should always keep in mind that the real axis 
$R \in ]-\infty,+\infty[$ is a periodic ring where the two  infinities $R=\pm \infty$ are glued together.
In addition, many integrals over $R$ that will appear in the analysis will only be convergent in the Cauchy principal value sense.
Whenever one is confused about the singularities or the physical meaning in the asymptotic regimes $R\to \pm \infty$, 
one can always return to the angle interpretation to clarify the problems.


\subsection{ Finite-time Lyapunov exponent $\lambda_T$ as an additive functional of the Riccati process $R(0 \leq t \leq T)$ }

\label{sub_lyap}

The finite-time Lyapunov exponent $\lambda_T$ characterizes the 
exponential growth of the initial 2D process $(y_1(t),y_2(t))$ of Eq. \ref{langevinmatrix}
during the time-window $0 \leq t \leq T$.
One can thus consider different definitions 
based either on the growth of the modulus $\sqrt{ y_1^2(t)+y_2^2(t) } = e^{\xi(t) }$,
or on the growth of the absolute value of the first component $\vert  y_1(t) \vert$
or on the growth of the absolute value of the second component $\vert  y_2(t) \vert$
\begin{eqnarray}
\lambda_T^{(modulus)} && \equiv \frac{1}{T} \ln \left( \frac{ \sqrt{ y_1^2(T)+y_2^2(T) } }{\sqrt{ y_1^2(0)+y_2^2(0) }}  \right)
= \frac{\xi(T) - \xi(0)}{T} =  \frac{1}{T}  \int_0^T dt \dot \xi(t)
\nonumber \\
\lambda_T^{(y_1)} && \equiv \frac{1}{T} \ln \left\vert \frac{   y_1(T) }{  y_1(0) }  \right\vert
=  \frac{1}{T}  \int_0^T dt \frac{ \dot y_1(t) }{y_1(t) } 
\nonumber \\
\lambda_T^{(y_2)} && \equiv \frac{1}{T} \ln \left\vert \frac{   y_2(T) }{  y_2(0) }  \right\vert
=  \frac{1}{T}  \int_0^T dt \frac{ \dot y_2(t) }{y_2(t) } 
 \label{deflambdaT12}
\end{eqnarray}
Since these three definitions will become equivalent for large $T$, 
one can choose the definition that is the most relevant for the physical application under study,
or the definition that is technically simpler.
In the applications to Anderson Localization models that we will consider in sections \ref{sec_lochalperin} and \ref{sec_locsusy},
it is standard to consider $\lambda_T^{(y_1)} $, so in the following we will choose this definition
\begin{eqnarray}
\lambda_T \equiv \lambda_T^{(y_1)} \equiv \frac{1}{T} \ln \left\vert \frac{   y_1(T) }{  y_1(0) }  \right\vert
=  \frac{1}{T}  \int_0^T dt \frac{ \dot y_1(t) }{y_1(t) } 
 \label{deflambdaT}
\end{eqnarray}
but if one prefers another definition in Eq \ref{deflambdaT12}
for other applications, it is straightforward to adapt the computations.
Eq. \ref{langevinmatrix} yields that the logarithmic growth
of the first component $y_1(t)$ reads in terms of the Riccati variable $R(t)  = \frac{y_2(t)}{y_1(t)} $
\begin{eqnarray}
 \frac{ \dot y_1(t) }{y_1(t) } = [M_{11}+M_{12}  R(t) ]  + \eta(t) [ W_{11} +W_{12} R(t)]
 \label{langevinlogy1}
\end{eqnarray}
It is convenient to eliminate the noise $\eta(t)$ via the Langevin Eq \ref{langevinric} for the Riccati variable
in order to rewrite Eq. \ref{langevinlogy1} as
\begin{eqnarray}
 \frac{ \dot y_1(t) }{y_1(t) } = [M_{11}+M_{12}  R(t) ]+ \left( \frac{ \dot R(t) -  a[R(t)]  }{b[R(t)]} \right) [ W_{11} +W_{12} R(t)]
 \equiv \alpha[R(t)] + \dot R(t) \beta[R(t)]
 \label{langevinxiricderi}
\end{eqnarray}
where we have introduced the two functions 
\begin{eqnarray}
 \alpha[R] && \equiv  
  [M_{11}+M_{12}  R ]  -  \frac{    a[R]  }{b[R]}   [ W_{11} +W_{12} R]
  \nonumber \\
  && =
     \frac{  (M_{11}W_{21}-M_{21}W_{11})
+ R (M_{11}W_{22}-M_{22}W_{11}
+ M_{12}W_{21}-M_{21}W_{12})
+ R^2 (M_{12}W_{22}-M_{22}W_{12})
  }{ W_{21}  + ( W_{22}-W_{11}) R  -W_{12} R^2  }
\nonumber \\
 \beta[R] && \equiv  
   \frac{   W_{11} +W_{12} R  }{b[R]} =
     \frac{W_{11} +W_{12} R }{ W_{21}  + ( W_{22}-W_{11}) R  -W_{12} R^2  }
 \label{alphabeta}
\end{eqnarray}
Plugging Eq. \ref{langevinxiricderi} into Eq. \ref{deflambdaT} yields
that the finite-time Lyapunov exponent $\lambda_T$
can be rewritten as the additive functional of the Riccati process $R(t)$ over the time-window $0 \leq t \leq T$
\begin{eqnarray}
\lambda_T  =  \frac{1}{T}  \int_0^T dt \left[\alpha[R(t)] + \dot R(t) \beta[R(t)] \right]
 \label{lambdaTadditive}
\end{eqnarray}


\subsection{ The other finite-time Lyapunov exponent $\lambda_T^{min}$  }

\label{sub_other}

The finite-time Lyapunov exponent $\lambda_T$ discussed in the previous subsection
is actually the maximum Lyapunov exponent
\begin{eqnarray}
\lambda_T  =  \lambda_T^{max}
 \label{lambdaTmax}
\end{eqnarray}
that governs the growth of any solution of Eq. \ref{langevinmatrix} considered independently.
However if one considers two solutions $(\pm)$ of Eq. \ref{langevinmatrix}
\begin{eqnarray}
\frac{d}{dt}
\begin{pmatrix} 
 y_1^{\pm}(t) 
 \\  y_2^{\pm}(t) 
  \end{pmatrix}
  = \left( M + \eta(t) W  \right)
  \begin{pmatrix} 
 y_1^{\pm}(t) 
 \\  y_2^{\pm}(t) 
  \end{pmatrix}
\label{langevinmatrixpm}
\end{eqnarray}
the determinant of these two solutions
\begin{eqnarray}
\Delta(t) \equiv 
 \begin{vmatrix} 
y_1^+(t) &  y_1^-(t) \\
y_2^+(t) & y_2^-(t)
 \end{vmatrix} 
 = y_1^+(t) y_2^-(t) -  y_2^+(t) y_1^-(t)
\label{deter}
\end{eqnarray}
allows to analyze the dynamics of the phase space volume :
the exponential growth of $\Delta(t) $ over the time-window $0 \leq t \leq T$
involves the sum of the two Lyapunov exponents
\begin{eqnarray}
  \lambda_T^{max} + \lambda_T^{min} 
  \equiv \frac{1}{T} \ln \left\vert \frac{   \Delta(T) }{  \Delta(0) }  \right\vert
=  \frac{1}{T}  \int_0^T dt \frac{ \dot \Delta(t) }{\Delta(t) } 
 \label{lambdaTmaxminsum}
\end{eqnarray}
Since the maximum Lyapunov exponent
$\lambda_T^{max}=\lambda_T $ has already been defined in Eq. \ref{deflambdaT},
Eq. \ref{lambdaTmaxminsum} defines the minimum Lyapunov exponent $\lambda_T^{min}  $.
The dynamics of the determinant of Eq. \ref{deter} is governed by the trace of the 
total matrix $(M+\eta(t)W)$ involved in Eq. \ref{langevinmatrix}
\begin{eqnarray}
\frac{ \dot \Delta(t) }{\Delta(t) } = {\rm Trace}  \left(M + \eta(t) W \right)
= (M_{11}+M_{22}) + \eta(t) (W_{11}+W_{22})
 \label{dyndeter}
\end{eqnarray}
So Eq. \ref{lambdaTmaxminsum}
only involves the traces of the two matrices $M$ and $W$ and 
the integral of the white noise $\eta(t)$ over the 
 time-window $0 \leq t \leq T$
 \begin{eqnarray}
  \lambda_T^{max} + \lambda_T^{min} 
 = (M_{11}+M_{22}) +  (W_{11}+W_{22}) \left[ \frac{1}{T}  \int_0^T dt   \eta(t) \right]
 \label{lambdaTmaxminsumres}
\end{eqnarray}

In many interesting applications, in particular in the Anderson Localization models considered in sections
\ref{sec_lochalperin} and \ref{sec_locsusy},
the traces of the matrices $M$ and $W$ vanish
\begin{eqnarray}
M_{11}+M_{22} =0 =W_{11}+W_{22}
 \label{tracezero}
\end{eqnarray}
 so the phase space volume is conserved  $\dot \Delta(t)=0$,
and the second Lyapunov exponent $ \lambda_T^{min}  $ is simply the opposite of $ \lambda_T^{max}=\lambda_T $ 
\begin{eqnarray}
\lambda_T^{min}  =  - \lambda_T^{max} = - \lambda_T
 \label{lambdaTminopposite}
\end{eqnarray}
For later purposes, it is useful to rewrite the two functions of Eq. \ref{alphabeta}
for the special case of vanishing traces (Eq. \ref{tracezero}) where we can eliminate $M_{22}=-M_{11}$
and $W_{22}=-W_{11}$
\begin{eqnarray}
 \alpha^{conserv}[R] &&  =
     \frac{  (M_{11}W_{21}-M_{21}W_{11})
+ R ( M_{12}W_{21}-M_{21}W_{12})
+ R^2 (M_{11}W_{12} - M_{12}W_{11})
  }{ W_{21}  - 2 W_{11} R  -W_{12} R^2  }
\nonumber \\
 \beta^{conserv}[R] && \equiv  
     \frac{W_{11} +W_{12} R }{ W_{21}  - 2 W_{11} R  -W_{12} R^2  }
 \label{alphabetaconser}
\end{eqnarray}


\subsection{ Example of another interesting additive functional of the Riccati process $R(0 \leq t \leq T)$ }

\label{sub_dos}

Another example of interesting observable is the density of zeros of the first component $y_1(t)= e^{\xi(t)} \cos \theta(t) $ 
during the time-window $0 \leq t \leq T$
\begin{eqnarray}
 N_T  \equiv \frac{1}{T}  \int_0^T dt \sum_{t_k : y_1(t_k)=0} \delta(t-t_k)= \frac{1}{T}  \int_0^T dt \sum_{t_k : \cos \theta(t_k)=0} \delta(t-t_k)
   \label{doszero}
\end{eqnarray}
Using the identities for the delta function
\begin{eqnarray}
\delta( \cos \theta(t) ) && =  \sum_{t_k : \cos \theta(t_k)=0} \frac{ \delta(t-t_k) } { \vert \dot \theta(t_k) \sin \theta(t_k)  \vert }
=  \sum_{t_k : \cos \theta(t_k)=0} \frac{ \delta(t-t_k) } { \vert \dot \theta(t_k)   \vert }
\nonumber \\
\delta\left( \frac{1}{R(t)}  \right) && =  \sum_{t_k : \frac{1}{R(t_k)} =0} \frac{ \delta(t-t_k) } { \left \vert \frac{ \dot R(t_k) }{ R^2(t_k)}   \right\vert }
   \label{cosid}
\end{eqnarray}
Eq. \ref{doszero} can be rewritten as the following functionals of the angle $\theta(t)$
or of the Riccati variable $R(t)$ 
\begin{eqnarray}
 N_T  = \frac{1}{T}  \int_0^T dt \ \vert \dot \theta(t) \vert \ \delta( \cos \theta(t) ) 
=  \frac{1}{T}  \int_0^T dt \ \frac{ \vert \dot R(t) \vert }{ R^2(t) } \ \delta\left( \frac{1}{R(t)}  \right) 
    \label{doszerotheta}
\end{eqnarray}
where the presence of the absolute values $\vert \dot \theta(t) \vert $ and $ \vert \dot R(t) \vert  $ 
make them different from the additive functionals of the form of Eq. \ref{lambdaTadditive}.
Writing the Langevin Eq. \ref{langevintheta} for $\theta(t)$ at times $t_k$ where $\cos \theta(t_k)=0 $
(and thus $\sin^2 \theta(t_k) =1 $)
\begin{eqnarray}
\dot \theta(t_k)  =  -M_{12}  - \eta(t) W_{12}
 \label{langevinthetazero}
\end{eqnarray}
or equivalently the Langevin Eq. \ref{langevinric} for $R(t)$ at times $t_k$ where $\frac{1}{R(t_k)} =0 $
\begin{eqnarray}
\frac{ \dot R(t_k) }{ R^2(t_k) } =  -M_{12}  - \eta(t) W_{12}
 \label{langevinriclarge}
\end{eqnarray}
one sees that the vanishing of the matrix element $W_{12}=0$
leads to the following simplifications in Eq. \ref{doszerotheta} 
\begin{eqnarray}
 N_T^{[W_{12}=0]}  = \frac{\vert M_{12} \vert }{T}  \int_0^T dt \  \ \delta( \cos \theta(t) ) 
=  \frac{\vert M_{12} \vert}{T}  \int_0^T dt \  \ \delta\left( \frac{1}{R(t)}  \right) 
    \label{doszerothetasimpli}
\end{eqnarray}
that is of the form of Eq. \ref{lambdaTadditive}
even if it is for the singular function $ \alpha[R]= \delta\left( \frac{1}{R}  \right)$.
It turns out that the Anderson Localization applications considered in section \ref{sec_lochalperin} and \ref{sec_locsusy}
will both correspond to this special case $W_{12}=0$.


\subsection{ Large deviations properties of the Lyapunov exponent $\lambda_T$ }

\label{sub_large}

For large $T$, the probability ${\cal P}_T(\lambda)$ 
to see the finite-size Lyapunov exponent $\lambda_T=\lambda$
 is expected to follow the large deviation form 
\begin{eqnarray}
{\cal P}_T(\lambda)  \opsimeq_{ T \to + \infty} e^{- T I(\lambda) }
\label{largedevlyapunov}
\end{eqnarray}
where the rate function $I(\lambda)$ is positive $I(\lambda) \geq 0$ and vanishes only 
at its minimum corresponding to the typical value $\lambda^{typ}$ that will be realized with probability one 
in the thermodynamic limit $T \to +\infty$
\begin{eqnarray}
I(\lambda^{typ}) =0 = I'(\lambda^{typ})
\label{typvanish}
\end{eqnarray}
All other values $\lambda \ne \lambda^{typ}$ appear with 
a probability ${\cal P}_T(\lambda) $ that is exponentially small in $T$ in Eq. \ref{largedevlyapunov},
but they are nevertheless important to understand the finite-size fluctuations as we now recall.
The generating function of $\lambda_T$
can be evaluated from Eq. \ref{largedevlyapunov}
via the Laplace saddle-point method for large $T$
\begin{eqnarray}
Z_T(k) \equiv  \int d \lambda \ {\cal P}_T(\lambda) \ e^{  T k \lambda  }
\opsimeq_{ LT\to + \infty} \int d \lambda \ e^{ T \left[ k \lambda   - I(\lambda) \right] } \opsimeq_{ T \to + \infty} e^{ T \mu(k) }
\label{multifz}
\end{eqnarray}
where the function $\mu(k)  $ 
 corresponds to the Legendre transform of the rate function $I(\lambda)$ 
as a consequence of the saddle-point evaluation of the integral in $\lambda$ in Eq. \ref{multifz}
\begin{eqnarray}
\mu(k)  && = \lambda k  - I(\lambda) 
\nonumber \\
0 && = k - I'(\lambda)
\label{legendre}
\end{eqnarray}
with the reciprocal Legendre transform
\begin{eqnarray}
I(\lambda)   && = \lambda k  - \mu(k) 
\nonumber \\
0 && = \lambda   - \mu'(k) 
\label{legendrereci}
\end{eqnarray}
The function $\mu(k)  $ is called the scaled cumulant generating function in the field of  large deviations.
Its power expansion in $k$ around $k=0$ where it vanishes $\mu(k=0) =0$ as a consequence of the normalization in Eq. \ref{multifz}
\begin{eqnarray}
 \mu(k) =  \sum_{n=1}^{+\infty}  \mu^{(n)}(0) \frac{k^n}{n!} 
 \label{muexpansion}
\end{eqnarray}
allows to evaluate the cumulants $c_n$ of the Lyapunov exponent $\lambda_T$ in terms of the derivative $\mu^{(n)}(0) $ of order $n$ at $k=0$
\begin{eqnarray}
 c_n = \frac{\mu^{(n)}(0)}{T^{n-1}} 
 \label{cumulants}
\end{eqnarray}
The first cumulant corresponds to the typical value $\lambda^{typ}$ where the rate function vanishes (Eq. \ref{typvanish})
\begin{eqnarray}
 c_1 = \mu'(0) =  \lambda^{typ}
 \label{c1}
\end{eqnarray}
The finite-size fluctuations around this typical value can be characterized by the next cumulants for $n=2,3,4...$
\begin{eqnarray}
 c_2 && = \frac{\mu^{''}(0)}{T} 
 \nonumber \\
  c_3 && = \frac{\mu^{'''}(0)}{T^2} 
   \nonumber \\
  c_4 && = \frac{\mu^{'''}(0)}{T^3} 
 \label{cumulants234}
\end{eqnarray}

In specific models, the exact computation of the whole large deviation rate function $I(\lambda)$ or of its Legendre transform $\mu(k)$
is usually not possible, but many results have been obtained in the references \cite{titov,zillmer,pumir,titovLloyd,ramola,fyodorov,vanneste,texier_jstat,texier_epl,texier_comtet}
already mentioned in the Introduction, with three main goals:

(1) compute explicitly the first cumulants via the perturbative analysis in $k$ of scaled cumulant generating function $\mu(k)$,
as described in particular in Refs \cite{titov,titovLloyd,ramola,texier_jstat,texier_epl,texier_comtet}.

(2) compute asymptotic forms of large deviations properties in various regions of the model parameters,
as described in particular in Refs \cite{titov,zillmer,pumir,titovLloyd,ramola,fyodorov,texier_jstat,texier_epl,texier_comtet}.

(3) compute exactly $\mu(k)$ for specific integer values of $k$,
as described in particular in Refs \cite{titov,zillmer,fyodorov}.


\subsection{ Gallavotti-Cohen symmetry of the rate function $I(\lambda)$ and the scaled cumulant generating function $\mu(k)$}

\label{sub_GCsym}

It turns out that symmetry relations for the large deviation rate function $I(\lambda)$ of the following form involving some constant $K$
\begin{eqnarray}
I(\lambda) =I(-\lambda) - K \lambda        
\label{symIlambda}
\end{eqnarray}
that corresponds to the following simple ratio for the probabilities $P_T (\pm \lambda) $ to observe the value $\lambda$ or its opposite $(-\lambda)$ (Eq. \ref{largedevlyapunov})
\begin{eqnarray}
\frac{ P_T (\lambda) }{ P_T (- \lambda) } \opsimeq_{ T \to + \infty}   e^{ K T \lambda } 
\label{ratioplambda}
\end{eqnarray}
or equivalently the translation of the symmetry of Eq. \ref{symIlambda}
for the Legendre transform $\mu(k)$ (Eqs \ref{legendre} and \ref{legendrereci}) representing the scaled cumulant generating function (Eq. \ref{muexpansion})
\begin{eqnarray}
\mu(k)= \mu(-K-k) 
\label{symmu}
\end{eqnarray}
have appeared independently in at least three different fields :

(i) in the field of non-equilibrium dynamics,
the symmetry relations of the form of Eqs \ref{symIlambda} and \ref{symmu}
are famous under the name 'Gallavotti-Cohen fluctuation relations'
(see \cite{galla,kurchan_langevin,Leb_spo,maes1999,jepps,derrida-lecture,harris_Schu,kurchan,searles,zia,maes2009,maes2017,chetrite_thesis,chetrite_HDR} and references therein).

(ii) in the field of multifractality of wavefunctions at Anderson transition critical points
(see the reviews \cite{janssenrevue,mirlinrevue} and references therein),
the symmetry of Eq. \ref{symIlambda} is written for the singularity spectrum $f(\alpha)$ in dimension $d$ as
\cite{mirlin06,Wegner_epsilon,sigma,milden07,qHall,vasquez,us_strongmultif}
\begin{eqnarray}
f(2d-\alpha)=f(\alpha)+d-\alpha
\label{symfaanderson}
\end{eqnarray}
The relation with the field (i) has been discussed in \cite{us_symmultif}.

(iii) in the field of products of random matrices, 
the symmetry of Eq. \ref{symmu} has been found for the case of real $2 \times 2$ random matrices of determinant unity
with the parameter $K=2$ \cite{vanneste,texier_jstat,texier_epl}
(see \cite{vanneste} for the proof and for the generalization to real $2n \times 2n$ symplectic matrices
with the parameter $K=2n$, as well as other properties in the general case).
For our present continuous-time model of Eq. \ref{langevinmatrix}, the condition of determinant unity
translates into the vanishing trace condition already discussed around Eq. \ref{tracezero}.
As a consequence, whenever the vanishing trace condition of Eq. \ref{tracezero} is satisfied,
the symmetry relations of Eqs \ref{symIlambda} and \ref{symmu} are expected to hold with the value $K=2$
\begin{eqnarray}
I(\lambda) && =I(-\lambda) - 2 \lambda        
\nonumber \\
\mu(k) && = \mu(-2-k) 
\label{symK2}
\end{eqnarray}


\section{Steady-state of the Riccati process $R(t)$ and typical Lyapunov exponent   }

\label{sec_steady}

In this section, the Riccati process is analyzed via its Fokker-Planck generator in subsection \ref{sub_FP}
and via the associated quantum supersymmetric Hamiltonian in subsection \ref{sub_susy}.
The corresponding steady-state discussed in subsection \ref{sub_steady}
is either an equilibrium steady-state without current $j_{st}=0$ as discussed in section \ref{sec_eqst}
or a non-equilibrium steady-state with a finite stationary current  $j_{st} \ne 0$
as discussed in section \ref{sec_noneqst},
and allows to compute the typical Lyapunov exponent as recalled in subsection \ref{sub_typ}.

\subsection{ Fokker-Planck operator for the Riccati process : force $F(R)$ and diffusion coefficient $D(R)$ }

\label{sub_FP}

The Stratonovich interpretation 
of the Langevin dynamics of Eq. \ref{langevinric}
leads to the Fokker-Planck equation for the probability $P_t(R)$ to see the Riccati value $R$ at time $t$
\begin{eqnarray}
\frac{ \partial P_{t}(R) }{\partial t }   = - \frac{ \partial  }{\partial R} 
\left( F(R)  P_{t}(R)  - D(R)   \frac{ \partial  P_{t}(R)}{\partial R}  \right) \equiv  {\cal F} P_t(.)
\label{fokkerplanck}
\end{eqnarray}
where the Fokker-Planck operator ${\cal F}$ involves 
the diffusion coefficient (see Eq \ref{ABric})
\begin{eqnarray}
 D(R) \equiv \frac{b^2[R]}{2} = \frac{\left[ Q_{21}  + ( Q_{22}-Q_{11}) R  -Q_{12} R^2 \right]^2}{2}
 \label{diffR}
\end{eqnarray}
and the force (see Eq \ref{ABric})
\begin{eqnarray}
F(R) && \equiv a[R]  - \frac{b[R]b'[R]}{2} 
 \nonumber \\ &&
 =  M_{21}  + ( M_{22}-M_{11}) R  -M_{12} R^2
 - \frac{ \left[ Q_{21}  + ( Q_{22}-Q_{11}) R  -Q_{12} R^2 \right]  \left[  ( Q_{22}-Q_{11})   - 2 Q_{12} R \right]}{2}
 \label{forceR}
\end{eqnarray}

\subsection{ Associated quantum supersymmetric Hamiltonian }

\label{sub_susy}

It is often useful to transform the non-symmetric generator of a dynamical process
into a symmetric operator via the appropriate similarity transformation
  (see the textbooks \cite{gardiner,vankampen,risken}).
In the present context, it is thus convenient to introduce the effective potential
\begin{eqnarray}
U(R) && \equiv - \int_0^R dR' \frac{F(R')}{D(R')} 
\nonumber \\
U'(R) && = - \frac{F(R)}{D(R)}
 \label{UR}
\end{eqnarray}
and to perform the change of variables
\begin{eqnarray}
P_{t}(R) = e^{- \frac{U(R)}{2} } \psi_t(R)
\label{ppsi}
\end{eqnarray}
The Fokker-Planck Eq \ref{fokkerplanck} for $ P_{t}(R) $
is then transformed 
 into the euclidean Schr\"odinger equation for $\psi_t(R)$
\begin{eqnarray}
- \frac{ \partial \psi_t(R) }{\partial t } = H \psi_t(R)
\label{schropsi}
\end{eqnarray}
where the quantum Hermitian Hamiltonian 
\begin{eqnarray}
 H = - \frac{ \partial  }{\partial R} D(R) \frac{ \partial  }{\partial R} +V(R)
\label{hamiltonien}
\end{eqnarray}
corresponds to an effective position-dependent 'mass' 
whenever the diffusion coefficient  $ D(R)=\frac{1}{2 m(R)}$ depends explicitly on the Riccati variable $R$.
The very specific structure of the scalar potential
\begin{eqnarray}
V(R) && \equiv D(R)  \frac{ [U'(R)]^2 }{4 }  -D(R) \frac{U''(R)}{2} -D'(R) \frac{U'(R)}{2}
\nonumber \\
&& = \frac{ F^2(R) }{4 D(R) } + \frac{F'(R)}{2}
\label{vfromu}
\end{eqnarray}
allows to factorize the Hamiltonian of Eq. \ref{hamiltonien}
into the supersymmetric form (see the review on supersymmetric quantum mechanics \cite{review_susyquantum} and references therein)
\begin{eqnarray}
H \equiv    Q^{\dagger} Q
\label{hsusy}
\end{eqnarray}
involving the two first-order operators 
\begin{eqnarray}
Q  && \equiv    \sqrt{ D(R) }  \left( \frac{ d }{ d R}  +\frac{ U'(R)}{2 } \right)
\nonumber \\
Q^{\dagger}  &&\equiv  \left(   - \frac{ d }{ d R}  +\frac{ U'(R)}{2 } \right)\sqrt{ D(R) }
\label{qsusy}
\end{eqnarray}
For later purposes, 
it is interesting to introduce the supersymmetric partner of the Hamiltonian of Eqs \ref{hamiltonien}
and \ref{hsusy}
\begin{eqnarray}
\breve{H } \equiv    Q Q^{\dagger} = - \frac{ \partial  }{\partial R} D(R) \frac{ \partial  }{\partial R} +\breve{V}(R)
\label{hsusypartner}
\end{eqnarray}
where the partner potential reads
\begin{eqnarray}
\breve{V }(R) && \equiv D(R)  \frac{ [U'(R)]^2 }{4 }  +D(R) \frac{U''(R)}{2}
 + \frac{ [D'(R)]^2 }{4 D(R) }-\frac{D''(R)}{2}
\nonumber \\
&& = \frac{ F^2(R) }{4 D(R) } - \frac{F'(R)}{2} +\frac{F(R) D'(R) }{2D(R)}+ \frac{ [D'(R)]^2 }{4 D(R) }-\frac{D''(R)}{2}
\label{vpartner}
\end{eqnarray}
In particular, the commutator between $ Q^{\dagger}$ and $Q$ corresponds to the difference between the two Hamiltonians, and thus to the difference between the two potentials
\begin{eqnarray}
[ Q^{\dagger}, Q]  && \equiv    Q^{\dagger} Q -  Q Q^{\dagger} = H- \breve{H } = V(R) - \breve{V }(R) 
\nonumber \\
&& =  F'(R) -\frac{F(R) D'(R) }{2D(R)}- \frac{ [D'(R)]^2 }{4 D(R) }+\frac{D''(R)}{2}
\label{commutateur}
\end{eqnarray}
In terms of the two functions $a[R]$ and $b[R]$ that appear in the Langevin Eq. \ref{langevinric}
and that have been used to define the diffusion coefficient (Eq. \ref{diffR}) and the force (Eq. \ref{forceR}),
this commutator reduces to
\begin{eqnarray}
&& [ Q^{\dagger}, Q]   = a'[R] - a[R] \frac{b'[R]}{b[R]}
\label{commutateurab}
 \\
&&= \frac{ ( M_{22}-M_{11}) W_{21} - M_{21} ( W_{22}-W_{11}) 
+ 2 (M_{21} W_{12} - M_{12} W_{21}) R
+ \left[ 
( M_{22}-M_{11}) W_{12} - M_{12} ( W_{22}-W_{11}) 
 \right] R^2
}{W_{21}  + ( W_{22}-W_{11}) R  -W_{12} R^2} 
\nonumber
\end{eqnarray}
where we have used the explicit expressions of Eq. \ref{ABric} in terms of the matrix elements of the initial model.

For the special case of vanishing traces (Eq. \ref{tracezero}) where we can eliminate $M_{22}=-M_{11}$
and $W_{22}=-W_{11}$, the commutator is directly related to the function $\alpha^{conserv}[R] $
of Eq. \ref{alphabetaconser}
\begin{eqnarray}
 [ Q^{\dagger}, Q]^{conserv}  &&= \frac{ 2 (-M_{11} W_{21} + M_{21} W_{11})
+ 2 (M_{21} W_{12} - M_{12} W_{21}) R
+ 2 (-M_{11} W_{12} + M_{12} W_{11}  R^2
}{W_{21}  -2 W_{11} R  -W_{12} R^2} 
\nonumber \\
&& = -2 \alpha^{conserv}[R]
\label{commutateurconser}
\end{eqnarray}


\subsection{ Steady-state $\rho_{st}(R) $ of the Fokker-Planck equation  }

\label{sub_steady}

The steady-state solution $\rho_{st}(R)$ of the Fokker-Planck dynamics Eq. \ref{fokkerplanck}
\begin{eqnarray}
0  = {\cal F} \rho_{st}(.) = - \frac{ \partial  }{\partial R} 
\left( F(R)  \rho_{st}(R)  - D (R)  \frac{ \partial  \rho_{st}(R)}{\partial R}  \right) 
\label{fokkerplanckst}
\end{eqnarray}
corresponds to the right eigenvector $r(R)$ of the Fokker-Planck operator ${\cal F} $ associated to the eigenvalue $\mu=0$.
The corresponding left eigenvector $l(R)$
\begin{eqnarray}
0  = {\cal F}^{\dagger} l(.) = \left( F(R)  \rho_{st}(R)  - D (R)  \frac{ \partial  \rho_{st}(R)}{\partial R}  \right)\frac{ \partial  }{\partial R} l(R)
\label{fokkerplanckadjoint}
\end{eqnarray}
is simply the constant function
\begin{eqnarray}
l(R)=1
 \label{lefttrivial}
\end{eqnarray}
as a consequence of the conservation of the total probability.
Eq. \ref{fokkerplanckst}
means that the corresponding steady-state current $j_{st}(R)=F(R)  \rho_{st}(R)  - D (R)   \rho_{st}'(R) $ cannot depend on $R$
\begin{eqnarray}
j_{st} =  F(R)  \rho_{st}(R)  - D(R)   \rho_{st}'(R) 
 \label{steady}
\end{eqnarray}
In terms of the effective potential $U(R)$ of Eq. \ref{steadydiff}, this equation for the steady state reads
\begin{eqnarray}
 \rho_{st}'(R) +    U'(R) \rho_{st}(R)  =-\frac{ j_{st} }{D(R)} 
 \label{steadydiff}
\end{eqnarray}
The integration constant will be determined by the requirement of
periodicity on the Riccati ring where $R=\pm \infty$ are glued together.
Since the density has to vanish at $R \to \pm \infty$ to be normalizable,
this periodicity constraint is somewhat tricky and it is clearer 
to first regularize the problem on a finite ring $R \in [R_{min},R_{max}]$.
(Note that if one chooses to work with the angle variable $\theta(t)$
instead of using the Riccati variable $R(t)=\tan \theta(t)$ (Eq. \ref{ricdef}),
one needs indeed to solve the corresponding finite ring model).
The general solution of Eq. \ref{steadydiff}
\begin{eqnarray}
 \rho^{Reg}_{st}(R) 
 =  e^{ -U(R)} 
 \left[ K - j_{st}  \int_{R_{min}}^{R} \frac{d R'}{D(R')} e^{ U(R') }   \right] 
\label{solReggene}
\end{eqnarray}
involves an integration constant $K$ that should be fixed by the requirement of periodicity
$  \rho^{Reg}_{st}(R_{min})=\rho^{Reg}_{st}(R_{max})$ leading to the equation for $K$
\begin{eqnarray}
   e^{ -U(R_{min})} K =
e^{ -U(R_{max})} 
 \left[ K - j_{st}  \int_{R_{min}}^{R_{max}} \frac{d R'}{D(R')} e^{ U(R') }   \right] 
\label{solReggeneperio}
\end{eqnarray}
So one needs to distinguish two very different cases as we now recall.


\subsection{ Case of periodic potential $U(R)$ with an equilibrium steady-state without current $j_{st}=0$  }

\label{sec_eqst}

When the effective potential $U(R)$ is periodic on the Riccati ring 
\begin{eqnarray}
   0 && =U(R_{max})  -U(R_{min})  = \int_{R_{min}}^{R_{max}} dR U'(R) = -  \int_{R_{min}}^{R_{max}} dR \frac{F(R)}{D(R)} 
\label{uperioreg}
\end{eqnarray}
i.e. in the limit $R_{max} \to + \infty$ and $R_{min} \to - \infty$, when the condition
\begin{eqnarray}
0=\int_{-\infty} ^{+\infty} dR U'(R) && = -  \int_{-\infty} ^{+\infty} dR \frac{F(R)}{D(R)} 
 \label{UReq}
\end{eqnarray}
is satisfied,
Eq. \ref{solReggeneperio} yields that the steady state current vanishes $j_{st}=0$.
So the steady state density
reduces to the analog of the Boltzmann distribution in the potential $U(R)$
\begin{eqnarray}
  \rho^{eq}_{st}(R) =  K e^{ -U(R)} 
 \label{steadyeq}
\end{eqnarray}
where the constant $K$ is fixed by the normalization
\begin{eqnarray}
1= \int_{-\infty}^{+\infty} dR \rho^{eq}_{st}(R)= K \int_{-\infty}^{+\infty} dR e^{ -  U(R) } 
 \label{partitioneq}
\end{eqnarray}
to be the inverse of the partition function in the potential $U(R)$.

In the quantum mechanical language of Eq. \ref{ppsi}, the groundstate wavefunction corresponding to $\rho^{eq}_{st}(R) $
\begin{eqnarray}
  \psi^{eq}_{st}(R) =  K e^{ - \frac{ U(R)}{2} } 
 \label{psieq}
\end{eqnarray}
is annihilated by the operator $Q$ of Eq. \ref{qsusy} 
\begin{eqnarray}
Q  \psi^{eq}_{st}(R)  =  \sqrt{ D(R) }\left(    \frac{ d }{ d R}  +\frac{ U'(R)}{2 } \right) K e^{ - \frac{ U(R)}{2}}=0
\label{qsusyeq}
\end{eqnarray}


\subsection{ Case of non-periodic potential with a non-equilibrium steady-state and a finite stationary current  $j_{st} \ne 0$  }

\label{sec_noneqst}

When the effective potential $U(R)$ 
is not periodic on the Riccati ring 
\begin{eqnarray}
   0 && \ne U(R_{max})  -U(R_{min})  = \int_{R_{min}}^{R_{max}} dR U'(R) = -  \int_{R_{min}}^{R_{max}} dR \frac{F(R)}{D(R)} 
\label{unonperioreg}
\end{eqnarray}
i.e. equivalently in the limit $R_{max} \to + \infty$ and $R_{min} \to - \infty$
\begin{eqnarray}
0 \ne \int_{-\infty} ^{+\infty} dR U'(R) && = -  \int_{-\infty} ^{+\infty} dR \frac{F(R)}{D(R)} 
 \label{URnoneq}
\end{eqnarray}
Eq. \ref{solReggeneperio} yields the value of the integration constant $K$,
and the steady state density of Eq. \ref{solReggene} reads
\begin{eqnarray}
 \rho^{neq}_{st}(R) 
 \equiv  \frac{(- j_{st}) e^{ -U(R)} 
 \left[ e^{ -U(R_{min})} \int_{R_{min}}^{R} \frac{d R'}{D(R')} e^{ U(R') } 
 +e^{ -U(R_{max}) }\int_{R}^{R_{max}} \frac{d R'}{D(R')} e^{ U(R')  } 
  \right] }
 {  \left[  e^{ -U(R_{min}) } 
 -  e^{ - U(R_{max}) } 
  \right] }
\label{noneqsolReg}
\end{eqnarray}
The steady-state current $j_{st}$ is then determined by
the normalization 
\begin{eqnarray}
1 && = \int_{R_{min}}^{R_{max}} dR \rho^{Reg}_{st}(R) 
\nonumber \\
&& = 
 \frac{(- j_{st})   \left[ e^{ -U(R_{min})}  \int_{R_{min}}^{R_{max}} dR e^{ -U(R)} \int_{R_{min}}^{R} \frac{d R'}{D(R')} e^{ U(R') } 
 +e^{ -U(R_{max}) }  \int_{R_{min}}^{R_{max}} dR e^{ -U(R)} \int_{R}^{R_{max}} \frac{d R'}{D(R')} e^{ U(R')  } 
  \right]
  } {  \left[  e^{ -U(R_{min}) }  -  e^{ - U(R_{max}) }   \right] }
\label{noneqsolRegnorma}
\end{eqnarray}
In the limit $R_{min} \to -\infty$ and $R_{max} \to +\infty$, one then needs to 
take into account the behavior of the potential $U(R)$ for $R \to \pm \infty$
to obtain the appropriate solution on the infinite Riccati ring.
As an example, let us describe the case 
where the potential difference $U(+\infty)-U(-\infty) =  +\infty $ diverges.


\subsubsection{ Non-equilibrium steady state when the potential difference diverges
$U(+\infty)-U(-\infty) =  +\infty $   }

\label{sec_ddpinfinite}

If the potential difference $U(+\infty)-U(-\infty)  $ diverges towards $+\infty$
\begin{eqnarray}
 U(+\infty)-U(-\infty) = \int_{-\infty} ^{+\infty} dR U'(R) =  - \int_{-\infty} ^{+\infty} dR \frac{F(R)}{D(R)}  = +\infty 
 \label{URneqplus}
\end{eqnarray}
only the left terms survive in the numerator and denominator of Eq. \ref{noneqsolReg}
in the limit $R_{min} \to -\infty$ and $R_{max} \to +\infty$, 
so the non-equilibrium solution on the infinite Riccati ring reduces to
\begin{eqnarray}
 \rho^{neq}_{st}(R) 
 =(- j_{st}) e^{ -U(R)} 
  \int_{-\infty}^{R} \frac{d R'}{D(R')} e^{ U(R') } 
\label{noneqsolRegplus}
\end{eqnarray}
and its normalization determines the negative steady-state current $j_{st}<0$
\begin{eqnarray}
1= \int_{-\infty}^{+\infty} dR  \rho^{neq}_{st}(R) 
 =(- j_{st})  \int_{-\infty}^{+\infty} dR  e^{ -U(R)} 
  \int_{-\infty}^{R} \frac{d R'}{D(R')} e^{ U(R') } 
\label{jstplus}
\end{eqnarray}
If $ U'(R) = -  \frac{F(R)}{D(R)} $ is large for $R \to \pm \infty$,
the saddle-point evaluation of Eq \ref{noneqsolRegplus}
yields that the asymptotic behaviors of the density at $R \to \pm \infty$
\begin{eqnarray}
\rho_{st}(R) && \opsimeq_{R \to \pm \infty}    - \frac{ j_{st} }{ D(R) U'(R) } = \frac{j_{st} }{F(R)}
\label{neqplusrhoinfinity}
\end{eqnarray}
only involves the steady state current $j_{st}$ and the force $F(R)$,
i.e. the diffusion contribution becomes negligible in the differential Eq. \ref{steady} for $R \to \pm \infty$.


\subsubsection{ Meaning of the non-equilibrium solution in the quantum mechanical language   }

In the quantum mechanical language of Eq. \ref{ppsi}, 
the right wavefunction corresponding to the non-equilibrium solution $ \rho^{neq}_{st}(R)$ of Eq. \ref{noneqsolReg}
\begin{eqnarray}
  \psi^{[r]}_{st}(R) =  e^{  \frac{ U(R) }{ 2} } \rho^{neq}_{st}(R)
   \label{psineqrho}
\end{eqnarray}
is not annihilated by the operator $Q$ of Eq. \ref{qsusy} (in contrast to Eq. \ref{qsusyeq} concerning the equilibrium case)
\begin{eqnarray}
Q  \psi^{[r]}_{st}(R)  && =
 \sqrt{ D(R) }\left(    \frac{ d }{ d R}  +\frac{ U'(R)}{2 } \right)\psi^{[r]}_{st}(R) =  \frac{(- j_{st})}{ \sqrt{ D(R) }} e^{  \frac{ U(R) }{ 2} } 
   \label{qsusynoneq}
\end{eqnarray}
but this state is annihilated by the operator $Q^{\dagger}$ of Eq. \ref{qsusy} 
\begin{eqnarray}
Q^{\dagger} \left( Q  \psi^{[r]}_{st}(R) \right)  =
\left(   - \frac{ d }{ d R}  +\frac{ U'(R)}{2 } \right)\sqrt{ D(R) } \left( \frac{(- j_{st})}{ \sqrt{ D(R) }} e^{  \frac{ U(R) }{ 2} }\right) =0
   \label{qsusynoneqdagger}
\end{eqnarray}
as it should to produce the ground-state of energy zero of the Hamiltonian of Eq. \ref{hsusy}.
Using the trivial left eigenvector $l(R)=1$ of Eq. \ref{lefttrivial}, the change of variable conjugated to Eq. \ref{psineqrho}
yields that the corresponding left wavefunction still corresponds to the equilibrium wavefunction (Eq. \ref{psieq})
\begin{eqnarray}
\psi^{[l]}(R)= e^{ - \frac{ U(R) }{ 2} } l(R) = e^{  - \frac{ U(R) }{ 2} }
 \label{leftpsi}
\end{eqnarray}
So even if the quantum Hamiltonian is Hermitian, 
the periodicity constraint on the Fokker-Planck right and left eigenvectors
can induce a difference between the associated left and right eigenwavefunctions.


\subsection{ Typical Lyapunov exponent in terms of the steady-state}

\label{sub_typ}

In the thermodynamics limit $T \to +\infty$, 
 the finite-time Lyapunov exponent $\lambda_T$ will converge towards its 
 typical value (Eq. \ref{typvanish}), where the additive functional of the Riccati process $R(t)$ of Eq. \ref{lambdaTadditive}
 can be evaluated from the steady-state $\rho_{st}(R)$ of the Riccati variable and the steady current $j_{st}$ discussed above
\begin{eqnarray}
\lambda^{typ} =\lambda_{(T \to +\infty)}   =  \int_{-\infty}^{+\infty} dR  \rho_{st}(R) \alpha(R) + j_{st} \beta^{tot}
 \label{lambdatypsteady}
\end{eqnarray}
with the notation
\begin{eqnarray}
\beta^{tot} \equiv  \int_{-\infty}^{+\infty} dR \beta(R) 
 \label{betatot}
\end{eqnarray}
Now that we have recalled how the typical value $\lambda^{typ} $
can be computed in terms of the steady state of the Riccati variable,
we will focus on its large deviations properties in the next sections.


\section{ Analysis via the large deviations at level 2.5 of the Riccati process }

\label{sec_optimization}

As recalled in the Introduction, the formulation of the Large Deviations at Level 2.5
has been a major achievement for various Markovian dynamics,
including Markov Chains (discrete-space and discrete-time) \cite{fortelle_thesis,fortelle_chain,review_touchette,c_largedevdisorder,c_reset},
Markov Jump processes (discrete-space and continuous-time)
\cite{fortelle_thesis,fortelle_jump,maes_canonical,maes_onandbeyond,wynants_thesis,chetrite_formal,BFG1,BFG2,chetrite_HDR,c_ring,c_interactions,c_open,barato_periodic,chetrite_periodic,c_reset}
and Diffusion processes (continuous-space and continuous-time) 
\cite{wynants_thesis,maes_diffusion,chetrite_formal,engel,chetrite_HDR,c_reset}.

In this section, the goal is to apply this Large Deviation analysis at Level 2.5 to the Riccati process over the large time window $0 \leq t \leq T$.
The empirical density and the empirical current are introduced in subsection \ref{sub_empi}
and allow to reconstruct the finite-size exponent $\lambda_T$.
As a consequence, the corresponding explicit rate function at level 2.5 recalled in subsection \ref{sub_2.5}
can be used to analyze the generating function of the Lyapunov exponent as explained in subsection \ref{sub_gene}.
The corresponding optimization problem leads to Euler-Lagrange equations in subsection \ref{sub_euler}
and to the scaled cumulant generating function $\mu(k)$ in subsection \ref{sub_muk}.
Finally, we describe how the optimization procedure can be decomposed in two steps in section \ref{sec_twosteps},
while the special case of an equilibrium steady-state requires some changes as discussed in subsection \ref{sec_optieq}.


\subsection{ Empirical density $\rho_T(R)$ and empirical current $j_T(R)$ of the Riccati process during $0 \leq t \leq T$}

\label{sub_empi}

The empirical density $ \rho_T(R)$ 
represents the histogram of the Riccati variable $R(t)$ seen during the time window $0 \leq t \leq T$
\begin{eqnarray}
 \rho_T(R)  \equiv \frac{1}{T} \int_0^T dt \  \delta( R(t)-R)  
\label{rhoempi}
\end{eqnarray}
with the normalization
\begin{eqnarray}
 \int_{-\infty}^{+\infty}  dR \rho_T(R)  = 1
\label{rhonorma}
\end{eqnarray}
The empirical current $j_T (R) $ measures the average of $\dot R(t)$ 
seen on the interval $[0,T]$ when the Riccati variable $R(t)$ at the same time $t$ takes the value $R$
\begin{eqnarray} 
j_T (R) \equiv   \frac{1}{T}  \int_0^T dt \  \dot R(t)   \delta( R(t)-R)  
\label{jempi}
\end{eqnarray}
The empirical density $ \rho_T(R)$ and the empirical current $j_T (R) $
allow to reconstruct the finite-size Lyapunov exponent of Eq. \ref{lambdaTadditive}
\begin{eqnarray}
\lambda_T  =  \frac{1}{T}  \int_0^T dt \left[\alpha[R(t)] + \dot R(t) \beta[R(t)] \right]
= \int_{-\infty}^{+\infty} dR \alpha(R) \rho_T(R) +  \int_{-\infty}^{+\infty} dR \beta(R) j_T(R)
 \label{lambdaTadditiveempi}
\end{eqnarray}
or any other additive functional of this form involving other functions $(\alpha(R),\beta(R))$.


\subsection{ Large deviations at level $2.5$ for the empirical density $\rho_T(R)$ and the empirical current $j_T$ }

\label{sub_2.5}

The derivative of the empirical current of Eq. \ref{jempi}
with respect to $R$ vanishes as $1/T$ for large $T$ and contains only boundary terms 
corresponding to the beginning $t=0$ and to the end $t=T$ of the time-window $[0,T]$
\begin{eqnarray} 
\frac{ dj_T(R)}{dR} \equiv  - \frac{1}{T}  \int_0^T dt \  \dot R(t)   \delta'( R(t)-R)
 = - \frac{1}{T}   \int_0^T dt \frac{d}{dt}  \delta( R(t)-R) = 
\frac{\delta( R(0)-R) - \delta( R(T)-R)}{T} 
\label{jempideri}
\end{eqnarray}
As a consequence, the empirical current $j_T(R) $ is independent of $R$ for large $T$
\begin{eqnarray}
j_T(R) \simeq j_T
\label{juniform}
\end{eqnarray}
and the large deviations at level 2.5 are formulated as follows
\cite{wynants_thesis,maes_diffusion,chetrite_formal,engel,chetrite_HDR,c_reset}.
For large $T$, the joint probability $P_T[ \rho(.), j]  $ to see the empirical density $\rho_T(R)=\rho(R)$
and the empirical current $j_T=j$
 satisfy the large deviation form 
\begin{eqnarray}
&& P_T[ \rho(.), j]   \opsimeq_{T \to +\infty}  \delta \left(\int_{-\infty}^{+\infty} dR \rho(R)  -1  \right)
 e^{- \displaystyle T  I_{2.5}[ \rho(.),j]  }
\label{ld2.5diff}
\end{eqnarray}
where the delta function imposes the normalization constraint of Eq. \ref{rhonorma},
while the explicit rate function for the Fokker-Planck dynamics of Eq. \ref{fokkerplanck}
\begin{eqnarray}
&& I_{2.5}[ \rho(.),j]  = \frac{1}{4} \int_{-\infty}^{+\infty} \frac{ dR } { D(R) \rho(R) }
 \left[ j - \rho(R) F(R) +D(R)  \rho'(R)     \right]^2
\label{i2.5diffusion}
\end{eqnarray}
vanishes only for the steady-state solution of Eq. \ref{steady}.


\subsection{ Generating function $Z_{T}(k)$ of the finite-size Lyapunov exponent $\lambda_T$ }

\label{sub_gene}

The finite-size Lyapunov exponent $\lambda_T$ 
can be rewritten in terms of the empirical density $\rho_T(R)$
and the empirical current $j_T$ as (Eqs \ref{lambdaTadditiveempi} and \ref{juniform})
\begin{eqnarray}
\lambda_T  = \int_{-\infty}^{+\infty} dR \alpha(R) \rho_T(R) +  j_T \beta^{tot}
 \label{lambdaTadditiveempij}
\end{eqnarray}
with the notation $ \beta^{tot}$ introduced in Eq. \ref{betatot}.
So its generating function (Eq \ref{multifz}) can be obtained from the joint probability of Eq. \ref{ld2.5diff} 
 \begin{eqnarray}
Z_T(k) && 
 = \int dj \int {\cal D} \rho(.) P_T[ \rho(.), j]  e^{ \displaystyle  T k 
 \left[ \int_{-\infty}^{+\infty} dR  \alpha(R) \ \rho(R) +   j \beta^{tot} \right] }
\nonumber \\
&&  \opsimeq_{T \to +\infty}  
 \int dj \int {\cal D} \rho(.) 
 \delta \left(\int_{-\infty}^{+\infty} dR \rho(R)  -1  \right)
 e^{ -  \displaystyle T \left[  I_{2.5}[ \rho(.),j] - k \int_{-\infty}^{+\infty} dR \ \alpha(R)  \ \rho(R) - k  j \beta^{tot}\right] }
\label{gene2.5}
\end{eqnarray}
For large $T$, the evaluation via the saddle-point method means that one needs to optimize the functional in the exponential
over the normalized density $\rho(R)$ and over the current $j$.


\subsection{ Optimization of the appropriate lagrangian function $ {\cal L}_k[ \rho(.),\rho'(.),j] $ }

\label{sub_euler}

In order to solve the optimization problem of Eq. \ref{gene2.5},
let us introduce the following Lagrangian function with the Lagrange multiplier $\omega(k)$ associated to the normalization constraint of the density
\begin{eqnarray}
 {\cal L}_k[ \rho(.),\rho'(.),j]   &&  \equiv
 \frac{1}{4  }  \int_{-\infty}^{+\infty} \frac{ dR } { D(R)  \rho(R) }
 \left[ j - \rho(R) F(R) +D(R)  \rho'(R)     \right]^2
 - k  \int_{-\infty}^{+\infty} dR \alpha(R) \rho(R)  - k  j \beta^{tot}
\nonumber \\
&&  +  \omega(k) \left(  \int_{-\infty}^{+\infty} dR \rho(R)  - 1 \right)
 \label{lagrangian}
\end{eqnarray}

The optimization with respect to the current $j$ yields  
\begin{eqnarray}
0 =  \frac{ \partial  {\cal L}_k[ \rho(.),\rho'(.),j] }{\partial j}  &&  =
  \int_{-\infty}^{+\infty} \frac{ dR } { 2 D(R)  }
 \left[ \frac{ j +D(R)  \rho'(R) }{\rho(R) } -  F(R)      \right]
  - k   \beta^{tot}
   \label{lagrangianderij}
\end{eqnarray}

Using the functional derivatives with respect to the density $\rho(R)$ and with respect to its derivative $ \rho'(R) $
\begin{eqnarray}
  \frac{\partial {\cal L}_k[ \rho(.),\rho'(.)]}{\partial  \rho(R)}
  && =   
\frac{ 1 } { 4 D(R) }
 \left[ F^2(R) - \left( \frac{j +D(R)  \rho'(R)    }{\rho(R)} \right)^2    \right] 
- k \alpha(R) + \omega(k)
\nonumber \\
   \frac{\partial {\cal L}_k[ \rho(.),\rho'(.)]}{\partial  \rho'(R)} 
&& = \frac{1}{2}   \left[  \frac{j +D(R)  \rho'(R)    }{\rho(R)}  - F(R)    \right]
 \label{derirr}
\end{eqnarray}
one obtains that 
the Euler-Lagrange equation
for the optimization with respect to the density $\rho(R)$ reads
\begin{eqnarray}
0 && =   \frac{\partial {\cal L}_k[ \rho(.),\rho'(.)]}{\partial  \rho(R)} 
- \frac{d}{d R} \left[   \frac{\partial {\cal L}_k[ \rho(.),\rho'(.)]}{\partial  \rho'(R)} \right]
\nonumber 
\\
&& = 
\frac{ 1 } { 4 D(R) }
 \left[ F^2(R) - \left( \frac{j +D(R)  \rho'(R)    }{\rho(R)} \right)^2    \right] 
- k \alpha(R) + \omega(k)
- \frac{1}{2} \frac{d}{d R} \left[   \frac{j +D(R)  \rho'(R)    }{\rho(R)}  - F(R)     \right]
\label{euler}
\end{eqnarray}
The physical interpretation is that 
\begin{eqnarray}
G(R) \equiv  \frac{j +D(R)  \rho'(R)    }{\rho(R)} 
\label{forceg}
\end{eqnarray}
represents the effective force that would be needed to make the density $\rho(R)$ and the current $j$ typical (Eq \ref{steady}).
The Euler-Lagrange Eq \ref{euler}
corresponds to the following Riccati differential equation for $G(R)$ 
\begin{eqnarray}
 \frac{ G'(R) }{2} + \frac{  G^2(R) }{4D(R)} =\frac{ F'(R) }{2} + \frac{  F^2(R) }{4D(R)}   -k \alpha(R)  + \omega(k)   
  \label{eulerg}
\end{eqnarray}
while the optimization of Eq \ref{lagrangianderij}
over the current $j$
corresponds to the condition
\begin{eqnarray}
0 =  \frac{1}{2}  \int_{-\infty}^{+\infty} \frac{ dR } { D(R)  }
 \left[ G(R) -  F(R)      \right]
  - k   \beta^{tot}
   \label{lagrangianderijg}
\end{eqnarray}


\subsection{ Scaled cumulant generating function $\mu(k)$ from the optimal value of the Lagrangian}

\label{sub_muk}

With the optimal solutions for the effective force $G(R)$, for the density $\rho(R)$ and for the current $j$,
one then needs to evaluate the corresponding optimal value of the Lagrangian of Eq. \ref{lagrangian} 
\begin{eqnarray}
 {\cal L}^{opt}_k  &&  \equiv {\cal L}_k[ \rho(.)=\rho_{opt}(.), \rho'(.)=\rho_{opt}'(.),j=j_{opt}]  
\nonumber \\
&& =  \int_{-\infty}^{+\infty}  dR \frac{ \rho(R) }{4 D(R) }
 \left[   \frac{j +D(R)  \rho'(R)    }{\rho(R)} -F(R)  \right]^2  
  - k  \int_{-\infty}^{+\infty} dR \alpha(R) \rho(R)  - k  j \beta^{tot}  
\nonumber \\
&&  = \int_{-\infty}^{+\infty}  dR \rho(R)  
\left( \frac{ \left[   G(R) -F(R)  \right]^2 }{4D(R)}   - k   \alpha(R)  \right) - k  j \beta^{tot}  
\label{lagrangianopt}
\end{eqnarray}
The Euler-Lagrange Eq. \ref{eulerg} can be used to replace $(-k \alpha(R))$ in order to obtain
\begin{eqnarray}
&& {\cal L}^{opt}_k  
  = 
 \int_{-\infty}^{+\infty}  dR \rho(R)  
\left( \frac{G^2(R)  - G(R) F(R)   }{2D(R)}  
+ \frac{ G'(R) - F'(R) }{2}   - \omega(k)   
  \right) 
- k  j \beta^{tot}  
 \nonumber \\
 &&  
  =  
 \int_{-\infty}^{+\infty}  dR \rho(R)  
\left( \frac{G^2(R)  - G(R) F(R)   }{2D(R)}  
   \right) 
+ \int_{-\infty}^{+\infty}  dR \rho(R)  
\left( \frac{ G'(R) - F'(R) }{2}      \right) 
- \omega(k)  - k  j \beta^{tot} 
 \label{lagrangianoptbis}
\end{eqnarray}
The integration by parts of the second contribution can be simplified via the use of Eq \ref{forceg}
to replace $\rho'(R)$
\begin{eqnarray}
&& \frac{1}{2} \int_{-\infty}^{+\infty}  dR \rho(R) \left( G'(R)-F'(R)  \right) 
 = - \frac{1}{2} \int_{-\infty}^{+\infty}  dR \rho'(R) \left( G(R)-F(R)  \right) 
\nonumber \\
&& = - \frac{1}{2} \int_{-\infty}^{+\infty}  dR \left[\frac{\rho(R) G(R) -j}{D(R)}  \right] \left( G(R)-F(R)  \right)
\nonumber \\
&& = -  \int_{-\infty}^{+\infty}  dR \rho(R)\left[\frac{ G^2(R)  -  G(R) F(R)  }{2D(R)}  \right]  
 + j \int_{-\infty}^{+\infty}  \frac{dR}{2 D(R) }  \left( G(R)-F(R)  \right)  
\label{lagrangianoptsimpli}
\end{eqnarray}
Plugging this result into Eq. \ref{lagrangianoptbis} and using Eq. \ref{lagrangianderijg},
leads to the simplifications
\begin{eqnarray}
&& {\cal L}^{opt}_k    
  =  - \omega(k)  + j \left[ \int_{-\infty}^{+\infty}  \frac{dR}{2 D(R) }  \left( G(R)-F(R)  \right)  
 - k  \beta^{tot} \right] = -  \omega(k)
 \label{lagrangianoptfinal}
\end{eqnarray}

The final result is thus that the generating function of Eq. \ref{gene2.5} displays the asymptotic behavior
 \begin{eqnarray}
Z_T(k)  \opsimeq_{T \to +\infty}  
 e^{ -  \displaystyle T  {\cal L}^{opt}_k     } = e^{   \displaystyle T  \omega(k)     }
\label{gene2.5res}
\end{eqnarray}
The identification with Eq. \ref{multifz} yields that the scaled cumulant generating function $\mu(k)$
simply corresponds to the Lagrange multiplier $\omega(k)$ in the optimization problem of Eq. \ref{lagrangian}
\begin{eqnarray}
\mu(k)=   \omega(k)
\label{lagrangianoptomega}
\end{eqnarray}


\subsection{ Summary of the optimization procedure in two steps}

\label{sec_twosteps}

Let us now summarize how the optimization problem should be solved in two steps.

\subsubsection{ First step concerning the effective force $G(R)$ and the scaled cumulant generating function $\mu(k)=   \omega(k)$ }

\label{subsub_first}

One should first compute together the effective force $G(R)$ and 
the scaled cumulant generating function $\mu(k)=\omega(k)$,
by solving the first-order differential Eq. \ref{eulerg} for $G(R)$ 
\begin{eqnarray}
 \frac{ G'(R) }{2} + \frac{  G^2(R) }{4D(R)} =\frac{ F'(R) }{2} + \frac{  F^2(R) }{4D(R)}   -k \alpha(R)  + \mu(k)   
  \label{eulergmu}
\end{eqnarray}
The condition of periodicity of $G(R)$ on the Riccati ring will select the appropriate solution,
while the condition of Eq. \ref{lagrangianderijg} will determine the value of 
the scaled cumulant generating function $\mu(k)$.

It is now interesting to discuss in more details the physical meaning of Eq. \ref{eulergmu}.
On the right handside of Eq. \ref{eulergmu}, one recognizes the supersymmetric potential of Eq. \ref{vfromu}
associated to the force $F(R)$
\begin{eqnarray}
V^{[F(.)]}(R)  \equiv   \frac{F'(R)}{2} + \frac{ F^2(R) }{4 D(R) } 
\label{vsusyF}
\end{eqnarray}
while the left handside of Eq. \ref{eulergmu} represents the supersymmetric potential 
that would be associated to the effective force $G(R)$
\begin{eqnarray}
V^{[G(.)]}(R)  \equiv    \frac{ G'(R) }{2} + \frac{  G^2(R) }{4D(R)}
\label{vsusyG}
\end{eqnarray}
so that Eq. \ref{eulergmu} can be rewritten as
\begin{eqnarray}
V^{[G(.)]}(R) =  V^{[F(.)]}(R)  -k \alpha(R)  + \mu(k)   
  \label{eulergmususy}
\end{eqnarray}
The physical meaning is that
 the modified potential $( V^{[F(.)]}(R)  -k \alpha(R) )$
should be rewritten as a new supersymmetric potential corresponding to some effective force $G$
up to the constant $\mu(k)$.
If one adds the kinetic term to obtain the full quantum supersymmetric Hamiltonian associated to $G$,
Eq. \ref{eulergmususy} yields in terms of the supersymmetric Hamiltonian $H$ of Eqs
\ref{hamiltonien} \ref{vfromu}
\begin{eqnarray}
H^{[G(.)]} \equiv  - \frac{ \partial  }{\partial R}  D(R) \frac{ \partial  }{\partial R} +V^{[G(.)]}(R) 
  && =  - \frac{ \partial  }{\partial R}  D(R) \frac{ \partial  }{\partial R} +V^{[F(.)]}(R) - k \alpha(R)+ \mu(k) 
  \nonumber \\
  && = H- k \alpha(R)+ \mu(k) 
  \label{gsusy}
\end{eqnarray}
Since the supersymmetric potential $H^{[G(.)]}  $ has by construction a vanishing ground-state energy,
Eq. \ref{gsusy} means that $(- \mu(k))$ is ground-state energy of the Hamiltonian
\begin{eqnarray}
H_k \equiv  H - k \alpha(R)   = - \frac{ \partial  }{\partial R}  D(R) \frac{ \partial  }{\partial R} +
 \frac{F'(R)}{2} + \frac{ F^2(R) }{4 D(R) } 
- k \alpha(R)
\label{hkalpha}
\end{eqnarray}

For the special case of vanishing traces (Eq. \ref{tracezero}) where the function $\alpha^{conserv}[R] $
is directly related to the commutator of Eq. \ref{commutateurconser}
\begin{eqnarray}
\alpha^{conserv}[R] = - \frac{Q^{\dagger} Q - Q Q^{\dagger}}{2} =- \frac{H - \breve{H }}{2}
\label{commutateurconserbis}
\end{eqnarray}
or equivalently to the difference between the supersymmetric Hamiltonian $H \equiv    Q^{\dagger} Q$ of Eq. \ref{hsusy} and its supersymmetric partner $\breve{H } \equiv    Q Q^{\dagger} $,
one obtains that the Hamiltonian of Eq. \ref{hkalpha} can be rewritten as the following linear combination
of $H \equiv    Q^{\dagger} Q$ and $\breve{H } \equiv    Q Q^{\dagger} $
of Eq. \ref{hsusypartner}
\begin{eqnarray}
H_k^{conserv} =  H - k \alpha^{conserv}(R) = \left( 1+\frac{k}{2} \right) H  - \frac{k}{2}  \breve{H } 
=  \left( 1+\frac{k}{2} \right) Q^{\dagger} Q  - \frac{k}{2} Q Q^{\dagger}
\label{hkalphaconserved}
\end{eqnarray}
So the Gallavotti-Cohen symmetry of Eq. \ref{symK2} for the ground-state energy $\mu(k) = \mu(-2-k)  $
corresponds at the level of the Hamiltonians
\begin{eqnarray}
H_{-2-k}^{conserv} = - \frac{k}{2}  H  +\left( 1+\frac{k}{2} \right) \breve{H } 
=  - \frac{k}{2} Q^{\dagger} Q  +\left( 1+\frac{k}{2} \right) Q Q^{\dagger}
\label{hkalphaconservedsym}
\end{eqnarray}
to an exchange of coefficients between the two Hamiltonians 
 $H \equiv    Q^{\dagger} Q$ and $\breve{H } \equiv    Q Q^{\dagger} $.


\subsubsection{ Second step concerning the density $\rho(R)$ and the current $j$ }

\label{subsub_second}

Once the effective force $G(R)$ has been found in the first step described above,
one should then compute together the density $\rho(R)$ and the current $j$
by solving the first-order differential Eq. \ref{forceg} for $\rho(R)$ 
\begin{eqnarray}
  \rho'(R)   - \frac{G(R)}{D(R)} \rho(R) = - \frac{j}{D(R)}
\label{forcegrho}
\end{eqnarray}
This equation corresponds to the steady state Eq \ref{steady}
where the force $F(R)$ has been replaced by the effective force $G(R)$.
The condition of periodicity for $\rho(R)$ on the Riccati ring will determine the appropriate solution,
while the current $j$ will be fixed by the normalization of the density $\rho(R)$.

\subsubsection{ Perturbative solution in the parameter $k$ of the optimization procedure }

In order to see more explicitly how the optimization procedure described above works,
its implementation at the level of the perturbation theory in the parameter $k$
is described in Appendix \ref{sec_pernoneq}.


\subsection{ Large deviations for a periodic potential $U(R)$ with an equilibrium steady-state  }

\label{sec_optieq}

Up to now in this section, we have considered the case of a non-periodic potential $U(R)$
 with a non-equilibrium steady-state $\rho_{st}^{neq}$ and a finite stationary current  $j_{st} \ne 0$ (see subsection \ref{sec_noneqst}).
In the present subsection, we focus on the case where the potential $U(R)$ is periodic on the Riccati ring and
where
the steady state $\rho_{st}$ is thus an equilibrium steady state $\rho^{eq}_{st}$ without current $j_{st}=0$
(subsection \ref{sec_eqst}).
We need to reconsider the whole optimization procedure 
 in order to mention the differences :
 
 (1) The large deviations of Eqs \ref{ld2.5diff} and \ref{i2.5diffusion} is now for the density $\rho(.)$ alone
\begin{eqnarray}
&& P_T[ \rho(.) ]   \opsimeq_{T \to +\infty}  \delta \left(\int_{-\infty}^{+\infty} dR \rho(R)  -1  \right)
 e^{- \displaystyle   \frac{T}{4} \int_{-\infty}^{+\infty} dR \frac{   \rho(R) } { D(R)}
 \left[ D(R)  \frac{ \rho'(R) }{ \rho(R)}  -  F(R)    \right]^2 }
\label{ld2.5diffrho}
\end{eqnarray}

(2) The Lagrangian of Eq. \ref{lagrangian} only contains the density $\rho(.)$ and its derivative $\rho'(.)$
\begin{eqnarray}
 {\cal L}_k[ \rho(.),\rho'(.)]  =
  \frac{1}{4} \int_{-\infty}^{+\infty} dR \frac{   \rho(R) } { D(R)}
 \left[ D(R)  \frac{ \rho'(R) }{ \rho(R)}  -  F(R)    \right]^2
 - k  \int_{-\infty}^{+\infty} dR \alpha(R) \rho(R)  
  +  \omega(k) \left(  \int_{-\infty}^{+\infty} dR \rho(R)  - 1 \right)
 \label{lagrangianrho}
\end{eqnarray}
So the Euler-Lagrange Equation for the optimization over the density is still given by Eq. \ref{eulerg}
\begin{eqnarray}
 \frac{ G'(R) }{2} + \frac{  G^2(R) }{4D(R)} =\frac{ F'(R) }{2} + \frac{  F^2(R) }{4D(R)}   -k \alpha(R)  + \omega(k)   
  \label{eulergeq}
\end{eqnarray}
for the effective force (Eq. \ref{forceg}) that does not contain the current $j$ anymore
\begin{eqnarray}
G(R) \equiv  D(R) \frac{  \rho'(R)    }{\rho(R)} 
\label{forcegeq}
\end{eqnarray}
However the condition of Eq. \ref{lagrangianderijg} that had 
been produced by the optimization over the current
is not present anymore.

(3) So here, with respect to the two steps defined above in subsection \ref{sec_twosteps},
one condition is missing in the first step for $G(R)$ described in subsection \ref{subsub_first}, 
while one variable (namely the current $j$) is missing in the second step described in subsection \ref{subsub_second}.
This means that one should consider the two steps in the opposite order.
One should first solve Eq. \ref{forcegeq} to obtain 
the density $\rho(R)$ in terms of the effective force $G(R)$.
It is convenient to introduce the potential associated to $G(R)$ as in Eq. \ref{UR}
\begin{eqnarray}
U_G(R) && \equiv - \int_0^R dR' \frac{G(R')}{D(R')} 
 \label{URg}
\end{eqnarray}
to write the solution of Eq. \ref{forcegeq}
for the normalized density as the equilibrium solution in the effective potential $U_G$ (Eqs \ref{steadyeq} \ref{partitioneq})
\begin{eqnarray}
  \rho(R) =  \frac{ e^{ -U_G(R)} }{\int_{-\infty}^{+\infty} dR' e^{ -  U_G(R') } } 
 \label{steadyeqg}
\end{eqnarray}
The periodicity requirement for $\rho(R)$ on the Riccati ring
yields that the potential $U_G(R)$ should be periodic on the Riccati ring
in the sense of Eq. \ref{UReq}
\begin{eqnarray}
0=\int_{-\infty} ^{+\infty} dR U_G'(R) && = -  \int_{-\infty} ^{+\infty} dR \frac{G(R)}{D(R)} 
 \label{UReqG}
\end{eqnarray}
This is the additional constraint on $G(R)$ 
that one should take into account to solve 
the Euler-Lagrange Eq. \ref{eulergeq} for $G(R)$ with the periodicity constraint on $G(R)$
in order to be able to determine the scaled cumulant generating function $\mu(k)$.

In order to see more explicitly how this optimization procedure works,
its implementation at the level of the perturbation theory in the parameter $k$
is described in Appendix \ref{sec_pereq}.


\section{ Analysis via the tilted dynamics of the Riccati process  }

\label{sec_tilted}

As recalled in the Introduction, the most standard method to analyze time-additive observables of stochastic processes consists in studying the appropriate 'tilted' dynamical process  \cite{derrida-lecture,sollich_review,lazarescu_companion,lazarescu_generic,jack_review,vivien_thesis,lecomte_chaotic,lecomte_thermo,lecomte_formalism,lecomte_glass,kristina1,kristina2,jack_ensemble,simon1,simon2,simon3,Gunter1,Gunter2,Gunter3,Gunter4,chetrite_canonical,chetrite_conditioned,chetrite_optimal,chetrite_HDR,touchette_circle,touchette_langevin,touchette_occ,touchette_occupation,derrida-conditioned,derrida-ring,bertin-conditioned,touchette-reflected,previousquantum2.5doob,quantum2.5doob}
and the corresponding 'conditioned' process defined via the generalization of Doob's h-transform.

To the best of our knowledge, all previous studies of large deviations properties of Lyapunov exponents
in various models
\cite{titov,zillmer,pumir,titovLloyd,ramola,fyodorov,vanneste,texier_jstat,texier_epl,texier_comtet}
are related to this tilted method despite different languages and different notations.

In this section, the goal is thus to describe this tilted dynamics method to 
analyze the large deviations properties of any additive functional
of the form of Eq. \ref{lambdaTadditive} and to make the link with the approach of the previous section.
The tilted Fokker-Planck generator is introduced in subsection \ref{sub_tilt},
while the associated conditioned process is discussed in subsection \ref{sub_condi}.
The corresponding tilted non-Hermitian quantum operator is introduced in \ref{sub_tiltH}.
After the descriptions of two special simpler cases in subsections \ref{sub_betazero} and \ref{sub_alphazero},
we return to the analysis of the general case in subsection \ref{sub_alphabeta}
and explain the correspondence with the Euler-Lagrange optimization approach in subsection \ref{sub_corresp}.


\subsection { Eigenvalue problem for the tilted Fokker-Planck operator ${\tilde{\cal F}}_k $}

\label{sub_tilt}

Using the the path-integral representation of the Fokker-Planck propagator of Eq. \ref{fokkerplanck}
\begin{eqnarray}
\langle  R_T  \vert e^{T {\cal F} } \vert R_0 \rangle
= \int_{R(0)=R_0}^{R(T)=R_T} {\cal D} [  R(.) ]  
 e^{ - \displaystyle 
  \int_0^{T} dt \left[  \frac{[ \dot R(t) - F(R(t)) ]^2 }{4D(R(t))}
 - \frac{  [D'(R(t)) ]^2}{16D(R(t))} + \frac{  D''(R(t)) }{4} + \frac{  F'(R(t)) }{2}
\right]
 }
\label{pathFP}
\end{eqnarray}
one obtains that the generating function of Eq. \ref{multifz}
 associated to the Lyapunov exponent given by the additive functional of Eq. \ref{lambdaTadditive}
 \begin{eqnarray}
Z_T(k) =  < e^{ \displaystyle  k \int_0^T dt \left(\alpha[R(t)] + \dot R(t) \beta[R(t)] \right) } > 
\label{multifzpath}
\end{eqnarray}
 can be rewritten as the path-integral corresponding
  to the following tilted Fokker-Planck operator ${\tilde{\cal F}}_k $ 
\begin{eqnarray}
 \frac{ \partial { \tilde P}_{t}(R) }{\partial t }   =
 {\tilde{\cal F}}_k {\tilde P}_{t}(.)
&& \equiv - \left( \frac{ \partial  }{\partial R} -k \beta(R) \right)
\left[ F(R)  { \tilde P}_{t}(R)  - D(R)   \left( \frac{ \partial  }{\partial R} -k \beta(R) \right) { \tilde P}_t(R) \right] + k \alpha(R) { \tilde P}_t(R)
\label{fokkerplancktilte}
\end{eqnarray}
that can be further rewritten as the continuity equation for the density ${ \tilde P}_{t}(R) $
\begin{eqnarray}
\frac{ \partial { \tilde P}_{t}(R) }{\partial t }   
= -  \frac{ \partial  { \tilde J}_{t}(R)  }{\partial R} + { \tilde \Sigma}_t(R)
\label{fokkerplancktiltecontinuity}
\end{eqnarray}
where the current ${ \tilde J}_{t}(R) $ involves the tilted force $ \left[  F(R)+ 2 k \beta(R) D(R)  \right] $
\begin{eqnarray}
 { \tilde J}_{t}(R)   = 
\left[  F(R)+ 2 k \beta(R) D(R)  \right] { \tilde P}_{t}(R)  - D(R)    \frac{ \partial { \tilde P}_t(R) }{\partial R} 
 \label{currenttilte}
\end{eqnarray}
while the creation term 
\begin{eqnarray}
 { \tilde \Sigma}_t(R) = \left[ k \beta(R) F(R) + k^2 \beta^2(R) D(R) + k \beta'(R) D(R)+ k \beta(R) D'(R)
   + k \alpha(R) \right] { \tilde P}_t(R)
\label{sigmacreation}
\end{eqnarray}
describes how the density ${ \tilde P}_t(R) $ can be created or destroyed. 
The scaled cumulant generating function $\mu(k)$ of Eq. \ref{multifz} then corresponds to the highest eigenvalue of
the tilted Fokker-Planck operator ${\tilde{\cal F}}_k$ that will dominate the propagator for large $T$
\begin{eqnarray}
\langle R_T \vert e^{ T {\tilde{\cal F}}_k } \vert R_0 \rangle 
\opsimeq_{T \to + \infty} e^{ T \mu(k) } {\tilde r}_k(R_T){\tilde l}_k(R_0)
\label{fokkerplanckspectral}
\end{eqnarray}
 with the corresponding positive right eigenvector ${\tilde r}_k(R)$
\begin{eqnarray}
\mu(k) {\tilde r}_k(R)   = {\tilde{\cal F}}_k {\tilde r}_k(.) 
 = - \left( \frac{ \partial  }{\partial R} -k \beta(R) \right)
\left[ F(R)    - D(R)   \left( \frac{ \partial  }{\partial R} -k \beta(R) \right)  \right] {\tilde r}_k(R)+ k \alpha(R) {\tilde r}_k(R)
\label{fptilteright}
\end{eqnarray}
and the corresponding positive left eigenvector ${\tilde l}_k(R)$ that is not trivial anymore (see Eq. \ref{lefttrivial})
\begin{eqnarray}
\mu(k) {\tilde l}_k(R)   = {\tilde{\cal F}}_k^{\dagger} {\tilde l}_k(.) 
 = \left[ F(R)    + D(R)   \left( \frac{ \partial  }{\partial R} +k \beta(R) \right) \right] 
\left( \frac{ \partial  }{\partial R} +k \beta(R) \right) {\tilde l}_k(R)
+ k \alpha(R) {\tilde l}_k(R)
\label{fptilteleft}
\end{eqnarray}
with the standard normalization 
\begin{eqnarray}
 \int_{-\infty}^{+\infty}  dR  {\tilde l}_k(R)  {\tilde r}_k(R) =1
 \label{fpeigennorma}
\end{eqnarray}
while the periodicity condition on the Riccati ring where $R=\pm \infty$ are glued together
can be imposed by first considering the regularized version on the finite ring $[R_{min},R_{max}]$
as done previously (see Eq  \ref{solReggene})
\begin{eqnarray}
 {\tilde r}_k(R_{min}) && =  {\tilde r}_k(R_{max}) 
\nonumber \\
 {\tilde l}_k(R_{min}) && =  {\tilde l}_k(R_{max}) 
 \label{fpeigenperio}
\end{eqnarray}


\subsection { Corresponding conditioned process constructed via the generalization of Doob's h-transform}

\label{sub_condi}

The probability to see the Riccati variable $R$ at some interior time $0 \ll t \ll T$ for the tilted dynamics
reads using the spectral asymptotic form of Eq. \ref{fokkerplanckspectral} for both time intervals $[0,t]$ and $[t,T]$
\begin{eqnarray}
{\tilde {\cal P}}_{t } (R) 
&& =  \frac{ \langle R_T \vert e^{ (T-t) {\tilde{\cal F}}_k } \vert R \rangle \langle R \vert e^{ t {\tilde{\cal F}}_k } \vert R_0 \rangle}
{ \int dR' \langle R_T \vert e^{ (T-t) {\tilde{\cal F}}_k } \vert R' \rangle \langle R' \vert e^{ t {\tilde{\cal F}}_k } \vert R_0 \rangle } 
 \opsimeq_{ 0 \ll t \ll T }
\frac{ e^{ (T-t) \mu(k) } {\tilde r}_k(R_T){\tilde l}_k(R) e^{ t \mu(k) } {\tilde r}_k(R){\tilde l}_k(R_0)}
{ \int dR'  e^{ (T-t) \mu(k) } {\tilde r}_k(R_T){\tilde l}_k(R') e^{ t \mu(k) } {\tilde r}_k(R'){\tilde l}_k(R_0)      } 
\nonumber \\
&&  \opsimeq_{ 0 \ll t \ll T } {\tilde l}_k(R)  {\tilde r}_k(R) 
\label{conditionnedspectralint}
\end{eqnarray}
Since it is independent of the interior time $t$ as long as $0 \ll t \ll T$,
 it is useful to introduce the notation
\begin{eqnarray}
{\tilde {\tilde \rho}}_k (R) \equiv  {\tilde l}_k(R)  {\tilde r}_k(R) 
 \label{rhokconditioned}
\end{eqnarray}
for the stationary density of the tilted dynamics in the interior time region $0 \ll t \ll T$,
and to construct the corresponding probability-preserving Fokker-Planck operator $ {\tilde {\tilde {\cal F}}}_k$
that has $ {\tilde {\tilde \rho}}_k (R) = {\tilde {\tilde r}}_k (R)  $ as right eigenvector for the eigenvalue zero,
while the corresponding left eigenvector ${\tilde {\tilde l}}_k (R) =1$ is the trivial one (Eq \ref{lefttrivial}),
via the generalization of Doob's h-transform  
\begin{eqnarray}
\frac{ \partial {\tilde {\tilde P}}_{t}(R) }{\partial t }   ={\tilde {\tilde {\cal F}}}_k  {\tilde {\tilde P}}_{t}(.) 
\equiv  
&&  {\tilde l}_k(R) {\tilde{\cal F}}_k   \frac{{\tilde {\tilde P}}_{t}(.)}{ {\tilde l}_k(.) }  - \mu(k) {\tilde {\tilde P}}_{t}(R)
\label{fokkerplancktiltedoubledoob}
\end{eqnarray}
Using the eigenvalue Eqs \ref{fptilteright} and \ref{fptilteleft},
one can indeed check that ${\tilde {\tilde \rho}}_k (R)= {\tilde l}_k(R)  {\tilde r}_k(R) $ is the right eigenvector of eigenvalue zero
\begin{eqnarray}
{\tilde {\tilde {\cal F}}}_k  {\tilde {\tilde \rho}}_k (.)
=  {\tilde l}_k(R) {\tilde{\cal F}}_k   \frac{{\tilde {\tilde \rho}}_k (.)}{ {\tilde l}_k(.) } 
 - \mu(k) {\tilde {\tilde \rho}}_k (R)
 =  {\tilde l}_k(R) \left[ {\tilde{\cal F}}_k   {\tilde r}_k(.)  - \mu(k) {\tilde r}_k(R) \right]
=0
\label{fokkerplancktiltedoubledoobright}
\end{eqnarray}
and that ${\tilde {\tilde l}}_k (R) =1 $ is the left eigenvector of eigenvalue zero
\begin{eqnarray}
{\tilde {\tilde {\cal F}}}_k^{\dagger}  {\tilde {\tilde l}}_k (.) 
=  \frac{1}{ {\tilde l}_k(R) } {\tilde{\cal F}}_k  {\tilde l}_k(.)   - \mu(k) =0
\label{fokkerplancktiltedoubledoobleft}
\end{eqnarray}
Using the eigenvalue Eq. \ref{fptilteleft}
for the left eigenvector ${\tilde l}_k(R)$, one obtains that
Eq. \ref{fokkerplancktiltedoubledoob} can be rewritten more concretely
as the probability-conserving Fokker-Planck Equation
\begin{eqnarray}
\frac{ \partial {\tilde {\tilde P}}_{t}(R) }{\partial t }   =  - \frac{ \partial  }{\partial R} 
\left[ F_{eff}(R) {\tilde {\tilde P}}_{t}(R) 
 - D(R)   \frac{ \partial  {\tilde {\tilde P}}_{t}(R)}{\partial R}  \right]
\label{fokkerplancktiltedouble}
\end{eqnarray}
where the effective force $ F_{eff}(R)$
contains explicitly the left eigenvector ${\tilde l}_k(R) $ of the tilted operator ${\tilde{\cal F}}_k $ (see Eq. \ref{fptilteleft})
\begin{eqnarray}
F_{eff}(R) \equiv F(R) +2 k \beta(R) D(R) + 2 D(R) \frac{{\tilde l}_k'(R)}{{\tilde l}_k(R)} 
 \label{forcetiltedouble}
\end{eqnarray}
As a consequence, the stationary density ${\tilde {\tilde \rho}}_k (R) =  {\tilde l}_k(R)  {\tilde r}_k(R)  $ of Eq. \ref{rhokconditioned}  of this process is associated to the stationary current
\begin{eqnarray}
{\tilde {\tilde j}}_k  = F_{eff}(R)  {\tilde {\tilde \rho}}_k(R)  - D(R)   {\tilde {\tilde \rho}}_k '(R) 
 \label{currenttiltedouble}
\end{eqnarray}
The physical interpretation 
is that ${\tilde {\tilde \rho}}_k (R)$ and ${\tilde {\tilde j}}_k (R) $ represent the density and the current conditioned to the Lyapunov
exponent value $\lambda=\mu'(\lambda)$ of the Legendre transform of Eq. \ref{legendrereci}.

The consistency with the section \ref{sec_optimization} suggests 
that the density $ {\tilde {\tilde \rho}}_k (R)$ and the current ${\tilde {\tilde j}}_k $ should coincide
 with the optimal density $\rho(R)$ and the optimal current $j$ of the Euler-Lagrange optimization,
 so that the effective force $F_{eff}(R)$ should coincide with the force $G(R)$ (see Eq. \ref{forceg}) 
 of the  section \ref{sec_optimization}.
 We will indeed reach this conclusion at the end of the present section after some transformations.

 
 \subsection { Eigenvalue problem for the tilted non-Hermitian quantum operator ${ \tilde H}_k$ }
 
 \label{sub_tiltH}

If one performs the same change of variables as in Eq. \ref{ppsi}
\begin{eqnarray}
{ \tilde P}_{t}(R) = e^{- \frac{U(R)}{2} } { \tilde \psi}_t(R)
\label{ppsitilt}
\end{eqnarray}
the tilted Fokker-Planck dynamics
of Eq. \ref{fokkerplancktilte}
becomes 
\begin{eqnarray}
- \frac{ \partial { \tilde \psi}_t(R) }{\partial t } = { \tilde H}_k { \tilde \psi}_t(R)
\label{schropsitilte}
\end{eqnarray}
with the non-Hermitian Hamiltonian containing the supersymmetric potential  $V(R)$ introduced in Eq. \ref{vfromu}
\begin{eqnarray}
{ \tilde H}_k  = - \left( \frac{ \partial  }{\partial R} -k \beta(R) \right) D(R)\left( \frac{ \partial  }{\partial R} -k \beta(R) \right)
+V(R) - k \alpha(R) 
\label{hamiltoniantilte}
\end{eqnarray}
while its adjoint reads
\begin{eqnarray}
{ \tilde H}^{\dagger}_k   = - \left( \frac{ \partial  }{\partial R} +k \beta(R) \right) D(R)\left( \frac{ \partial  }{\partial R} +k \beta(R) \right)
+V(R) - k \alpha(R) 
\label{hamiltoniantiltedagger}
\end{eqnarray}

In this language, the scaled cumulant generating function $\mu(k)$ of Eq. \ref{multifz}
corresponding to the highest eigenvalue of
the tilted Fokker-Planck operator ${\tilde{\cal F}}_k$ (Eq. \ref{fokkerplanckspectral})
can be determined as follows :
 $[-\mu(k)]$ represents the smallest eigenvalue of
the tilted Hamiltonian $ { \tilde H}_k$
that will dominate the propagator for large $T$
\begin{eqnarray}
\langle R_T \vert e^{ - T {\tilde H}_k } \vert R_0 \rangle \opsimeq_{T \to + \infty} e^{ T \mu(k) } {\tilde \psi}^{[r]}_k(R_T)  {\tilde \psi}^{[l]}_k(R_0)
\label{hamiltonspectral}
\end{eqnarray}
 with the corresponding positive right and left eigenvectors 
\begin{eqnarray}
- \mu(k)  {\tilde \psi}^{[r]}_k(R) && = { \tilde H}_k {\tilde \psi}^{[r]}_k(R)
\nonumber \\
- \mu(k) {\tilde \psi}^{[l]}_k(R)  && = { \tilde H}_k^{\dagger} {\tilde \psi}^{[l]}_k(R)
\label{htilteigen}
\end{eqnarray}
The change of variables of Eq. \ref{ppsitilt} between the Fokker-Planck and the quantum eigenvectors
\begin{eqnarray}
{\tilde r}_k(R) && = e^{- \frac{U(R)}{2} } {\tilde \psi}^{[r]}_k(R)
\nonumber \\
{\tilde l}_k(R) && = e^{+ \frac{U(R)}{2} } {\tilde \psi}^{[l]}_k(R)
\label{changeleftright}
\end{eqnarray}
yields that the boundary conditions of Eq. \ref{fpeigenperio} become for the right and the left quantum eigenstates
\begin{eqnarray}
e^{- \frac{U(R_{min})}{2} } {\tilde \psi}^{[r]}_k(R_{min}) && =e^{- \frac{U(R_{max})}{2} } {\tilde \psi}^{[r]}_k(R_{max})
\nonumber \\
e^{+ \frac{U(R_{min})}{2} } {\tilde \psi}^{[l]}_k(R_{min})   && = e^{+ \frac{U(R_{max})}{2} } {\tilde \psi}^{[l]}_k(R_{max})
 \label{psieigenperio}
\end{eqnarray}

In this quantum language, the stationary density of the conditioned process of Eq. \ref{rhokconditioned} reads
\begin{eqnarray}
{\tilde {\tilde \rho}}_k (R) \equiv  {\tilde l}_k(R)  {\tilde r}_k(R) =  {\tilde \psi}^{[l]}_k(R) {\tilde \psi}^{[r]}_k(R)
 \label{rhokconditionedpsi}
\end{eqnarray}
while the effective force of Eq. \ref{currenttiltedouble} becomes using $U'(R)= - \frac{F(R)}{D(R)}$ (Eq. \ref{UR})
\begin{eqnarray}
F_{eff}(R) && = F(R) +2 k \beta(R) D(R) + 2 D(R) \frac{ d \ln {\tilde l}_k(R)}{dR} 
= F(R) +2 k \beta(R) D(R) + 2 D(R) \left[\frac{U'(R)}{2} + \frac{ d \ln {\tilde \psi}^{[l]}_k(R)}{dR}   \right]
\nonumber \\
&& = 2 k \beta(R) D(R) + 2 D(R)  \frac{ d \ln {\tilde \psi}^{[l]}_k(R)}{dR}   
 \label{currenttiltedoublepsi}
\end{eqnarray}

Before discussing the general case, it is now useful to discuss two special simpler cases.


 \subsection { Special case $ \beta(R) \equiv 0$ : Hermitian Hamiltonian with a tilted scalar potential}
 
  \label{sub_betazero}

If $\beta(R) \equiv 0 $, i.e. if the additive functional of Eq. \ref{lambdaTadditive} contains only the first contribution 
involving $\alpha(R)$
\begin{eqnarray}
\lambda^{[\beta(R) \equiv 0 ]}_T  =  \frac{1}{T}  \int_0^T dt \alpha[R(t)] 
 \label{lambdaTadditivebetazero}
\end{eqnarray}
 then the tilted Hamiltonian of Eq. \ref{hamiltoniantilte} 
\begin{eqnarray}
{ \tilde H}^{[\beta(R) \equiv 0 ]}_k  
= - \frac{ \partial  }{\partial R}  D(R) \frac{ \partial  }{\partial R} +V(R) - k \alpha(R) = H- k \alpha(R) \equiv H_k
\label{hamiltoniantiltebetazero}
\end{eqnarray}
corresponds to the Hamiltonian $H_k= H- k \alpha(R)$ already introduced in Eq. \ref{hkalpha},
that only contains the additional
contribution $(- k \alpha(R) )$ for the scalar potential with respect to the initial Hamiltonian of Eq. \ref{hamiltonien}.
So this corresponds to the standard Feynman-Kac formula for 
functionals associated to scalar potentials \cite{feynman,kac,c_these,review_maj}.


 \subsection { Special case $ \alpha(R) \equiv 0$ : Hermitian Hamiltonian with an electromagnetic vector potential}
 
   \label{sub_alphazero}

If $\alpha(R) \equiv 0 $, i.e. if the additive functional of Eq. \ref{lambdaTadditive} contains only the second contribution 
involving $\beta(R)$ 
\begin{eqnarray}
\lambda_T^{[\alpha(R) \equiv 0 ]}  =  \frac{1}{T}  \int_0^T dt  \dot R(t) \beta[R(t)] 
 \label{lambdaTadditivealphazero}
\end{eqnarray}
then one can consider $k=i q$ with real $q$ in the generating function of Eq. \ref{multifz}
in order to generate the cumulants via the Fourier transform 
\begin{eqnarray}
Z_T^{[\alpha(R) \equiv 0 ]}(k=i q) \equiv  \int d \lambda \ {\cal P}_T(\lambda) \ e^{  i T q \lambda  } = <e^{ \displaystyle  i  q \int_0^T dt  \dot R(t) \beta[R(t)]   } > 
\label{multifzfourier}
\end{eqnarray}
The tilted Hamiltonian of Eq. \ref{hamiltoniantilte} is then Hermitian
\begin{eqnarray}
{ \tilde H}^{[\alpha(R) \equiv 0 ]}_{k=iq}  =  \left( -i \frac{ \partial  }{\partial R} - q  \beta(R) \right) D(R)
\left( -i \frac{ \partial  }{\partial R} -  q \beta(R) \right)
+V(R) = \left({ \tilde H}^{[\alpha(R) \equiv 0 ]}_{k=iq}  \right)^{\dagger}
\label{hamiltoniantiltealphazero}
\end{eqnarray}
The tilt $(- q  \beta(R) )$ with respect to the quantum canonical momentum operator $p=-i \frac{ \partial  }{\partial R}$
can be interpreted as the presence of the electromagnetic potential vector
\begin{eqnarray}
A(R) \equiv q  \beta(R) 
\label{vectorpotential}
\end{eqnarray}
This generalization of the Feynman-Kac formula 
to electromagnetic potential vectors has been much used 
in the context of polymer physics to take into account topological constraints \cite{edw67,wiegel}
and to analyze in detail the winding properties of Brownian paths \cite{orsay_winding1,orsay_winding2,orsay_winding3,orsay_winding4,c_these,winding1,winding2}.

Since the Riccati variable $R \in ]-\infty,+\infty[$ lives on the periodic ring, the quantum problem corresponds to the famous 
Aharonov-Bohm effect, so the important parameter is the global phase accumulated during a lap around the ring
\begin{eqnarray}
\Phi = \int_{-\infty}^{+\infty} dR A(R)= q  \int_{-\infty}^{+\infty} dR \beta(R) = q \beta^{tot}
\label{phaseAB}
\end{eqnarray}
where one recognizes the parameter $\beta^{tot}$ introduced in Eq. \ref{betatot}.
At the level of the eigenvalue Eq. \ref{htilteigen} 
\begin{eqnarray}
\mu(iq)  {\tilde \psi}^{[r]}_{iq}(R) && = { \tilde H}^{[\alpha(R) \equiv 0 ]}_{k=iq} {\tilde \psi}^{[r]}_{iq}(R)
\label{htilteiq}
\end{eqnarray}
the fact that only the Aharonov-Bohm flux of Eq. \ref{phaseAB}
matters can be seen via the gauge transformation that redefines the phase of the wavefunction
\begin{eqnarray}
  {\tilde \psi}^{[r]}_{iq}(R)   = \phi^{[r]}_{iq}(R)  e^{ \displaystyle iq \int_{-\infty}^R dR' \beta(R')}
\label{gauge}
\end{eqnarray}
 Eq. \ref{htilteiq} 
yields
that the eigenvalue equation 
for the new wavefunction $\phi^{[r]}_{iq}(R) $
involves the initial Hamiltonian $H$ of
Eq. \ref{hamiltonien}
\begin{eqnarray}
-\mu(iq)  \phi^{[r]}_{iq}(R)   && = 
\left[ -  \frac{ \partial  }{\partial R}  D(R) \frac{ \partial  }{\partial R}  
+V(R)  \right] \phi^{[r]}_{iq}(R) = H \phi^{[r]}_{iq}(R)
\label{htiltegauge}
\end{eqnarray}
so that the parameter $q$ will only survive in the boundary condition
 for the new wavefunction $\phi^{[r]}_{iq}(R)$
derived from Eq. \ref{psieigenperio}
\begin{eqnarray}
\phi^{[r]}_{iq}(R_{min})  e^{ \displaystyle - \frac{U(R_{min})}{2}  }
 = \phi^{[r]}_{iq}(R_{max})  e^{ \displaystyle - \frac{U(R_{max})}{2} +i  q \beta^{tot} }
 \label{psieigenperiophiab}
\end{eqnarray}
via the Aharonov-Bohm flux $\Phi = q \beta^{tot} = q \int_{R_{min}}^{R_{max}} dR \beta(R)$ of Eq. \ref{phaseAB}.


 \subsection { General case with the two contributions }
 
 \label{sub_alphabeta}
 
When the additive functional of Eq. \ref{lambdaTadditive} contains the two contributions in $\alpha(R)$ and $\beta(R)$,
the scalar potential $V(r)$ is shifted by $(-k \alpha(r))$, while the kinetic part contains
the imaginary vector potential of Eq. \ref{vectorpotential} when one returns to the notation $k=iq$
 \begin{eqnarray}
A(R) \equiv -i k  \beta(R) 
\label{vectorpotentiali}
\end{eqnarray}
 It is still useful to perform the gauge transformations analogous to Eq. \ref{gauge}, even if the factors in the exponential are not phases anymore \cite{derrida-ring}.
 The transformations for the right and for the left eigenvectors
 \begin{eqnarray}
  {\tilde \psi}_{k}^{[r]}(R)  && = \phi^{[r]}_{k}(R)  e^{ \displaystyle k \int_{-\infty}^R dR' \beta(R')}
  \nonumber \\
   {\tilde \psi}_{k}^{[l]}(R)  && = \phi^{[l]}_{k}(R)  e^{ - \displaystyle k \int_{-\infty}^R dR' \beta(R')}
\label{gaugerl}
\end{eqnarray}
lead to the eigenvalue equations (Eqs \ref{htilteigen}) for the new wavefunctions
\begin{eqnarray}
- \mu(k)  \phi^{[r]}_k(R) && = \left[H- k \alpha(R) \right] \phi^{[r]}_k(R)
\nonumber \\
- \mu(k) \phi^{[l]}_k(R)  && = \left[H- k \alpha(R) \right] \phi^{[l]}_k(R)
\label{htiltephieigen}
\end{eqnarray}
 that involve the Hermitian Hamiltonian $H_k= \left[H- k \alpha(R) \right] $ of Eq. \ref{hkalpha},
 while $\beta(R)$ only survives via the global parameter $ \beta^{tot} $ of Eq. \ref{betatot}
 in the new boundary conditions derived from Eqs \ref{psieigenperio}
  \begin{eqnarray}
e^{- \frac{U(R_{min})}{2} }  \phi^{[r]}_{k}(R_{min}) && 
=e^{- \frac{U(R_{max})}{2} + k \beta_{tot}}  \phi^{[r]}_{k}(R_{max})
\nonumber \\
e^{+ \frac{U(R_{min})}{2} }  \phi^{[l]}_{k}(R_{min})   && 
= e^{+ \frac{U(R_{max})}{2} -  k \beta_{tot}}  \phi^{[l]}_{k}(R_{max})
 \label{phieigenperiophi}
\end{eqnarray}

In this language, the stationary density of the conditioned process of Eq. \ref{rhokconditionedpsi} reads
\begin{eqnarray}
{\tilde {\tilde \rho}}_k (R)  =  {\tilde \psi}^{[l]}_k(R) {\tilde \psi}^{[r]}_k(R) = \phi^{[l]}_k(R) \phi^{[r]}_k(R)
 \label{rhokconditionedphi}
\end{eqnarray}
while the effective force of Eq. \ref{currenttiltedoublepsi} becomes 
\begin{eqnarray}
F_{eff}(R) &&  = 2 k \beta(R) D(R) + 2 D(R)  \frac{ d \ln {\tilde \psi}^{[l]}_k(R)}{dR}   
=  2 k \beta(R) D(R) + 2 D(R) \left[ - k \beta(R) + \frac{ d \ln \phi^{[l]}_k(R)}{dR}   \right]
\nonumber \\
&& = 2 D(R)  \frac{ d \ln \phi^{[l]}_k(R)}{dR}
 \label{currenttiltedoublephi}
\end{eqnarray}


 \subsection { Equations for the effective force $F_{eff}(R) $ and correspondence with the Euler-Lagrange optimization}

\label{sub_corresp}
 
Via the change of variables of Eq. \ref{currenttiltedoublephi} between $\phi^{[l]}_k(R) $ and $ F_{eff}(R)$,
 the eigenvalue Schr\"odinger Eq. \ref{htiltephieigen} for $\phi^{[l]}_k(R) $
 \begin{eqnarray}
- \mu(k)  \phi^{[l]}_k(R) && = \left[H- k \alpha(R) \right] \phi^{[l]}_k(R)
= \left[ - \frac{ \partial  }{\partial R}  D(R) \frac{ \partial  }{\partial R} +V(R) - k \alpha(R)\right]\phi^{[l]}_k(R)
\nonumber \\
&& = -D(R) \frac{ d^2   \phi^{[l]}_k(R)  }{dR^2}  -D'(R)\frac{ d   \phi^{[l]}_k(R)  }{dR}
+  \left[ V(R) - k \alpha(R)\right]\phi^{[l]}_k(R)
\label{erwin}
\end{eqnarray}
 translates into the following Riccati equation for $F_{eff}(R)$ using Eq. \ref{vfromu}
   \begin{eqnarray}
\frac{ F_{eff}'(R)}{2}  + \frac{F_{eff}^2(R) }{4D(R)} && =   V(R) - k \alpha(R)+ \mu(k)  
= \frac{ F^2(R) }{4 D(R) } + \frac{F'(R)}{2}- k \alpha(R)+ \mu(k)  
\label{Gphi}
\end{eqnarray}
 that coincides with the Euler-Lagrange Eq. \ref{eulergmu} for $G(R)$ 
 of the section \ref{sec_optimization}.

The boundary condition of Eq. \ref{phieigenperiophi} for $\phi^{[l]}_k(R) $
becomes for $F_{eff}(R)$ using $U'(R)= - \frac{F(R)}{D(R)}$ (Eq. \ref{UR})
 \begin{eqnarray}
  0 && 
  =\int_{R_{min}} ^{R_{max}} dR \left[ \frac{ d \ln \phi^{[l]}_k(R)}{dR}  + \frac{U'(R)}{2}\right] - k \beta^{tot}
  = \int_{R_{min}} ^{R_{max}} dR \left[ \frac{ F_{eff}(R)}{2 D(R)}  - \frac{F(R)}{2 D(R)}\right] - k \beta^{tot}
 \label{boundaryfeff}
\end{eqnarray}
that coincides with the condition of Eq. \ref{lagrangianderijg} for the function $G(R)$ of the section
\ref{sec_optimization}.
 
 In summary, the effective force $F_{eff}(R)$ of the conditioned process introduced in Eq. \ref{forcetiltedouble}
 satisfies the same Riccati equation (Eq \ref{Gphi}), should be also periodic on the Riccati ring,
  and should satisfy the same boundary condition (Eq. \ref{boundaryfeff})
 as the force $G(R)$ appearing in the Euler-Lagrange optimization of the section \ref{sec_optimization}.
 So we can at last conclude that the effective force $F_{eff}(R)$ coincides with $G(R)$
 \begin{eqnarray}
  F_{eff}(R) = G(R)
 \label{identification}
\end{eqnarray}
as it should by consistency between the interpretations of the two points of view.
The density $ {\tilde {\tilde \rho}}_k (R)$ (Eq. \ref{rhokconditioned}) and the current ${\tilde {\tilde j}}_k $
(Eq. \ref{currenttiltedouble}) coincide
 with the optimal density $\rho(R)$ and the optimal current $j$ of the Euler-Lagrange optimization.
The perturbative solutions in $k$ for the effective force $F_{eff}(R) = G(R) $, for the 
density $ {\tilde {\tilde \rho}}_k (R)=\rho(R)$  and the current ${\tilde {\tilde j}}_k =j$
are given in the Appendices \ref{sec_pernoneq} and \ref{sec_pereq}.


\section{ Application to Anderson Localization in a random scalar potential   }

\label{sec_lochalperin}

The phenomenon of Anderson Localization \cite{anderson} has attracted a continuous interest
since its introduction (see the books \cite{bougerol,lifbook,luckbook,crisantibook}
and the reviews \cite{50years,janssenrevue,mirlin_revue2000,mirlinrevue}).
Among the exactly soluble one-dimensional models for the typical Lyapunov exponent (see the recent overviews \cite{tourigny_scatter,tourigny_Lyapunov,tourigny_continuous,tourigny_houches}
and references therein), the Halperin model \cite{halperin} is 
based on the one-dimensional Schr\"odinger equation at energy $E$
for the wave function $\psi(x)$ 
\begin{eqnarray}
E \psi(x) = - \psi''(x)+v(x) \psi(x)
 \label{schrodinger}
\end{eqnarray}
when the scalar potential $v(x)$ is a gaussian white noise of strength $\sigma$ (Eq. \ref{whitenoise})
\begin{eqnarray}
v(x) = \sigma \eta(x) 
 \label{vgauss}
\end{eqnarray}

For the large deviations properties of the Lyapunov exponent,
the Halperin model is also one of the most studied model 
\cite{titov,zillmer,ramola,fyodorov,texier_jstat,texier_epl}
with detailed results concerning the first cumulants \cite{titov,ramola,texier_jstat,texier_epl},
asymptotic forms of large deviations in various regions of the model parameters \cite{zillmer,fyodorov,texier_jstat,texier_epl},
as well as exact results for $\mu(k)$ for even integer values of $k$ \cite{titov,zillmer,fyodorov}.

In this section, it is thus interesting to revisit the Halperin model in order to apply the general formalism described in the previous sections. 
In subsection \ref{sub_2dH}, we recall how the second-order Schr\"odinger Eq. \ref{schrodinger}
can be recast into a two-dimensional matrix Langevin dynamics, while the corresponding dynamics
for the Riccati variable is described in subsection \ref{sub_ricH} with its corresponding Halperin steady state \cite{halperin}.
We then describe how the Riccati process allows to obtain the finite-size Lyapunov exponent $\lambda_L$
in subsection \ref{sub_lyapH} and the finite-size density of states $N_L$ in subsection \ref{sub_dosH}.
Finally, the large deviations properties of the finite-size Lyapunov exponent $\lambda_L$
are discussed in subsection \ref{sub_largedevlyapH}, while the iterative procedure to compute the first cumulants is given in subsection \ref{sub_cumH}.


\subsection{ Corresponding two-dimensional matrix Langevin dynamics  }

\label{sub_2dH}

Using the notations
\begin{eqnarray}
y_1(x) && \equiv \psi(x) 
\nonumber \\
y_2(x) && \equiv \psi'(x) 
 \label{schrodingery12}
\end{eqnarray}
Eq \ref{schrodinger} can be recast into the matrix form
\begin{eqnarray}
\frac{d}{dx}
\begin{pmatrix} 
 y_1(x) 
 \\  y_2(x) 
  \end{pmatrix}
  =
\begin{pmatrix} 
\psi'(x) 
 \\  \psi''(x)
  \end{pmatrix}
= \begin{pmatrix} 
0 &  1 \\
v(x)-E & 0 
 \end{pmatrix} 
 \begin{pmatrix} 
\psi(x) 
 \\  \psi'(x)
  \end{pmatrix}
 = \begin{pmatrix} 
0 &  1 \\
\sigma \eta(x) -E & 0 
 \end{pmatrix} 
\begin{pmatrix} 
 y_1(x) 
 \\  y_2(x) 
  \end{pmatrix}
\label{matrixschrodinger}
\end{eqnarray}
One recognizes the form of Eq. \ref{langevinmatrix} if the spatial coordinate $x$ of the Anderson Localization model
is identified with the time $t$ of the dynamical model of Eq. \ref{langevinmatrix}.
The two matrices $M$ and $W$ 
have only a few non-vanishing matrix elements
\begin{eqnarray}
M= \begin{pmatrix} 
0 &  1 \\
-E & 0 
 \end{pmatrix} 
\ \ \ \ \ \ \ \ \ 
W = \begin{pmatrix} 
0 &  0 \\
\sigma & 0 
 \end{pmatrix} 
\label{mqhalperin}
\end{eqnarray}
In particular, the two traces vanish (Eq. \ref{tracezero}).
The conserved determinant of Eq. \ref{deter} corresponds to the wronskian
of two independent solutions $\psi_{\pm}(x)$ 
of the Schr\"odinger Eq. \ref{schrodinger}
\begin{eqnarray}
\Delta(x) \equiv 
 \begin{vmatrix} 
y_1^+(x) &  y_1^-(x) \\
y_2^+(x) & y_2^-(x)
 \end{vmatrix} 
 =  \begin{vmatrix} 
\psi_+(x) &  \psi_-(x) \\
\psi_+'(x) & \psi_-'(x)
 \end{vmatrix} 
 = \psi_+(x)   \psi_-'(x) -  \psi_+'(x)   \psi_-(x)
\label{deterwronskien}
\end{eqnarray}
and the two Lyapunov exponents are opposite (Eq. \ref{lambdaTminopposite}).

Finally, the vanishing trace condition of Eq. \ref{tracezero} yields
that the symmetry relations of Eq \ref{symK2} are expected to hold for the Halperin model.


\subsection{ Dynamics for the Riccati variable $R(x)=  \frac{\psi'(x)}{\psi(x)}$ }

\label{sub_ricH}

The Riccati variable of Eq. \ref{ricdef}
\begin{eqnarray}
R(x) \equiv \frac{y_2(x)}{y_1(x)} =\frac{\psi'(x)}{\psi(x)} 
 \label{riccontinuous}
\end{eqnarray}
follows the Langevin dynamics of Eqs \ref{langevinric} and \ref{ABric}
\begin{eqnarray}
\frac{dR(x)}{dx} = -E - R^2(x)+ \sigma \eta(x)
 \label{schrodingerric}
\end{eqnarray}
So the corresponding Fokker-Planck Eq. \ref{fokkerplanck}
involves 
the constant diffusion coefficient (see Eq \ref{diffR})
\begin{eqnarray}
 D =  \frac{\sigma^2}{2} 
 \label{diffRh}
\end{eqnarray}
and the quadratic force (see Eq \ref{forceR})
\begin{eqnarray}
F(R) = -E - R^2
 \label{forceRh}
\end{eqnarray}
The  supersymmetric Hamiltonian introduced in Eqs \ref{hamiltonien} \ref{vfromu} \ref{hsusy} reads
\begin{eqnarray}
 H = - D \frac{ d^2  }{d R^2}   +\frac{ ( E+R^2)^2 }{4 D} -R
 =  Q^{\dagger} Q
\label{hamiltonienh}
\end{eqnarray}
with the operators of Eq. \ref{qsusy}
\begin{eqnarray}
Q  && \equiv    \sqrt{ D }  \left( \frac{ d }{ d R} +\frac{E+R^2}{ 2 D}  \right)  
\nonumber \\
Q^{\dagger}  &&\equiv \sqrt{ D} \left(   - \frac{ d }{ d R} +\frac{E+R^2}{ 2 D}  \right)
\label{qsusyh}
\end{eqnarray}
The corresponding cubic potential of Eq. \ref{UR} 
\begin{eqnarray}
U(R) = - \int_0^R dR' \frac{F(R')}{D}  = \frac{E}{D} R  + \frac{R^3}{3 D}
 \label{URh}
\end{eqnarray}
displays the asymptotic behaviors
\begin{eqnarray}
U(R) \opsimeq_{R \to +\infty} +\infty
\nonumber \\
U(R) \opsimeq_{R \to -\infty} -\infty
 \label{URhasymp}
\end{eqnarray}
So the Halperin steady state \cite{halperin} of the Fokker-Planck dynamics (Eq. \ref{steadydiff})
corresponds to the non-equilibrium case of Eq. \ref{noneqsolRegplus}
\begin{eqnarray}
 \rho^{neq}_{st}(R)  =(- j_{st}) e^{ -U(R)}   \int_{-\infty}^{R} \frac{d R'}{D} e^{ U(R') } 
   =  ( - j_{st} )  e^{   - \frac{E}{D} R - \frac{ R^3}{3 D} }
   \int_{-\infty}^{R} \frac{d R'}{D}
 e^{   \frac{E}{D}  R' +\frac{  (R')^3}{3 D} }
\label{noneqsolh}
\end{eqnarray}
with the asymptotic behaviors of Eq. \ref{neqplusrhoinfinity}
\begin{eqnarray}
\rho_{st}^{neq}(R) && \opsimeq_{R \to \pm \infty}   - \frac{ j_{st} }{ D(R) U'(R) }= \frac{j_{st} }{F(R)} =  - \frac{ j_{st} }{ E+R^2} 
\label{neqplusrhoinfinityh}
\end{eqnarray}
while the finite stationary current  $j_{st}$ is fixed by the normalization 
\begin{eqnarray}
1= \int_{-\infty}^{+\infty} dR  \rho^{neq}_{st}(R) 
=  ( - j_{st} ) \int_{-\infty}^{+\infty} dR e^{   - \frac{E}{D} R - \frac{ R^3}{3 D} }
   \int_{-\infty}^{R} \frac{d R'}{D}
 e^{   \frac{E}{D}  R' +\frac{  (R')^3}{3 D} }
\label{jstplush}
\end{eqnarray}


\subsection{ Finite-size Lyapunov exponent $\lambda_L$ as an additive functional of the Riccati process $R(0 \leq x \leq L)$ }

\label{sub_lyapH}

The exponential growth of the absolute value of the
wavefunction $\psi(x)=y_1(x)$ on the interval $0 \leq x \leq L$
\begin{eqnarray}
 \lambda_L     \equiv \frac{1}{L} \ln \left \vert \frac{\psi(L) }{\psi(0)} \right\vert
 =  \frac{1}{L} \int_0^L dx \ \frac{\psi'(x)}{\psi(x)} 
 =  \frac{1}{L} \int_0^L dx \ R(x)
  \label{lyapunovpsi}
\end{eqnarray}
corresponds to the additive functional of the form of Eq. \ref{lambdaTadditive} with the simple functions
\begin{eqnarray}
\alpha[R] && =R
\nonumber \\
\beta[R] && =0
 \label{alphabetah}
\end{eqnarray}
and can be thus obtained from the empirical density $\rho_L(R)$ of the Riccati variable (Eq. \ref{rhoempi}) via Eq. \ref{lambdaTadditiveempi}
\begin{eqnarray}
\lambda_L = \int_{-\infty}^{+\infty} dR R \rho_L(R) 
 \label{lambdaTadditiveempih}
\end{eqnarray}
In particular, its typical value (Eq. \ref{lambdatypsteady}) can be obtained from the steady-state of Eq. \ref{noneqsolh}
\begin{eqnarray}
\lambda^{typ}  = \int_{-\infty}^{+\infty} dR R \rho_{st}^{neq}(R) 
 = ( - j_{st} ) 
 \int_{-\infty}^{+\infty} dR R
 e^{   - \frac{E}{D} R - \frac{ R^3}{3 D} }
   \int_{-\infty}^{R} \frac{d R'}{D}
 e^{   \frac{E}{D}  R' +\frac{  (R')^3}{3 D} }
 \label{lambdaTadditiveempihtyp}
\end{eqnarray}


\subsection{ Finite-size density of states $N_L$   }

\label{sub_dosH}

In Anderson Localization models, another interesting observable is 
the finite-size density of states of energy smaller than $E$
that can be obtained from the density of zeros of the solution $\psi(x)$ of Eq. \ref{schrodinger} at energy $E$
\begin{eqnarray}
 N_L  \equiv \frac{1}{L}  \int_0^L dx \sum_{x_k : \psi(x_k)=0} \delta(x-x_k)
   \label{doscontinuous}
\end{eqnarray}
Since $W_{12}=0$ and $M_{12}=1$ (Eq. \ref{mqhalperin}), Eq. \ref{doszerothetasimpli}
yields that
\begin{eqnarray}
 N_L    =  \frac{1}{L}  \int_0^L dx \  \ \delta\left( \frac{1}{R(x)}  \right) \equiv \hat \rho_L (\zeta=0 )
    \label{doszerothetasimplih}
\end{eqnarray}
corresponds to the empirical density $ \hat \rho_L (\zeta )$ at the origin $\zeta=0$ of the variable $\zeta=\frac{1}{R} $ on the interval $0 \leq x \leq L$. Via the change of variables $ \hat \rho_L (\zeta ) d \zeta = \rho_L(R) dR$,
Eq. \ref{doszerothetasimplih} yields that the finite-size density of state $N_L$
can be obtained from the asymptotic behavior for $R \to \pm \infty$
of the empirical density $\rho_L(R) $ of the Riccati variable $R$
as
\begin{eqnarray}
 N_L  = \lim_{R \to \pm \infty} \left[ R^2 \rho_L(R)  \right]
   \label{doscontinuousempi}
\end{eqnarray}
In particular, its typical value 
from the asymptotic behavior (Eq. \ref{neqplusrhoinfinity}) of the steady-state  of Eq. \ref{noneqsolh}
\begin{eqnarray}
N^{typ} = \lim_{R \to \pm \infty} \left[ R^2 \rho_{st}^{neq}(R)  \right] = - j_{st}
 \label{dostyp}
\end{eqnarray}
corresponds to the steady-state current $(-j_{st})$.


\subsection{ Large deviations for the finite-size Lyapunov exponent $\lambda_L$}

\label{sub_largedevlyapH}

The scaled cumulant generating function $\mu(k)$ 
of the finite-size Lyapunov exponent $\lambda_L$ (Eq. \ref{multifz})
can be analyzed via the optimization procedure summarized in subsection \ref{sec_twosteps}.
Let us write the corresponding explicit equations for the present Halperin model.

The Euler-Lagrange Eq. \ref{eulergmu} for the effective force $G(R)$ reads for the present model
(see Eqs \ref{diffRh} \ref{forceRh} \ref{alphabetah})
\begin{eqnarray}
 \frac{ G'(R) }{2} + \frac{  G^2(R) }{4D} =  \frac{ (E+R^2)^2 }{4D} -(1+k)R   + \mu(k)   
  \label{eulergh}
\end{eqnarray}
while the condition of Eq. \ref{lagrangianderijg} becomes
\begin{eqnarray}
0 =   \int_{-\infty}^{+\infty} dR \left[ G(R) + (E+R^2)     \right]
   \label{lagrangianderijgh}
\end{eqnarray}

Since the present model corresponds to the case of vanishing traces (Eq. \ref{tracezero})
the quantum Hamiltonian of Eqs \ref{hkalpha} 
\begin{eqnarray}
H_k \equiv  H - k \alpha(R)   = - D \frac{ \partial^2  }{\partial R^2} +
  \frac{ (E+R^2)^2 }{4D} -(1+k)R  
\label{hkalphah}
\end{eqnarray}
can be rewritten in terms of the operators $Q$ and $Q^{\dagger}$ of Eqs \ref{qsusyh}
\begin{eqnarray}
H_k =  \left( 1+\frac{k}{2} \right) Q^{\dagger} Q  - \frac{k}{2} Q Q^{\dagger}
\label{hkalphaconservedh}
\end{eqnarray}
as a linear combination of the initial supersymmetric Hamiltonian
$H= Q^{\dagger} Q$ and of its partner $\breve{H} = Q Q^{\dagger}$.
So the Gallavotti-Cohen symmetry of Eq. \ref{symK2} for the ground-state energy $\mu(k) = \mu(-2-k)  $
corresponds at the level of the Hamiltonians to the exchange of coefficients as discussed in Eq. \ref{hkalphaconservedsym}.


\subsection{ First cumulants of the finite-size Lyapunov exponent $\lambda_L$}

\label{sub_cumH}

Since the present model corresponds to the case where
the potential difference diverges $U(+\infty)-U(-\infty) =  +\infty $,
the perturbative solution described in the Appendix subsection \ref{sec_perturbativeddp}
yields that the coefficients $\mu_m$ of the series expansion of $\mu(k)$ (Eq. \ref{gmuk})
are given by Eq. \ref{resmukh}
in terms of the non-equilibrium steady state $\rho^{neq}_{st}(R) $
of Eq. \ref{noneqsolh}
\begin{eqnarray}
\mu_1 && =    \int_{-\infty}^{+\infty}d R  R \rho^{neq}_{st}(R)  = \lambda^{typ}
 \nonumber   \\
\mu_2 && =\int_{-\infty}^{+\infty}d R    \left[  \frac{     G_1^2(R)    }{4D} \right]  \rho^{neq}_{st}(R)
  \nonumber \\
\mu_3 && = \int_{-\infty}^{+\infty}d R \left[  \frac{     G_1(R)  G_2(R)  }{2D} \right]  \rho^{neq}_{st}(R)
  \nonumber \\
 \mu_4  && 
= \int_{-\infty}^{+\infty}d R  \left[    \frac{     G_1(R)  G_3(R)  }{2D} - \frac{     G_2^2(R)    }{4D}  \right]  \rho^{neq}_{st}(R)
\label{resmukhh}
\end{eqnarray}
where the functions $G_m(R)$ are given by Eq. \ref{Gmh} in terms of the potential $U(R)$ of Eq. \ref{URh}
\begin{eqnarray}
 G_{1}(R)  && = -  e^{ U(R) } \int_{R}^{+\infty} d R'  \left[ 2 \mu_1 - 2 R' \right] e^{ -U(R') } 
 \nonumber \\
  G_{2}(R)  && = -  e^{ U(R) } \int_{R}^{+\infty} d R'  \left[ 2 \mu_2  - \frac{     G_1^2(R')    }{2D}\right]   e^{ -U(R') } 
   \nonumber \\
  G_{3}(R)  && = -  e^{ U(R) } \int_{R}^{+\infty} d R'  \left[  2 \mu_3  - \frac{     G_1(R')  G_2(R')  }{D}  \right]   e^{ -U(R') } 
     \nonumber \\
  G_{4}(R)  && = -  e^{ U(R) } \int_{R}^{+\infty} d R'  \left[   2 \mu_4  - \frac{     G_1(R')  G_3(R')  }{D} - \frac{     G_2^2(R')    }{2D}   \right]   e^{ -U(R') } 
\label{Gmhh}
\end{eqnarray}
The iterative procedure goes as follows : one plugs $ \mu_1  = \lambda^{typ}$
into \ref{Gmhh} to compute $G_1(R)$ that can be then plugged into Eq \ref{resmukhh} to compute $\mu_2$,
that can be then plugged into Eq \ref{Gmhh} to compute $G_2(R)$, and so on.


\section{ Anderson Localization in a random supersymmetric potential   }

\label{sec_locsusy}

In the field of exactly soluble one-dimensional Anderson Localization models for the typical Lyapunov exponent,
another much studied model (besides the case of the random scalar potential considered in the previous section) involves a random supersymmetric potential \cite{susy_erikmann,susy_bou,susy_desbois,susy_oshanin,susy_review}:
one considers the Schr\"odinger equation at energy $E$
for the supersymmetric Hamiltonian 
\begin{eqnarray}
E \psi(x) = - \psi''(x)+\left[ w^2(x)+w'(x)  \right] \psi(x) =  \left(\frac{d}{dx} + w(x)  \right)\left(- \frac{d}{dx} + w(x)  \right)\psi(x) 
 \label{schrodingersusy}
\end{eqnarray}
where $w(x)$ involves the white noise $\eta(x)$ of strength $g$
\begin{eqnarray}
w(x) = \nu g^2 + g \eta(x) 
 \label{wgauss}
\end{eqnarray}
while the parameter $\nu \geq 0$ parametrizes the averaged value $<w(x)>= \nu g^2 $ 
and replaces the standard notation $\mu$ of the literature on the Anderson Localization 
model of Eq \ref{schrodingersusy} because the notation $\mu(k)$ is already used in the present article
for the scaled cumulant generating function of Eq. \ref{multifz}. 

For the large deviations properties of the Lyapunov exponent,
many results have been obtained
\cite{ramola,texier_jstat} in particular for the first cumulants and for asymptotic forms of large deviations in various regions of the model parameters.

In this section, we thus revisit this supersymmetric Anderson Localization model in order to apply the general formalism described in the previous sections. 
We mention the reformulation as a two-dimensional matrix Langevin dynamics in subsection \ref{sub_2ds}, 
and the corresponding Riccati dynamics in subsection \ref{sub_rics}.
We then describe how the Riccati process allows to obtain the finite-size density of states $N_L$ in subsection \ref{sub_doss} and the finite-size Lyapunov exponent $\lambda_L$
in subsection \ref{sub_lyaps}.
The steady state of the Riccati process is given in subsection \ref{sub_negE} for the region of negative energy
and in subsection \ref{sub_posE} for the region of positive energy.
Finally, we describe how the large deviations properties of the finite-size Lyapunov exponent $\lambda_L$
can be analyzed in subsection \ref{sub_largedevlyaps}, while the iterative procedure to compute the first cumulants is given in subsection \ref{sub_cumpos} for the region of positive energy and in subsection \ref{sub_cumneg} for the region of negative energy.


\subsection{Corresponding two-dimensional matrix Langevin dynamics   }

\label{sub_2ds}

Using the notations
\begin{eqnarray}
y_1(x) && \equiv \psi(x) 
\nonumber \\
y_2(x) && \equiv \psi'(x) - w(x) \psi(x)
 \label{schrodingery12susy}
\end{eqnarray}
Eq \ref{schrodingersusy} can be recast into the matrix form
\begin{eqnarray}
\frac{d}{dx}
\begin{pmatrix} 
 y_1(x) 
 \\  y_2(x) 
  \end{pmatrix}
  =
\begin{pmatrix} 
\psi'(x) 
 \\  (w^2(x)-E) \psi(x)-w(x) \psi'(x)
  \end{pmatrix}
 = \begin{pmatrix} 
w(x) &  1 \\
 -E & -w(x) 
 \end{pmatrix} 
\begin{pmatrix} 
 y_1(x) 
 \\  y_2(x) 
  \end{pmatrix}
  = \begin{pmatrix} 
\nu g^2+ g \eta(x) &  1 \\
 -E & - \nu g^2 -g \eta(x) 
 \end{pmatrix} 
\begin{pmatrix} 
 y_1(x) 
 \\  y_2(x) 
  \end{pmatrix}
   \nonumber 
\end{eqnarray}
One recognizes the form of Eq. \ref{langevinmatrix} where the two matrices $M$ and $W$ 
read
\begin{eqnarray}
M= \begin{pmatrix} 
\nu g^2 &  1 \\
-E & -\nu g^2 
 \end{pmatrix} 
\ \ \ \ \ \ \ \ \ 
W = \begin{pmatrix} 
g &  0 \\
0 & - g 
 \end{pmatrix} 
\label{mqsusy}
\end{eqnarray}
Again, the two traces vanish (Eq. \ref{tracezero}), the conserved
 determinant of Eq. \ref{deter} corresponds to the wronskian
of two independent solutions $\psi_{\pm}(x)$ 
of the Schr\"odinger Eq. \ref{schrodingersusy}
\begin{eqnarray}
\Delta(x) \equiv 
 \begin{vmatrix} 
y_1^+(x) &  y_1^-(x) \\
y_2^+(x) & y_2^-(x)
 \end{vmatrix} 
 =  \begin{vmatrix} 
\psi_+(x) &  \psi_-(x) \\
\psi_+'(x)- w(x) \psi_+(x) & \psi_-'(x)- w(x) \psi_-(x)
 \end{vmatrix} 
 = \psi_+(x)   \psi_-'(x) -  \psi_+'(x)   \psi_-(x)
\label{deterwronskienbis}
\end{eqnarray}
and the two Lyapunov exponents are opposite (Eq. \ref{lambdaTminopposite}).
 
 Finally, the vanishing trace condition of Eq. \ref{tracezero} yields
that the symmetry relations of Eq \ref{symK2} are also expected to hold for the present model.


\subsection{ Dynamics for the Riccati variable $R(x)=  \frac{\psi'(x)}{\psi(x)}-w(x)$ }

\label{sub_rics}

The Riccati variable of Eq. \ref{ricdef}
\begin{eqnarray}
R(x) \equiv \frac{y_2(x)}{y_1(x)} =\frac{\psi'(x)}{\psi(x)} -w(x)
 \label{riccontinuouss}
\end{eqnarray}
satisfies the Langevin dynamics (see Eqs \ref{langevinric} and \ref{ABric})
\begin{eqnarray}
\frac{dR}{dx} = -E - R^2(x)-2   R(x) w(x)
=  -E -2 \nu g^2 R(x)- R^2(x)-2  g R(x) \eta(x)
 \label{schrodingerrics}
\end{eqnarray}
So the corresponding Fokker-Planck Eq. \ref{fokkerplanck}
involves 
the quadratic diffusion coefficient (see Eq \ref{diffR})
\begin{eqnarray}
 D(R) =  2g^2 R^2 
 \label{diffRs}
\end{eqnarray}
and the quadratic force (see Eq \ref{forceR})
\begin{eqnarray}
F(R) = -E - 2g^2(\nu+1) R- R^2
 \label{forceRs}
\end{eqnarray}

The supersymmetric Hamiltonian of Eq \ref{hamiltonien} \ref{vfromu}
reads
\begin{eqnarray}
 H = Q^{\dagger} Q = - \frac{ \partial  }{\partial R} ( 2g^2 R^2 )  \frac{ \partial  }{\partial R} 
 + \frac{ \left[ E + 2g^2(\nu+1) R+  R^2\right]^2 }{8 g^2 R^2} - g^2(\nu+1)  - R
\label{hamiltoniens}
\end{eqnarray}

Since the derivative $U'(R)$ of Eq. \ref{UR}
\begin{eqnarray}
U'(R) = -  \frac{F(R)}{D(R)}  = 
\frac{E }{2g^2 R^2 } 
+ \frac{\left( 1+\nu \right)}{ R } 
+ \frac{ 1}{2g^2  } 
 \label{URderi}
\end{eqnarray}
is singular at the origin $R \to 0$, one needs to change the arbitrary constant in the definition of 
the potential $U(R)$ of Eq. \ref{UR}.
So we will choose
\begin{eqnarray}
U(R) \equiv - \frac{E }{2g^2 R } + \left( 1+\nu \right)\ln \vert R \vert+ \frac{ R}{2g^2  } 
 \label{URs}
\end{eqnarray}
Since the signs of the leading singularities for $R \to 0^{\pm}$ depend on the sign of the energy $E$,
the global structure of the potential $U(R)$ will be completely different for positive energies $E>0$ and 
for negative energies $E<0$. This complete change in behavior is of course natural for the density of states
as we now recall.


\subsection{ Finite-size density of states $N_L$ }

\label{sub_doss}

Since one has again $W_{12}=0$ and $M_{12}=1$, the analysis of the finite-size density of states $N_L$
is the same as in the previous section (Eqs \ref{doscontinuous} \ref{doszerothetasimplih})
with the same final result (Eq. \ref{doscontinuousempi})
\begin{eqnarray}
 N_L  = \lim_{R \to \pm \infty} \left[ R^2 \rho_L(R)  \right]
   \label{doscontinuousempis}
\end{eqnarray}
Here from the factorized structure of the supersymmetric Hamiltonian 
of Eq. \ref{schrodingersusy}, one knows that the density of states vanishes
in the whole region of negative energies, while it will be finite for positive energy $E>0$
\begin{eqnarray}
 N_L^{[E<0]}  && =0
 \nonumber \\
 N_L^{[E>0]}  && > 0 
   \label{doscontinuousempisnegpos}
\end{eqnarray}


\subsection{ Finite-size Lyapunov exponent $\lambda_L$  }

\label{sub_lyaps}

Let us first recall that at zero energy $E=0$, the ground-state solution of Eq. \ref{schrodingersusy}
\begin{eqnarray}
 \psi^{[E=0]}(x) = e^{ \int_0^x dy w(y)} =  e^{ \nu g^2 x + g \int_0^x dy \eta(y)}
 \label{psizero}
\end{eqnarray}
involves the finite-size Lyapunov exponent
\begin{eqnarray}
 \lambda_L^{[E=0]} \equiv \frac{1}{L} \ln \left \vert \frac{\psi^{[E=0]}(L)}{\psi^{[E=0]}(0)}  \right \vert= \frac{1}{L}  \int_0^L dy w(y)
 =   \nu g^2 +  \frac{g}{L}  \int_0^L dy \eta(y)
 \label{lambdapsizero}
\end{eqnarray}
Since it depends only on the integral of the noise $\eta(y)$ over the region $0 \leq y \leq L$,
$\lambda_L^{[E=0]}  $ is simply Gaussian around its typical value $(\nu g^2)$.

Let us now turn to the other cases with non-zero energy $E \ne 0$.
The exponential growth of the absolute value of the
wavefunction $\psi(x)=y_1(x)$ on the interval $0 \leq x \leq L$ 
is measured by the finite-size Lyapunov exponent
\begin{eqnarray}
 \lambda_L   &&   \equiv \frac{1}{L} \ln \left \vert \frac{\psi(L) }{\psi(0)} \right\vert
 =  \frac{1}{L} \int_0^L dx \ \frac{\psi'(x)}{\psi(x)} 
 =  \frac{1}{L} \int_0^L dx \left[ R(x)+w(x) \right] 
 =  \frac{1}{L} \int_0^L dx \left[ R(x)-  \frac{ R^2(x)+  E  + R'(x)   }{2   R(x) } \right] 
  \nonumber \\
 && =  \frac{1}{L} \int_0^L dx \left[ \frac{R(x)}{2}  -  \frac{   E    }{2   R(x) }-  \frac{   R'(x)   }{2   R(x) } \right] 
 \equiv \frac{1}{L}  \int_0^L dx \left[\alpha[R(x)] +  R'(x) \beta[R(x)] \right]
   \label{lyapunovpsis}
\end{eqnarray}
that corresponds to the general form of Eq. \ref{lambdaTadditive}
with the two functions
\begin{eqnarray}
\alpha[R] && = \frac{R}{2} -  \frac{E}{2 R}
\nonumber \\
\beta[R] && = -  \frac{1}{2 R}
 \label{alphabetahs}
\end{eqnarray}

The important parameter $\beta^{tot}$ of Eq. \ref{betatot}
\begin{eqnarray}
\beta^{tot} \equiv  \int_{-\infty}^{+\infty} dR \beta[R]
 \label{betatots}
\end{eqnarray}
thus requires to use the Cauchy principal value definition of the integral 
both for $R \to 0^{\pm}$ and for $R\to \pm \infty$.
With a small cut-off $\epsilon$ and a large cut-off ${\cal R}$,
one obtains
\begin{eqnarray}
-2 \beta^{tot} && =\lim_{\substack{\epsilon \to 0^+ \\ {\cal R } \to +\infty}} 
\left( \int_{- {\cal R}}^{-\epsilon}  \frac{dR}{R} + \int_{\epsilon}^{ {\cal R}}  \frac{dR}{R}\right)
= \lim_{\substack{\epsilon \to 0^+ \\ {\cal R } \to +\infty}} 
\left(  \ln \vert -\epsilon \vert   -  \ln \vert - {\cal R} \vert
+  \ln \vert  {\cal R} \vert - \ln \vert -\epsilon \vert   \right) =0
 \label{betatotcauchy}
\end{eqnarray}
So $ \beta^{tot}$ actually vanishes, and one needs to consider only the contribution 
in $\alpha[R]$ in Eq. \ref{lambdaTadditiveempij}
\begin{eqnarray}
 \lambda_L   =  \int_{-\infty}^{+\infty} dR \alpha[R] \rho_L(R)
   \label{lyapunovpsisalpha}
\end{eqnarray}


\subsection{ Steady state and typical Lyapunov exponent in the region of negative energies $E<0$  }

\label{sub_negE}

In the region of negative energies $E= - \vert E \vert<0$,
the asymptotic behaviors for $R \to \pm \infty$ and for $R \to 0^{\pm}$ of the potential of Eq. \ref{URs}
\begin{eqnarray}
U(R) && \opsimeq_{R \to +\infty} +\infty
\nonumber \\
U(R) && \opsimeq_{R \to 0^+} + \infty
\nonumber \\
U(R) && \opsimeq_{R \to 0^-}  -\infty
\nonumber \\
U(R) && \opsimeq_{R \to -\infty} -\infty
 \label{URhasympseneg}
\end{eqnarray}
yields that the Fokker-Planck dynamics will converge towards equilibrium in the region 
of positive Riccati variable $R \in ]0,+\infty[$ (see Eq. \ref{steadyeq})
\begin{eqnarray}
  \rho^{eq}_{st}(R) = \frac{e^{ -U(R)} }{ {\cal Z} } =  \frac{ R^{-1-\nu} 
   e^{ - \frac{ 1}{2g^2  } \left( \frac{ \vert E \vert}{ R } +R \right) }
   }{  {\cal Z} } \ \ \ \ \ \ \ {\rm for } \ \ \ \  \ 0<R<+\infty
 \label{steadyeqs}
\end{eqnarray}
with the corresponding partition function
\begin{eqnarray}
 {\cal Z} = \int_{0}^{+\infty} dR e^{ -  U(R) }  = 
 \int_{0}^{+\infty} \frac{dR }{R^{1+\nu }} e^{ - \frac{ 1}{2g^2  } \left( \frac{ \vert E \vert}{ R } +R \right) }
 = 2 \vert E \vert^{-\frac{\nu}{2}} K_{\nu} \left( \frac{ \sqrt{ \vert E \vert } }{g^2} \right)
 \label{partitioneqs}
\end{eqnarray}
in terms of the Bessel function $K_{\nu}(z)$.
If the initial condition happens to be on the negative side $R<0$, the particle will flow towards $U=-\infty$
and will be reinjected at $U=+\infty$ on the positive side $R>0$ where it will be trapped forever with the equilibrium distribution
of Eq. \ref{steadyeqs}.

Accordingly, the density of states vanishes in the whole region of negative energies $E<0$
\begin{eqnarray}
 N^{typ}  = \lim_{R \to \pm \infty} \left[ R^2 \rho_{st}^{eq}(R)  \right] =0
   \label{doscontinuoustyps}
\end{eqnarray}
as already discussed around Eq. \ref{doscontinuousempisnegpos}.

The typical value of the Lyapunov exponent of Eq. \ref{lyapunovpsisalpha}
\begin{eqnarray}
\lambda^{typ} = \int_{0}^{+\infty} dR \alpha[R] \rho_{st}^{eq}(R)   
 \label{lambdaTadditiveempihtypeq}
\end{eqnarray}
involves the function of Eq. \ref{alphabetahs} that reads
 in the region of negative energy $E= - \vert E \vert<0$
\begin{eqnarray}
\alpha[R] && = \frac{R}{2} -  \frac{E}{2 R} = \frac{1}{2} \left( \frac{\vert E \vert}{ R} + R  \right) 
 \label{alphaforneg}
\end{eqnarray}
One recognizes the combination $\frac{1}{2} \left( \frac{\vert E \vert}{ R} + R  \right) $
that appear in the exponential part of the integral defining the partition function of Eq. \ref{partitioneqs}.
As a consequence, the derivative of the partition function ${\cal Z} $ with respect to 
the variable $(\frac{1}{g^2})$
corresponds to the integral
\begin{eqnarray}
- \frac{ \partial  {\cal Z} }{ \partial (\frac{1}{g^2})} &&  = 
 \int_{0}^{+\infty} \frac{dR }{R^{1+\nu }} 
  \frac{ 1}{2  } \left( \frac{ \vert E \vert}{ R } +R \right)
 e^{ - \frac{ 1}{2g^2  } \left( \frac{ \vert E \vert}{ R } +R \right) }
 =  \int_{0}^{+\infty} \frac{dR }{R^{1+\nu }} 
  \alpha[R]
 e^{ - \frac{ 1}{2g^2  } \left( \frac{ \vert E \vert}{ R } +R \right) }
=  {\cal Z} \int_{0}^{+\infty} dR \alpha[R]  \rho^{eq}_{st}(R)
\nonumber \\
&& =  {\cal Z}\lambda^{typ} 
  \label{partitioneqsderi}
\end{eqnarray}
So the typical Lyapunov exponent  reads using the Bessel function $K_{\nu}(z)$ of Eq. \ref{partitioneqs}
\begin{eqnarray}
\lambda^{typ} = -  \frac{ \partial  \ln ({\cal Z}) }{ \partial (\frac{1}{g^2})} 
= - \frac{\sqrt{ \vert E \vert } K_{\nu}' \left( \frac{ \sqrt{ \vert E \vert } }{g^2} \right)}{K_{\nu} \left( \frac{ \sqrt{ \vert E \vert } }{g^2} \right)}
  \label{typfrompartitioneqsderi}
\end{eqnarray}
and one recovers the typical value of Eq. \ref{lambdapsizero} for $E \to 0^-$.


\subsection{ Steady state and typical Lyapunov exponent in the region of positive energies $E>0$  }

\label{sub_posE}

In the region of positive energies $E>0$,
the asymptotic behaviors for $R \to \pm \infty$ and for $R \to 0^{\pm}$ of the potential of Eq. \ref{URs}
\begin{eqnarray}
U(R) && \opsimeq_{R \to +\infty} +\infty
\nonumber \\
U(R) && \opsimeq_{R \to 0^+} - \infty
\nonumber \\
U(R) && \opsimeq_{R \to 0^-}  +\infty
\nonumber \\
U(R) && \opsimeq_{R \to -\infty} -\infty
 \label{URhasympsepos}
\end{eqnarray}
will produce a non-equilibrium steady state where the particle follows the descent of the potential from
$U(R=+\infty)=+\infty$ towards $U(R=0^+)=-\infty$,
  it is then reinjected at $ U(R=0^-)=+\infty$ and follows the descent of the potential towards 
  $U(R=-\infty)=-\infty$ where it is reinjected at $U(R=+\infty)=+\infty$ to begin another cycle.
  As for Eq. \ref{noneqsolReg}, one can regularize the problem on $[R_{min},-\epsilon] \cup [+\epsilon,R_{max}]$ to write the appropriate periodic steady state solution on this Ring
  where $(R_{min}, ,R_{max}) $ are glued together as in Eq. \ref{noneqsolReg},
  and where $(-\epsilon,+\epsilon)$ are glued together.
  Then one takes the limit $R_{min} \to -\infty$, $R_{max} \to +\infty$ and $\epsilon \to 0$
  and one obtain the steady state defined by the two expressions for negative $R<0$ and for positive $R>0$ respectively
\begin{eqnarray}
 \rho^{neq}_{st}(R<0) && =(- j_{st}) e^{ -U(R)}   \int_{-\infty}^{R} \frac{d R'}{D(R')} e^{ U(R') } 
 = 
\frac{  (- j_{st}) } {2 g^2 \vert R \vert^{1+\nu} } 
 e^{  - \frac{ 1}{2g^2  } \left( R- \frac{  E }{ R }  \right) }
   \int_{-\infty}^{R} d R' \vert R' \vert^{\nu-1} e^{  \frac{ 1}{2g^2  } \left( R'-\frac{  E }{ R' }  \right)  }  
 \nonumber \\
  \rho^{neq}_{st}(R>0) && =(- j_{st}) e^{ -U(R)}   \int_{0}^{R} \frac{d R'}{D(R')} e^{ U(R') } 
  = \frac{  (- j_{st}) } {2 g^2 R^{1+\nu} } 
 e^{  - \frac{ 1}{2g^2  } \left( R- \frac{  E }{ R }  \right) }
   \int_{0}^{R} d R' (R' )^{\nu-1} e^{  \frac{ 1}{2g^2  } \left( R'-\frac{  E }{ R' }  \right)  }  
    \label{noneqsols}
\end{eqnarray}
while the steady state current $j_{st}$ is determined by the normalization condition
\begin{eqnarray}
1= \int_{-\infty}^{+\infty} dR  \rho^{neq}_{st}(R) 
 =(- j_{st})  \left[ \int_{-\infty}^{0} dR e^{ -U(R)}   \int_{-\infty}^{R} \frac{d R'}{D(R')} e^{ U(R') } 
 + \int_0^{+\infty} dR e^{ -U(R)}   \int_{0}^{R} \frac{d R'}{D(R')} e^{ U(R') } 
  \right]
 \label{jstmoins}
\end{eqnarray}
This solution is thus very similar to Eq. \ref{noneqsolh}, except that there are now two regions $R<0$
and $R>0$ that are glued together with the following properties at the two junctions on the ring :

(i) For $R \to -\infty$ and for $R \to +\infty$, the saddle-point evaluation of the two integrals 
of Eq. \ref{noneqsols} as in Eq. \ref{neqplusrhoinfinity} and \ref{neqplusrhoinfinityh}
leads to the asymptotic behavior
\begin{eqnarray}
\rho_{st}^{neq}(R) && \opsimeq_{R \to \pm \infty}    \frac{ (-j_{st}) }{ D(R) U'(R) }
= \frac{j_{st} }{F(R)} =   \frac{ (-j_{st}) }{ R^2 + 2g^2(\nu+1) R +E} 
\label{neqplusrhoinfinityhs}
\end{eqnarray}
so that the steady state current $(-j_{st})$ directly corresponds to 
the typical density of states (Eq. \ref{neqplusrhoinfinity}) as in Eq. \ref{dostyp}
\begin{eqnarray}
N^{typ} = \lim_{R \to \pm \infty} \left[ R^2 \rho_{st}^{neq}(R)  \right] = - j_{st}
 \label{dostyps}
\end{eqnarray}

(ii) For $R \to 0^-$ and for $R \to 0^+$, the saddle-point evaluation of the two integrals 
of Eq. \ref{noneqsols} leads to the asymptotic behavior
\begin{eqnarray}
\rho_{st}^{neq}(R) && \opsimeq_{R \to 0^{\pm}}  
= \frac{j_{st} }{F(0)} =   \frac{ (-j_{st}) }{ E} 
\label{neqzero}
\end{eqnarray}

The typical value of the Lyapunov exponent can be obtained 
from the non-equilibrium steady state $\rho_{st}^{neq}(R)  $
of Eq. \ref{noneqsols}
\begin{eqnarray}
 \lambda^{typ} = \int_{-\infty}^{+\infty} dR \alpha[R] \rho_{st}^{neq}(R)  
&& =  \int_{-\infty}^{0} dR \left( \frac{R}{2} -  \frac{E}{2 R}\right) 
\frac{  (- j_{st}) } {2 g^2 \vert R \vert^{1+\nu} } 
 e^{  - \frac{ 1}{2g^2  } \left( R- \frac{  E }{ R }  \right) }
   \int_{-\infty}^{R} d R' \vert R' \vert^{\nu-1} e^{  \frac{ 1}{2g^2  } \left( R'-\frac{  E }{ R' }  \right)  } 
\nonumber \\
&& + \int_{0}^{+\infty} dR \left( \frac{R}{2} -  \frac{E}{2 R}\right) 
\frac{  (- j_{st}) } {2 g^2 R^{1+\nu} } 
 e^{  - \frac{ 1}{2g^2  } \left( R- \frac{  E }{ R }  \right) }
   \int_{0}^{R} d R' (R' )^{\nu-1} e^{  \frac{ 1}{2g^2  } \left( R'-\frac{  E }{ R' }  \right)  } 
 \label{lambdaTadditiveempihtypeqs}
\end{eqnarray}


\subsection{ Large deviations for the finite-size Lyapunov exponent $\lambda_L$  }

\label{sub_largedevlyaps}

The scaled cumulant generating function $\mu(k)$ 
of the finite-size Lyapunov exponent $\lambda_L$ (Eq. \ref{multifz})
is determined by
the Euler-Lagrange Eq. \ref{eulergmu} for the effective force $G(R)$ that reads for the present model
(see Eqs \ref{diffRs} \ref{forceRs} \ref{alphabetahs})
\begin{eqnarray}
 \frac{ G'(R) }{2} + \frac{  G^2(R) }{4 2g^2 R^2 } = 
  \frac{ \left[ E + 2g^2(\nu+1) R+  R^2\right]^2 }{8 g^2 R^2} - g^2(\nu+1)  - R
   -k \left( \frac{R}{2} -  \frac{E}{2 R} \right)  + \mu(k)   
  \label{eulerghs}
\end{eqnarray}
while the condition of Eq. \ref{lagrangianderijg} becomes
\begin{eqnarray}
0 =   \int_{-\infty}^{+\infty} \frac{dR}{R^2}  \left[ G(R) 
+E + 2g^2(\nu+1) R+ R^2
     \right]
   \label{lagrangianderijghs}
\end{eqnarray}

Since the present model corresponds to the case of vanishing traces (Eq. \ref{tracezero})
the quantum Hamiltonian of Eqs \ref{hkalpha} can be rewritten in terms of the operators $Q$ and $Q^{\dagger}$ of Eqs \ref{qsusyh}
\begin{eqnarray}
H_k && = - \frac{ \partial  }{\partial R} ( 2g^2 R^2 )  \frac{ \partial  }{\partial R} 
 + \frac{ \left[ E + 2g^2(\nu+1) R+  R^2\right]^2 }{8 g^2 R^2} - g^2(\nu+1)  - R
  -k \left( \frac{R}{2} -  \frac{E}{2 R} \right) 
 \nonumber \\ &&
 =  \left( 1+\frac{k}{2} \right) Q^{\dagger} Q  - \frac{k}{2} Q Q^{\dagger}
\label{hkalphaconserveds}
\end{eqnarray}
as a linear combination of the initial supersymmetric Hamiltonian
$H= Q^{\dagger} Q$ and of its partner $\breve{H} = Q Q^{\dagger}$.
So the Gallavotti-Cohen symmetry of Eq. \ref{symK2} for the ground-state energy $\mu(k) = \mu(-2-k)  $
corresponds at the level of the Hamiltonians to the exchange of coefficients as discussed in Eq. \ref{hkalphaconservedsym}.


\subsection{ Explicit first cumulants of the finite-size Lyapunov exponent in the region of positive energies $E>0$  }

\label{sub_cumpos}

In the region of positive energy $E>0$, the steady-state is a non-equilibrium steady state $\rho^{neq}_{st}$
  with a steady current $j_{st}$ (see Eq. \ref{noneqsols}), so one can apply the perturbative procedure in two steps described in the Appendix \ref{sec_pernoneq}.
  To take into account the singularities of the potential $U(R)$ both at $R \to 0^{\pm}$
  and at $R \to \pm \infty$ (see Eq. \ref{URhasympsepos}),
  one can first regularize the periodic Riccati ring by considering  $[R_{min},-\epsilon] \cup [+\epsilon,R_{max}]$ to write the appropriate periodic solutions
  where $(R_{min}, ,R_{max}) $ are glued together
  and where $(-\epsilon,+\epsilon)$ are glued together.
  Then one takes the limit $R_{min} \to -\infty$, $R_{max} \to +\infty$ and $\epsilon \to 0$
  to obtain the final solution as for Eq. \ref{noneqsols}.
  
  \subsubsection{ Perturbative solution in $k$ for the effective force $G(R)$ and the scaled cumulant generating function $\mu(k)$}

One obtains that the coefficients $\mu_m$ of the expansion of Eq. \ref{gmuk}
for the scaled cumulant generating function $\mu(k)$ are given by
\begin{eqnarray}
\mu_1 && =    \int_{-\infty}^{+\infty}d R   \alpha(R)  \rho^{neq}_{st}(R)  = \lambda^{typ}
 \nonumber   \\
\mu_2 && =\int_{-\infty}^{+\infty}d R    \left[  \frac{     G_1^2(R)    }{4D} \right]  \rho^{neq}_{st}(R)
  \nonumber \\
\mu_3 && = \int_{-\infty}^{+\infty}d R \left[  \frac{     G_1(R)  G_2(R)  }{2D} \right]  \rho^{neq}_{st}(R)
  \nonumber \\
 \mu_4  && 
= \int_{-\infty}^{+\infty}d R  \left[    \frac{     G_1(R)  G_3(R)  }{2D} - \frac{     G_2^2(R)    }{4D}  \right]  \rho^{neq}_{st}(R)
\label{resmukhs}
\end{eqnarray}
while the functions $G_m(R)$
 that appear in the perturbative expansion of the effective force $G(R)$ (Eq. \ref{gmuk}) read
for the scaled cumulant generating function $\mu(k)$
\begin{eqnarray}
 G_{1}(R)  && = -  e^{ U(R) } \int_{R}^{B^+(R)} d R'  \left[ 2 \mu_1 - 2 \alpha[R' ] \right] e^{ -U(R') } 
 \nonumber \\
  G_{2}(R)  && = -  e^{ U(R) } \int_{R}^{B^+(R)} d R'  \left[ 2 \mu_2  - \frac{     G_1^2(R')    }{2D}\right]   e^{ -U(R') } 
   \nonumber \\
  G_{3}(R)  && = -  e^{ U(R) } \int_{R}^{B^+(R)} d R'  \left[  2 \mu_3  - \frac{     G_1(R')  G_2(R')  }{D}  \right]   e^{ -U(R') } 
     \nonumber \\
  G_{4}(R)  && = -  e^{ U(R) } \int_{R}^{B^+(R)} d R'  \left[   2 \mu_4  - \frac{     G_1(R')  G_3(R')  }{D} - \frac{     G_2^2(R')    }{2D}   \right]   e^{ -U(R') } 
\label{Gmhs}
\end{eqnarray}
where we have introduced the following notation for the upper boundary of the integral
\begin{eqnarray}
B^+(R) && = 0 \ \ \ \ \ \ \ \ \ \ {\rm for } \ \ \ \ \  -\infty <R<0
\nonumber \\
B^+(R)  && = + \infty \ \ \ \ \ \   {\rm for } \ \ \ \ \ \ 0 <R<+\infty
  \label{upperb}
\end{eqnarray}
in order to avoid the writing of two separate definitions for the two regions $R<0$ and $R>0$ as in Eq. \ref{noneqsols}.


  \subsubsection{Perturbative solution in $k$ for the optimal density $\rho(R)$ and the optimal current $j$ }

Here the normalized auxiliary function (Eq. \ref{auxiliaire}) involves 
the same upper boundary of Eq. \ref{upperb} for the integral
\begin{eqnarray}
\Pi(R) && \equiv  (- j_{st} ) \frac{ e^{ U(R) } }{ D(R) }
\int_{R}^{B^+(R)} d R'     e^{ -U(R') } 
   \label{auxiliaires}
\end{eqnarray}
while the $\rho_m(R) $ that appear in the perturbative expansion of the density (Eq. \ref{perrhoj})
\begin{eqnarray}
 \rho_1(R) && = e^{ -U(R)}   \int_{B^-(R)}^{R} dR' e^{ U(R') } 
\left[ \frac{ G_1(R') \rho^{neq}_{st}(R') -   j_1   }{D(R)} \right]
    \nonumber \\
 \rho_2(R) && = e^{ -U(R)}   \int_{B^-(R)}^{R} dR' e^{ U(R') }   
\left[   \frac{ G_2(R') \rho^{neq}_{st}(R') + G_1(R') \rho_1(R')-   j_2   }{D(R')}  \right]
         \nonumber \\
  \rho_3(R) && = e^{ -U(R)}   \int_{B^-(R)}^{R} dR' e^{ U(R') } 
\left[  \frac{ G_3(R') \rho^{neq}_{st}(R') + G_2(R') \rho_1(R)+ G_1(R') \rho_2(R')-   j_3   }{D(R')}  \right]
             \nonumber \\
 \rho_4(R) && = e^{ -U(R)}   \int_{B^-(R)}^{R} dR' e^{ U(R') } 
\left[ \frac{ G_4(R') \rho^{neq}_{st}(R') + G_3(R') \rho_1(R')+ G_2(R') \rho_2(R')+ G_1(R') \rho_3(R')-   j_4   }{D(R')}  \right]
     \label{noneqsolsrho}
\end{eqnarray}
 involve 
the same lower boundary for the integrals as for the non-equilibrium steady state 
$\rho^{neq}_{st}$ of Eq. \ref{noneqsols}
\begin{eqnarray}
B^-(R)  && = - \infty \ \ \ \ \ \   {\rm for } \ \ \ \ \ \  -\infty <R<0      
\nonumber \\
B^-(R) && = 0 \ \ \ \ \ \ \ \ \ \ {\rm for } \ \ \ \ \  0 <R<+\infty
  \label{lowerb}
\end{eqnarray}

Finally, the coefficients $j_m$ that appear in the perturbative expansion of the current (Eq. \ref{perrhoj})
are given by
\begin{eqnarray}
j_1 && =\int_{-\infty}^{+\infty}d R \left[G_1(R) \rho^{neq}_{st}(R)    \right]\Pi(R) 
  \nonumber \\
  j_2 && =\int_{-\infty}^{+\infty}d R \left[G_2(R) \rho^{neq}_{st}(R) + G_1(R) \rho_1(R) \right]\Pi(R) 
   \nonumber \\
 j_3 && =\int_{-\infty}^{+\infty}d R' \left[  G_3(R) \rho^{neq}_{st}(R) + G_2(R) \rho_1(R)+ G_1(R) \rho_2(R)- \right]\Pi(R) 
   \nonumber \\
 j_4 && =\int_{-\infty}^{+\infty}d R \left[ G_4(R) \rho^{neq}_{st}(R) + G_3(R) \rho_1(R)+ G_2(R) \rho_2(R)+ G_1(R) \rho_3(R)  \right]\Pi(R) 
   \label{solujmhs}
\end{eqnarray}


\subsection{ Explicit first cumulants of the finite-size Lyapunov exponent in the region of negative energies $E<0$  }

\label{sub_cumneg}

In the region of negative energies $E<0$ where the steady state is 
an equilibrium steady state $\rho^{eq}_{st}(R)$ on $0<R<+\infty$ (see Eq. \ref{steadyeqs}),
one cannot apply the procedure described in the subsection \ref{sec_optieq} and in the Appendix \ref{sec_pereq}
that were meant for an equilibrium on the periodic Riccati ring.
Here the equilibrium concerns an interval $0<R<+\infty$ with two boundaries at $R=0$ and $R=+\infty$.
As a consequence, one needs to reconsider the perturbative procedure 
and one obtains that it can be decomposed in two steps as follows.

\subsubsection{ Perturbative solution in $k$ for the effective force $G(R)$ and the scaled cumulant generating function $\mu(k)$}

The general solutions of Eq. \ref{eulergper} for the functions $G_m(R)$
that appear in the series expansion 
 of Eq. \ref{gmuk} for $G(R)$
 involves some integration constant $K_m$
\begin{eqnarray}
 G_m(R)  =  e^{ U(R)} 
 \left[ K_m +  \int_{0}^{R} dR' \Omega_m(R')  e^{ - U(R') }   \right] 
\label{solReggeneGmbis}
\end{eqnarray}
In order to avoid the strong divergence as $ e^{ U(R)} $ for $R \to 0$, one needs to select the value
\begin{eqnarray}
 K_m=0
 \label{Kmzero}
\end{eqnarray}
Then in order to avoid the strong divergence as $ e^{ U(R)} $ for $R \to +\infty$,
one needs to impose the vanishing of the integral
\begin{eqnarray}
 0 =  \int_{0}^{+\infty} dR' \Omega_m(R')  e^{ - U(R') }  
\label{integzero}
\end{eqnarray}
So the solution of Eq. \ref{solReggeneGmbis} becomes
\begin{eqnarray}
 G_m(R)  =  e^{ U(R)}   \int_{0}^{R} dR' \Omega_m(R')  e^{ - U(R') }   
 =  - e^{ U(R)}   \int_{R}^{+\infty} dR' \Omega_m(R')  e^{ - U(R') }   
\label{solReggeneGmbisfin}
\end{eqnarray}
i.e. more explicitly for $m=1,2,3,4$ with the explicit expressions 
 of Eq. \ref{eulergper} for $ \Omega_m(R) $
\begin{eqnarray}
G_1(R)  && =  e^{ U(R)}   \int_{0}^{R} dR'   e^{ - U(R') }     \left[  2 \mu_1 - 2 \alpha(R')   \right]
  \nonumber \\
G_2(R)  && =  e^{ U(R)}   \int_{0}^{R} dR'   e^{ - U(R') }  \left[   2 \mu_2  - \frac{     G_1^2(R')    }{2D(R')}  \right]
  \nonumber \\
G_3(R)  && = e^{ U(R)}   \int_{0}^{R} dR'   e^{ - U(R') }  \left[      2 \mu_3  - \frac{     G_1(R')  G_2(R')  }{D(R')}  \right]
  \nonumber \\
G_4(R)  && = e^{ U(R)}   \int_{0}^{R} dR'   e^{ - U(R') }  \left[      2 \mu_4  - \frac{     G_1(R')  G_3(R')  }{D(R')} - \frac{     G_2^2(R')    }{2D(R')}   \right]
  \label{eulergpersoleq}
\end{eqnarray}

The condition of Eq. \ref{integzero} can be rewritten in terms of $\rho_{st}^{eq} (R)$ Eq. \ref{steadyeqs}
\begin{eqnarray}
 0 =  \int_{0}^{+\infty} dR \Omega_m(R)  \rho_{st}^{eq}(R)
\label{integzerorho}
\end{eqnarray}
With the explicit expressions of Eq. \ref{eulergper} for $ \Omega_m(R) $,
one obtains that the coefficients $\mu_m$ of the perturbative expansion of Eq. \ref{gmuk}
for the scaled cumulant generating function $\mu(k)$
are given by
\begin{eqnarray}
   \mu_1 && = \int_{0}^{+\infty} dR    \alpha(R)   \rho_{st}^{eq}(R)
  \nonumber \\
 \mu_2  && =  \int_{0}^{+\infty} dR \frac{     G_1^2(R)    }{4D(R)} \rho_{st}^{eq}(R)
   \nonumber \\
 \mu_3  && =  \int_{0}^{+\infty} dR \frac{     G_1(R)  G_2(R)  }{2D(R)}  \rho_{st}^{eq}(R)
  \nonumber \\
 \mu_4  && = \int_{0}^{+\infty} dR  \left[ \frac{     G_1(R)  G_3(R)  }{2D(R)} - \frac{     G_2^2(R)    }{4D(R)}  \right]\rho_{st}^{eq}(R)
  \label{mueqfin}
\end{eqnarray}


\subsubsection{ Perturbative solution in $k$ for the optimal density $\rho(R)$ }

Even if the perturbative solution for the scaled cumulant generating function $\mu(k)$ has already been obtained in Eq. \ref{mueqfin},
it is nevertheless interesting to derive the corresponding 
perturbative solution in $k$ for the optimal density $\rho(R)$.

The general solution of Eq. \ref{forcegrj} for  the functions $\rho_m(R)$
that appear in the series expansion 
 of Eq. \ref{perrhoj}  for $\rho(R)$
 involves some integration constant $C_m$
\begin{eqnarray}
 \rho_m(R)  =  e^{ - U(R)}  \left[ C_m +  \int_{1}^{R} dR' \Upsilon_m(R')  e^{  U(R') }   \right] 
      \label{solrhoms}
\end{eqnarray}
The normalization condition for the density $\rho(R)$ yields the following condition for any order $m \geq 1$
\begin{eqnarray}
0 =\int_{0}^{+\infty} dR \rho_m(R)  
      \label{solrhomsc}
\end{eqnarray}
This condition determines the constant $C_m$, 
and the solution of Eq. \ref{solrhoms} can be then rewritten with the
notation ${\cal Z}$ de Eq. \ref{partitioneqs}
\begin{eqnarray}
 \rho_m(R)  
 =  \frac{ e^{ - U(R)} }{ {\cal Z}}  \int_{0}^{+\infty} dR'' e^{ -  U(R'') } 
 \int_{R''}^{R} dR' e^{  U(R') }   \Upsilon_m(R')  
       \label{solrhomsfin}
\end{eqnarray}
i.e. more explicitly for $m=1,2,3,4$ with the explicit expressions 
 of Eq. \ref{forcegrj} for $ \Upsilon_m(R)$ 
\begin{eqnarray}
 \rho_1(R)   && =  
  \frac{ e^{ - U(R)} }{ {\cal Z}}  \int_{0}^{+\infty} dR'' e^{ -  U(R'') } 
  \int_{R''}^{R} dR'   e^{  U(R') }
 \left[  \frac{ G_1(R') \rho^{eq}_{st}(R')    }{D(R')} \right]
 \label{rhomeqsolu}    \\
 \rho_2(R)   && = 
  \frac{ e^{ - U(R)} }{ {\cal Z}}  \int_{0}^{+\infty} dR'' e^{ -  U(R'') } 
  \int_{R''}^{R} dR'   e^{  U(R') }
\left[    \frac{ G_2(R') \rho^{eq}_{st}(R') + G_1(R') \rho_1(R')   }{D(R')}  \right]
         \nonumber \\
 \rho_3(R)   && =  
  \frac{ e^{ - U(R)} }{ {\cal Z}}  \int_{0}^{+\infty} dR'' e^{ -  U(R'') } 
 \int_{R''}^{R} dR'   e^{  U(R') }
\left[    \frac{ G_3(R') \rho^{eq}_{st}(R') + G_2(R') \rho_1(R')+ G_1(R') \rho_2(R')   }{D(R')}  \right]
             \nonumber \\
 \rho_4(R)   && =  
  \frac{ e^{ - U(R)} }{ {\cal Z}}  \int_{0}^{+\infty} dR'' e^{ -  U(R'') } 
 \int_{R''}^{R} dR'   e^{  U(R') }
\left[  \frac{ G_4(R') \rho^{eq}_{st}(R') + G_3(R') \rho_1(R')+ G_2(R') \rho_2(R')+ G_1(R') \rho_3(R')   }{D(R')}   \right]    \nonumber
\end{eqnarray}


\section{ Conclusion  }

\label{sec_conclusion}

In this paper, we have revisited the large deviations properties of the finite-time Lyapunov exponent
for the 2D matrix Langevin dynamics in relation with the recent progresses made in the field of 
large deviations for non-equilibrium stochastic processes, where additive functionals can be analyzed from two points of view.
In the first approach, one starts from the large deviations at level 2.5 for the joint probability of the empirical density and of the empirical current of the Riccati process in order to compute
the cumulant generating function of the Lyapunov exponent via some Euler-Lagrange optimization.
We have described in detail how this optimization procedure can be solved perturbatively
in order to obtain explicitly the first cumulants, both when the Riccati steady state is 
a non-equilibrium state with current or an equilibrium state without current.
We have then discussed the second approach, where the cumulant generating function 
is obtained via the spectral analysis of the appropriate tilted Fokker-Planck operator. We have explained how the associated conditioned process constructed via the generalization of Doob's h-transform is useful to clarify the equivalence with the first approach.
Finally, we have applied this general framework to 
one-dimensional Anderson Localization models with random scalar potential
and with random supersymmetric potential.


\section*{Acknowledgements}

It is a pleasure to thank Christophe Texier for his explanations \cite{texier_private}
on the relations between the numerous results obtained in his papers 
\cite{ramola,fyodorov,texier_jstat,texier_epl,texier_comtet}
concerning the large deviations properties of Lyapunov exponents in various models.


\appendix


\section{Explicit first cumulants for the case of a Riccati non-equilibrium steady-state }

\label{sec_pernoneq}

In this Appendix, we consider the case where the potential $U(R)$ introduced in Eq. \ref{UR}
 is non-periodic on the Riccati ring (see Eq. \ref{unonperioreg}) and
where
the steady state $\rho_{st}$ is a non-equilibrium steady state $\rho^{neq}_{st}$ with a steady current $j_{st}$
(see subsection \ref{sec_noneqst}). We describe how the optimization procedure in two steps 
of subsection \ref{sec_twosteps} can be implemented at the level of the perturbation theory
in the parameter $k$.

\subsection{ Perturbative solution in $k$ for the effective force $G(R)$ and the scaled cumulant generating function $\mu(k)$}

\label{sec_perG}

Plugging the perturbative expansions for the effective force $ G(R) $ and for the scaled cumulant generating function $\mu(k)$
\begin{eqnarray}
G(R) && = F(R)+ k G_1(R)+ k^2 G_2(R) + k^3 G_3(R)+ k^4 G_4(R)+O(k^5)
\nonumber \\
\mu(k) && = k \mu_1 + k^2 \mu_2 + k^3 \mu_3+ k^4 \mu_4 +O(k^5)
  \label{gmuk}
\end{eqnarray}
into the Euler-Lagrange Eq. \ref{eulerg} yields 
the following differential equations order by order 
 in terms of the notation $U'(R)=- \frac{F(R)}{D(R)} $ of Eq \ref{UR}
\begin{eqnarray}
  G_1'(R)  - U'(R)  G_1(R) && =  2 \mu_1 - 2 \alpha(R)   \equiv \Omega_1(R)
  \nonumber \\
G_2'(R)  -   U'(R) G_2(R)  && =      2 \mu_2  - \frac{     G_1^2(R)    }{2D(R)}  \equiv \Omega_2(R)
  \nonumber \\
G_3'(R)  -   U'(R)   G_3(R)  && =      2 \mu_3  - \frac{     G_1(R)  G_2(R)  }{D(R)}  \equiv \Omega_3(R)
  \nonumber \\
G_4'(R)  -   U'(R)   G_4(R)  && =      2 \mu_4  - \frac{     G_1(R)  G_3(R)  }{D(R)} - \frac{     G_2^2(R)    }{2D(R)}  \equiv \Omega_4(R)
  \label{eulergper}
\end{eqnarray}
while the condition of Eq. \ref{lagrangianderijg} gives 
\begin{eqnarray}
\beta^{tot} && =   \int_{-\infty}^{+\infty} dR \frac{ G_1(R) } { 2D(R)  } 
  \nonumber \\
0 && =  \int_{-\infty}^{+\infty} dR \frac{ G_m(R) } { 2D(R)  }   \ \ \    \ \ \   \ \ \ {\rm for } \ \ m \geq 2
   \label{lagrangianderijgperint}
\end{eqnarray}
For any $m \geq 1$, the solution $G_m(R) $ for the effective force at order $k^m$ should be periodic on the infinite Riccati ring.
Since we wish to write the generic solution for any case, 
 let us first regularize the problem on a finite ring $R \in [R_{min},R_{max}]$ as in Eq. \ref{noneqsolReg}
to write the solution for $G_m(R)$ as a function of the inhomogeneous term $\Omega_m(R) $ of Eq. \ref{eulergper} as
\begin{eqnarray}
 G_{m}(R) 
 = \frac{ e^{ U(R)} 
 \left[  e^{ U(R_{min})} \int_{R_{min}}^{R} d R'  \Omega_m(R') e^{ -U(R') } 
 +e^{ U(R_{max}) }\int_{R}^{R_{max}} d R'  \Omega_m(R') e^{ -U(R') } 
  \right] }
 {  \left[  e^{ U(R_{min}) }  -  e^{  U(R_{max}) } 
  \right] }
\label{GmReg}
\end{eqnarray}
The conditions of Eq. \ref{lagrangianderijgperint} then involve integrals of the form
\begin{eqnarray}
&&   \int_{R_{min}}^{R_{max}} dR \frac{ G_m(R) } { 2D(R)  }   
  =  \frac{  \int_{R_{min}}^{R_{max}} dR\frac{ e^{ U(R)} }{2 D(R) }
 \left[ e^{ U(R_{min})} \int_{R_{min}}^{R} d R'  \Omega_m(R')   e^{ -U(R') } 
 +e^{ U(R_{max}) }\int_{R}^{R_{max}} d R'  \Omega_m(R') e^{ -U(R') }   \right] }
 {  \left[  e^{ U(R_{min}) }  -  e^{  U(R_{max}) }   \right] }
  \nonumber \\
  && =
 \frac{   \int_{R_{min}}^{R_{max}}d R'  \frac{ \Omega_m(R') }{2} e^{ -U(R') }
 \left[ e^{ U(R_{min})} \int_{R'}^{R_{max}}dR\frac{ e^{ U(R)} }{ D(R) } 
 +e^{ U(R_{max}) } \int_{R_{min}}^{R'}dR\frac{ e^{ U(R)} }{ D(R) }
   \right] }
 {  \left[  e^{ U(R_{min}) }  -  e^{  U(R_{max}) }   \right] }  
   \nonumber \\
  && =
    \frac{  \int_{R_{min}}^{R_{max}}d R'  \frac{ \Omega_m(R') }{2} e^{ -U(R') }
 \left[ e^{ -U(R_{max})} \int_{R'}^{R_{max}}dR\frac{ e^{ U(R)} }{ D(R) } 
 +e^{ -U(R_{min}) } \int_{R_{min}}^{R'}dR\frac{ e^{ U(R)} }{ D(R) }
   \right] }
 {  \left[  e^{ - U(R_{max}) }  -  e^{  U(R_{min}) }   \right] }  
   \nonumber \\
  && =
   \int_{R_{min}}^{R_{max}}d R'  \frac{ \Omega_m(R') }{2} \left( \frac{ \rho^{neq}_{st}(R')}{j_{st}} \right)
   \label{simpliintegg}
\end{eqnarray}
where we have recognized the steady state solution $\rho^{neq}_{st}(R) $ of Eq. \ref{noneqsolReg} and the corresponding steady state current $j_{st}$.

So the coefficients $\mu_m$ of the expansion of Eq. \ref{gmuk}
for the scaled cumulant generating function $\mu(k)$ that appear in the inhomogeneous terms $\Omega_m(R) $ of Eq. \ref{eulergper}
are determined order by order by Eq. \ref{lagrangianderijgperint} using Eq \ref{simpliintegg}
\begin{eqnarray}
\beta^{tot} && =    \int_{R_{min}}^{R_{max}} dR \frac{ G_1(R) } { 2D(R)  } 
=   \int_{R_{min}}^{R_{max}}d R  \frac{ \Omega_1(R) }{2} \left( \frac{ \rho^{neq}_{st}(R)}{j_{st}} \right)
=   \int_{R_{min}}^{R_{max}}d R \left[ \mu_1 -  \alpha(R)
\right] 
\left( \frac{ \rho^{neq}_{st}(R)}{j_{st}} \right)
   \label{eqsformum}   \\
0 && =    \int_{R_{min}}^{R_{max}} dR \frac{ G_2(R) } { 2D(R)  } 
=   \int_{R_{min}}^{R_{max}}d R  \frac{ \Omega_2(R) }{2} \left( \frac{ \rho^{neq}_{st}(R)}{j_{st}} \right)
= \int_{R_{min}}^{R_{max}}d R  \left[  \mu_2  - \frac{     G_1^2(R)    }{4D(R)}
\right] 
\left( \frac{ \rho^{neq}_{st}(R)}{j_{st}} \right)
  \nonumber \\
0 && =    \int_{R_{min}}^{R_{max}} dR \frac{ G_3(R) } { 2D(R)  } 
=   \int_{R_{min}}^{R_{max}}d R \frac{ \Omega_3(R) }{2} \left( \frac{ \rho^{neq}_{st}(R)}{j_{st}} \right) 
= \int_{R_{min}}^{R_{max}}d R \left[ \mu_3  - \frac{     G_1(R)  G_2(R)  }{2D(R)} 
\right] 
\left( \frac{ \rho^{neq}_{st}(R)}{j_{st}} \right)
  \nonumber \\
0 && =    \int_{R_{min}}^{R_{max}} dR \frac{ G_4(R) } { 2D(R)  } 
=   \int_{R_{min}}^{R_{max}}d R \frac{ \Omega_4(R) }{2} \left( \frac{ \rho^{neq}_{st}(R)}{j_{st}} \right)
= \int_{R_{min}}^{R_{max}}d R  \left[   \mu_4  - \frac{     G_1(R)  G_3(R)  }{2D(R)} - \frac{     G_2^2(R)    }{4D(R)} 
\right] 
\left( \frac{ \rho^{neq}_{st}(R)}{j_{st}} \right)
\nonumber
\end{eqnarray}
The normalization of the steady state solution $\rho^{neq}_{st}(R) $ on $[R_{min},R_{max}]$ allows to rewrite these equations as
\begin{eqnarray}
\mu_1 && =    \int_{R_{min}}^{R_{max}}d R   \alpha(R) \rho^{neq}_{st}(R) + \beta^{tot}j_{st} = \lambda^{typ}
 \nonumber   \\
\mu_2 && = \int_{R_{min}}^{R_{max}}d R  \left[  \frac{     G_1^2(R)    }{4D(R)} \right] 
 \rho^{neq}_{st}(R)
  \nonumber \\
\mu_3 && = \int_{R_{min}}^{R_{max}}d R \left[  \frac{     G_1(R)  G_2(R)  }{2D(R)} \right] 
 \rho^{neq}_{st}(R)
  \nonumber \\
 \mu_4  && 
= \int_{R_{min}}^{R_{max}}d R  \left[    \frac{     G_1(R)  G_3(R)  }{2D(R)} - \frac{     G_2^2(R)    }{4D(R)}  \right] 
 \rho^{neq}_{st}(R)
\label{resmuk}
\end{eqnarray}
The first equation corresponds to the typical value $\lambda^{typ} $ of Eq. \ref{lambdatypsteady} as it should (Eq \ref{c1}).
The other equations allow to compute $\mu_{m=2,3,4}$ in terms of the solutions $G_m(R)$ given in Eq \ref{GmReg}.

In the limit $R_{min} \to -\infty$ and $R_{max} \to +\infty$, one then needs to 
take into account the behavior of the potential $U(R)$ for $R \to \pm \infty$
to obtain the appropriate solution on the infinite Riccati ring.
One example is described below in subsection \ref{sec_perturbativeddp}.


\subsection{ Perturbative solution in $k$ for the optimal density $\rho(R)$ and the optimal current $j$}

\label{sec_perrho}

Even if the perturbative solution for the scaled cumulant generating function $\mu(k)$ has already been obtained in Eq. \ref{resmuk},
it is nevertheless interesting to derive the corresponding 
perturbative solutions in $k$ for the optimal density $\rho(R)$ and the optimal current $j$
\begin{eqnarray}
\rho(R) && = \rho^{neq}_{st}(R)+ k \rho_1(R)+ k^2 \rho_2(R)+ k^3 \rho_3(R)+ k^4 \rho_4(R)+O(k^5)
\nonumber \\
j && = j_{st} + k j_1 +k^2 j_2+ k^3 j_3 +k^4 j_4 +O(k^5)
  \label{perrhoj}
\end{eqnarray}
Plugging these perturbative expansions and the perturbative expansion for the optimal force $ G(R) $ (Eq. \ref{gmuk})
into Eq. \ref{forceg} with the notation $U'(R)=- \frac{F(R)}{D(R)} $ of
Eq \ref{UR} leads to the following differential equations 
 order by order 
\begin{eqnarray}
    \rho_1'(R) + U'(R) \rho_1(R) && = \frac{ G_1(R) \rho^{neq}_{st}(R) -   j_1   }{D(R)}\equiv \Upsilon_1(R)
    \nonumber \\
     \rho_2'(R) + U'(R) \rho_2(R) && = \frac{ G_2(R) \rho^{neq}_{st}(R) + G_1(R) \rho_1(R)-   j_2   }{D(R)}  \equiv \Upsilon_2(R)
         \nonumber \\
     \rho_3'(R) + U'(R) \rho_3(R) && = \frac{ G_3(R) \rho^{neq}_{st}(R) + G_2(R) \rho_1(R)+ G_1(R) \rho_2(R)-   j_3   }{D(R)}  \equiv \Upsilon_3(R)
             \nonumber \\
\rho_4'(R) + U'(R) \rho_4(R) && = \frac{ G_4(R) \rho^{neq}_{st}(R) + G_3(R) \rho_1(R)+ G_2(R) \rho_2(R)+ G_1(R) \rho_3(R)-   j_4   }{D(R)}  \equiv \Upsilon_4(R)
      \label{forcegrj}
\end{eqnarray}
For any $m \geq 1$, the solution $\rho_m(R) $ for the density at order $k^m$
should be periodic on the Riccati ring, so 
one can write the solution in terms of the inhomogeneous terms $ \Upsilon_m(R)$ of Eq. \ref{forcegrj} 
 in the form analogous to Eq. \ref{noneqsolReg}
\begin{eqnarray}
 \rho^{neq}_{m}(R) 
 = \frac{ e^{ -U(R)} 
 \left[ e^{ -U(R_{min})} \int_{R_{min}}^{R} d R'  \Upsilon_m(R') e^{ U(R') } 
 +e^{ -U(R_{max}) }\int_{R}^{R_{max}} d R'  \Upsilon_m(R')  e^{ U(R')  } 
  \right] }
 {  \left[  e^{ -U(R_{min}) }  -  e^{ - U(R_{max}) }   \right] }
\label{noneqsolRegm}
\end{eqnarray}
 The normalization condition for the density $\rho(R)$ yields the following condition for any order $m \geq 1$ 
\begin{eqnarray}
0 && =     \int_{R_{min}}^{R_{max}} dR \rho_m(R)
= \frac{  \int_{R_{min}}^{R_{max}} dR e^{ -U(R)} 
 \left[ e^{ -U(R_{min})} \int_{R_{min}}^{R} d R'  \Upsilon_m(R') e^{ U(R') } 
 +e^{ -U(R_{max}) }\int_{R}^{R_{max}} d R'  \Upsilon_m(R')  e^{ U(R')  } 
  \right] }
 {  \left[  e^{ -U(R_{min}) }  -  e^{ - U(R_{max}) }   \right] }
 \nonumber \\
 && 
= \frac{ \int_{R_{min}}^{R_{max}}d R'  \Upsilon_m(R') e^{ U(R') } 
 \left[ e^{ -U(R_{min})}  \int_{R'}^{R_{max}} dR e^{ -U(R)} 
 +e^{ -U(R_{max}) } \int_{R_{min}}^{R'} dR e^{ -U(R)}
  \right] }
 {  \left[  e^{ -U(R_{min}) }  -  e^{ - U(R_{max}) }   \right] }
  \nonumber \\
 && 
=\int_{R_{min}}^{R_{max}}d R'  \Upsilon_m(R') e^{ U(R') } 
\left[  \frac{  e^{ U(R_{max})}  \int_{R'}^{R_{max}} dR e^{ -U(R)} 
 +e^{ U(R_{min}) } \int_{R_{min}}^{R'} dR e^{ -U(R)} }
 {    e^{ U(R_{max}) }  -  e^{  U(R_{min}) }    }  \right]
   \label{normarhom}
\end{eqnarray}
These equations determine order by order the current coefficients $j_m$ 
that appear in the inhomogeneous terms $\Upsilon_m(R) $ of Eq. \ref{forcegrj}.
It is thus convenient to introduce the probability distribution
\begin{eqnarray}
\Pi(R') && \equiv  (- j_{st} ) \frac{ e^{ U(R') } }{ D(R') }
\left[  \frac{  e^{ U(R_{max})}  \int_{R'}^{R_{max}} dR e^{ -U(R)} 
 +e^{ U(R_{min}) } \int_{R_{min}}^{R'} dR e^{ -U(R)} }
 {    e^{ U(R_{max}) }  -  e^{  U(R_{min}) }    }  \right]
   \label{auxiliaire}
\end{eqnarray}
that is normalized as a consequence of Eq. \ref{noneqsolRegnorma} defining the stationary current $j_{st}$
\begin{eqnarray}
&& \int_{R_{min}}^{R_{max}}d R' \Pi(R') 
 =   (- j_{st} )  
\left[  \frac{  e^{ U(R_{max})} \int_{R_{min}}^{R_{max}}d R' \frac{ e^{ U(R') } }{ D(R') } \int_{R'}^{R_{max}} dR e^{ -U(R)} 
 +e^{ U(R_{min}) } \int_{R_{min}}^{R_{max}}d R' \frac{ e^{ U(R') } }{ D(R') }\int_{R_{min}}^{R'} dR e^{ -U(R)} }
 {    e^{ U(R_{max}) }  -  e^{  U(R_{min}) }    }  \right]
 \nonumber \\
 && =   (- j_{st} )  
\left[  \frac{  e^{ U(R_{max})} 
\int_{R_{min}}^{R_{max}} dR e^{ -U(R)} \int_{R_{min}}^{R}d R' \frac{ e^{ U(R') } }{ D(R') }  
 +e^{ U(R_{min})  }
\int_{R_{min}}^{R_{max}} dR e^{ -U(R)}  \int_{R}^{R_{max}}d R' \frac{ e^{ U(R') } }{ D(R') } }
 {    e^{ U(R_{max}) }  -  e^{  U(R_{min}) }    }  \right]
 \nonumber \\
&& = 
 \frac{(- j_{st})   \left[ e^{ -U(R_{min})}  \int_{R_{min}}^{R_{max}} dR e^{ -U(R)} \int_{R_{min}}^{R} \frac{d R'}{D(R')} e^{ U(R') } 
 +e^{ -U(R_{max}) }  \int_{R_{min}}^{R_{max}} dR e^{ -U(R)} \int_{R}^{R_{max}} \frac{d R'}{D(R')} e^{ U(R')  } 
  \right]
  } {  \left[  e^{ -U(R_{min}) }  -  e^{ - U(R_{max}) }   \right] } =1
   \label{auxiliairenorma}
\end{eqnarray}
The solutions of Eqs \ref{normarhom} for the current coefficients $j_m$ 
can be then rewritten as
\begin{eqnarray}
j_1 && =\int_{R_{min}}^{R_{max}}d R \left[G_1(R) \rho^{neq}_{st}(R)    \right]\Pi(R) 
  \nonumber \\
  j_2 && =\int_{R_{min}}^{R_{max}}d R \left[G_2(R) \rho^{neq}_{st}(R) + G_1(R) \rho_1(R) \right]\Pi(R) 
   \nonumber \\
 j_3 && =\int_{R_{min}}^{R_{max}}d R' \left[  G_3(R) \rho^{neq}_{st}(R) + G_2(R) \rho_1(R)+ G_1(R) \rho_2(R)- \right]\Pi(R) 
   \nonumber \\
 j_4 && =\int_{R_{min}}^{R_{max}}d R \left[ G_4(R) \rho^{neq}_{st}(R) + G_3(R) \rho_1(R)+ G_2(R) \rho_2(R)+ G_1(R) \rho_3(R)  \right]\Pi(R) 
   \label{solujm}
\end{eqnarray}

In the limit $R_{min} \to -\infty$ and $R_{max} \to +\infty$, one then needs to 
take into account the behavior of the potential $U(R)$ for $R \to \pm \infty$
to obtain the appropriate solution on the infinite Riccati ring.
Let us now describe one example.


\subsection{ Perturbative solution when the potential difference diverges
$U(+\infty)-U(-\infty) =  +\infty $ }

\label{sec_perturbativeddp}

When the potential difference diverges $U(+\infty)-U(-\infty) =  +\infty $,
we have described the non-equilibrium steady state in the subsection \ref{sec_ddpinfinite}, 
and it is thus interesting to write the corresponding perturbative solution in $k$.

In the solution of Eq. \ref{GmReg}, only the right terms survive in the numerator and denominator in the limit $R_{min} \to -\infty$ and $R_{max} \to +\infty$ 
\begin{eqnarray}
 G_{m}(R) 
 = - e^{ U(R) }
 \int_{R}^{+\infty} d R'  \Omega_m(R') e^{ -U(R') } 
\label{GmRegddp}
\end{eqnarray}
i.e. more explicitly for $m=1,2,3,4$
\begin{eqnarray}
 G_{1}(R)  && = -  e^{ U(R) } \int_{R}^{+\infty} d R'  \left[ 2 \mu_1 - 2 R' \right] e^{ -U(R') } 
 \nonumber \\
  G_{2}(R)  && = -  e^{ U(R) } \int_{R}^{+\infty} d R'  \left[ 2 \mu_2  - \frac{     G_1^2(R')    }{2D}\right]   e^{ -U(R') } 
   \nonumber \\
  G_{3}(R)  && = -  e^{ U(R) } \int_{R}^{+\infty} d R'  \left[  2 \mu_3  - \frac{     G_1(R')  G_2(R')  }{D}  \right]   e^{ -U(R') } 
     \nonumber \\
  G_{4}(R)  && = -  e^{ U(R) } \int_{R}^{+\infty} d R'  \left[   2 \mu_4  - \frac{     G_1(R')  G_3(R')  }{D} - \frac{     G_2^2(R')    }{2D}   \right]   e^{ -U(R') } 
\label{Gmh}
\end{eqnarray}
while Eq \ref{resmuk} read
\begin{eqnarray}
\mu_1 && =    \int_{-\infty}^{+\infty}d R  R \rho^{neq}_{st}(R)  = \lambda^{typ}
 \nonumber   \\
\mu_2 && =\int_{-\infty}^{+\infty}d R    \left[  \frac{     G_1^2(R)    }{4D} \right]  \rho^{neq}_{st}(R)
  \nonumber \\
\mu_3 && = \int_{-\infty}^{+\infty}d R \left[  \frac{     G_1(R)  G_2(R)  }{2D} \right]  \rho^{neq}_{st}(R)
  \nonumber \\
 \mu_4  && 
= \int_{-\infty}^{+\infty}d R  \left[    \frac{     G_1(R)  G_3(R)  }{2D} - \frac{     G_2^2(R)    }{4D}  \right]  \rho^{neq}_{st}(R)
\label{resmukh}
\end{eqnarray}
So the iterative procedure goes as follows : one plugs $ \mu_1  = \lambda^{typ}$
into \ref{GmReg} to compute $G_1(R)$ that can be then plugged into Eq \ref{resmukh} to compute $\mu_2$,
that can be then plugged into Eq \ref{GmReg} to compute $G_2(R)$, and so on.

In the solution of Eq. \ref{noneqsolRegm}, only the left terms survive in the numerator and denominator in the limit $R_{min} \to -\infty$ and $R_{max} \to +\infty$ 
\begin{eqnarray}
 \rho_{m}(R) 
 = e^{ -U(R)} 
 \int_{-\infty}^{R} d R'  \Upsilon_m(R') e^{ U(R') } 
\label{noneqsolRegmdd}
\end{eqnarray}
i.e. more explicitly for $m=1,2,3,4$
\begin{eqnarray}
 \rho_1(R)  && = e^{ -U(R)}   \int_{-\infty}^{R} d R'  \left[\frac{ G_1(R) \rho^{neq}_{st}(R) -   j_1   }{D} \right]  e^{ U(R') } 
 \nonumber \\
  \rho_2(R)  && = e^{ -U(R)}   \int_{-\infty}^{R} d R' \left[  \frac{ G_2(R) \rho^{neq}_{st}(R) + G_1(R) \rho_1(R)-   j_2   }{D} \right]  e^{ U(R') } 
 \nonumber \\
  \rho_3(R)  && = e^{ -U(R)}   \int_{-\infty}^{R} d R'  \left[\frac{ G_3(R) \rho^{neq}_{st}(R) + G_2(R) \rho_1(R)+ G_1(R) \rho_2(R)-   j_3   }{D} \right] e^{ U(R') } 
 \nonumber \\
  \rho_4(R)  && = e^{ -U(R)}   \int_{-\infty}^{R} d R' \left[ \frac{ G_4(R) \rho^{neq}_{st}(R) + G_3(R) \rho_1(R)+ G_2(R) \rho_2(R)+ G_1(R) \rho_3(R)-   j_4   }{D} \right]  e^{ U(R') } 
 \label{noneqsolRegmh}
\end{eqnarray}
In terms of the function of Eq. \ref{auxiliaire}
\begin{eqnarray}
\Pi(R') && \equiv  (-j_{st}) \frac{ e^{ U(R') } }{ D(R') } \int_{R'}^{+\infty} dR e^{ -U(R)} 
   \label{auxiliaireh}
\end{eqnarray}
the current coefficients read
\begin{eqnarray}
j_1 && =\int_{-\infty}^{+\infty}d R \left[G_1(R) \rho^{neq}_{st}(R)    \right]\Pi(R) 
  \nonumber \\
  j_2 && =\int_{-\infty}^{+\infty}d R \left[G_2(R) \rho^{neq}_{st}(R) + G_1(R) \rho_1(R) \right]\Pi(R) 
   \nonumber \\
 j_3 && =\int_{-\infty}^{+\infty}d R' \left[  G_3(R) \rho^{neq}_{st}(R) + G_2(R) \rho_1(R)+ G_1(R) \rho_2(R)- \right]\Pi(R) 
   \nonumber \\
 j_4 && =\int_{-\infty}^{+\infty}d R \left[ G_4(R) \rho^{neq}_{st}(R) + G_3(R) \rho_1(R)+ G_2(R) \rho_2(R)+ G_1(R) \rho_3(R)  \right]\Pi(R) 
   \label{solujmh}
\end{eqnarray}
So here the iterative procedure goes as follows : one first compute $j_1$ to plug it into
into Eq. \ref{noneqsolRegmh} to obtain $\rho_1(R)$ that can be then plugged into Eq. \ref{solujmh} to compute $j_2$,
that can be then plugged into Eq \ref{noneqsolRegmh} to compute $\rho_2(R)$, and so on.


\section{Explicit first cumulants for the case of a Riccati equilibrium steady-state }

\label{sec_pereq}

In this Appendix, we focus on the case where the potential $U(R)$ introduced in Eq. \ref{UR} 
is periodic on the Riccati ring (see Eq. \ref{uperioreg}) and
where
the steady state $\rho_{st}$ is thus an equilibrium steady state $\rho^{eq}_{st}$ without current $j_{st}=0$
(see subsection \ref{sec_eqst}). We describe the perturbative solution in $k$
of the optimization procedure of subsection \ref{sec_optieq}.

The general solutions of Eq. \ref{eulergper} for the functions $G_m(R)$
that appear in the series expansion 
 of Eq. \ref{gmuk} for $G(R)$
 involves some integration constant $K_m$
\begin{eqnarray}
 G_m(R) 
 =  e^{ U(R)} 
 \left[ K_m +  \int_{R_{min}}^{R} dR' \Omega_m(R')  e^{ - U(R') }   \right] 
\label{solReggeneGm}
\end{eqnarray}
The requirement of periodicity on the Riccati ring
$  G_m(R_{min})=G_m(R_{max})$ leads to the equation
\begin{eqnarray}
   e^{ U(R_{min})} K_m =
e^{ U(R_{max})} 
 \left[ K_m +  \int_{R_{min}}^{R_{max}} dR' \Omega_m(R')  e^{ -U(R') }   \right] 
\label{solReggeneperioGm}
\end{eqnarray}
i.e. when one takes into account the periodicity of the potential $ U(R_{min})= U(R_{max})$,
one obtains that $K_m$ disappears and one is left with the conditions
that involve only the inhomogeneous terms $ \Omega_m(R)$ of Eqs \ref{eulergper}
\begin{eqnarray}
 0 =  \int_{R_{min}}^{R_{max}} dR' \Omega_m(R')  e^{ -U(R') }   
\label{solReggeneperioGO}
\end{eqnarray}
In terms of the equilibrium steady state of Eq. \ref{steadyeq},
Eq. \ref{solReggeneperio}
can be rewritten in the limit $R_{min} \to -\infty$ and $R_{max} \to +\infty$ as
\begin{eqnarray}
 0 =  \int_{-\infty}^{+\infty} dR \Omega_m(R) \rho_{st}^{eq}(R)
\label{solReggeneperiosteq}
\end{eqnarray}
More explicitly for $m=1,2,3,4$ when one takes into account the explicit expressions of the
$ \Omega_m(R)$ of Eqs \ref{eulergper} and the normalization of $\rho_{st}^{eq}(R)
 $, one obtains that the coefficients $\mu_m$ of the perturbative
 expansion of Eq. \ref{gmuk} for $\mu(k)$ 
 are given by
 \begin{eqnarray}
  \mu_1 && =  \int_{-\infty}^{+\infty} dR \alpha(R)    \rho_{st}^{eq}(R)  = \lambda^{typ}
  \nonumber \\
   \mu_2  && = \int_{-\infty}^{+\infty} dR \frac{     G_1^2(R)    }{4D(R)}   \rho_{st}^{eq}(R)
     \nonumber \\
     \mu_3  && =  \int_{-\infty}^{+\infty} dR\frac{     G_1(R)  G_2(R)  }{2D(R)}   \rho_{st}^{eq}(R)
  \nonumber \\
     \mu_4  && =  \int_{-\infty}^{+\infty} dR \left[ \frac{     G_1(R)  G_3(R)  }{2D(R)} - \frac{     G_2^2(R)    }{4D(R)}  \right] \rho_{st}^{eq}(R)
  \label{eulergpereq}
\end{eqnarray}
In order to determine the integration constants $K_m$ for the $G_m(R)$ of Eq. \ref{solReggeneGm},
one needs to consider the constraints of Eq. \ref{UReqG} in their regularized form involving $R_{min}$ and $R_{max}$
\begin{eqnarray}
0 && =  \int_{R_{min}} ^{R_{max}} dR \frac{G_m(R)}{D(R)} 
=  \int_{R_{min}} ^{R_{max}} dR \frac{e^{ U(R)}  }{D(R)} 
 \left[ K_m +  \int_{R_{min}}^{R} dR' \Omega_m(R')  e^{ -U(R') }   \right] 
 \nonumber \\
 && = K_m \int_{R_{min}} ^{R_{max}} dR \frac{e^{ U(R)}  }{D(R)} 
 + 
\int_{R_{min}} ^{R_{max}} dR \frac{e^{ U(R)}  }{D(R)} 
  \int_{R_{min}}^{R} dR' \Omega_m(R')  e^{ -U(R') } 
 \label{UReqGm}
\end{eqnarray}
Plugging this value of $K_m$ into Eq. \ref{solReggene}, one obtains 
the final result for the regularized solution $G_m(R)$
\begin{eqnarray}
 G_m(R) 
 =  e^{ U(R)} \frac{
 \int_{R_{min}} ^{R_{max}}  \frac{ dR''}{D(R'')}e^{ U(R'')} 
  \int_{R''}^{R} dR' \Omega_m(R')  e^{ - U(R') }    
 }{\int_{R_{min}} ^{R_{max}}   \frac{ dR''}{D(R'')}e^{ U(R'')} }  
   \label{solReggenefin}
\end{eqnarray}
More explicitly for $m=1,2,3,4$ when one takes into account the explicit expressions of the
$ \Omega_m(R)$ of Eqs \ref{eulergper}
\begin{eqnarray}
 G_1(R) 
&& =  e^{ U(R)} \frac{
 \int_{R_{min}} ^{R_{max}}  \frac{ dR''}{D(R'')}e^{ U(R'')} 
  \int_{R''}^{R} dR'  \left[  2 \mu_1 - 2 \alpha(R')  \right] e^{ - U(R') }    
 }{\int_{R_{min}} ^{R_{max}}   \frac{ dR''}{D(R'')}e^{ U(R'')} }  
  \nonumber \\
 G_2(R) 
&& =  e^{ U(R)} \frac{
 \int_{R_{min}} ^{R_{max}}  \frac{ dR''}{D(R'')}e^{ U(R'')} 
  \int_{R''}^{R} dR'   \left[     2 \mu_2  - \frac{     G_1^2(R')    }{2D(R')} \right]  e^{ - U(R') }    
 }{\int_{R_{min}} ^{R_{max}}   \frac{ dR''}{D(R'')}e^{ U(R'')} }  
  \nonumber \\
   G_3(R) 
&& =  e^{ U(R)} \frac{
 \int_{R_{min}} ^{R_{max}}  \frac{ dR''}{D(R'')}e^{ U(R'')} 
  \int_{R''}^{R} dR'   \left[      2 \mu_3  - \frac{     G_1(R')  G_2(R')  }{D(R')}  \right] e^{ - U(R') }    
 }{\int_{R_{min}} ^{R_{max}}   \frac{ dR''}{D(R'')}e^{ U(R'')} }  
    \nonumber \\
     G_4(R) 
&& =  e^{ U(R)} \frac{
 \int_{R_{min}} ^{R_{max}}  \frac{ dR''}{D(R'')}e^{ U(R'')} 
  \int_{R''}^{R} dR'  \left[       2 \mu_4  - \frac{     G_1(R')  G_3(R')  }{D(R')} - \frac{     G_2^2(R')    }{2D(R')}  \right] e^{ - U(R') }    
 }{\int_{R_{min}} ^{R_{max}}   \frac{ dR''}{D(R'')}e^{ U(R'')} }  
  \label{eulergperg}
\end{eqnarray}

So here the iterative procedure goes as follows : one plugs $\mu_1=\lambda^{typ}$ into
into Eq. \ref{eulergperg} to obtain $G_1(R)$ that can be then plugged into Eq. \ref{eulergpereq} to compute $\mu_2$,
that can be then plugged into Eq \ref{eulergperg} to compute $G_2(R)$, and so on.


\end{document}